\documentclass[10pt]{iopart}
\pdfoutput=1
\usepackage{fullpage}
\usepackage{iopams,graphicx,color}
\usepackage{setstack}
\usepackage[colorlinks=true]{hyperref} 
\hypersetup{urlcolor=blue,linkcolor=blue,citecolor=blue,colorlinks=true}

\usepackage{color}
\usepackage{pgf,tikz}
\usepackage{verbatim}
\usepackage{marginnote}
\usepackage{enumitem}

\usetikzlibrary{fadings,decorations.pathmorphing}

\newcommand{\ccvec}{\left(\begin{array}{c}c^\dag(k)\\ c^\dag(k+\pi)\end{array}\right)}
\newcommand{\cpcmvec}{\left(\begin{array}{c}c_{+}^\dag(k)\\ c_{-}^\dag(k)\end{array}\right)}
\newcommand{\twomat}[4]{\left(\begin{array}{cc} #1 & #2 \\ #3 & #4 \end{array}\right)}

\newcommand{\Hl}[1]{{#1}_l}
\newcommand{\Hr}[1]{{#1}_r}

\usepackage{braket}

\begin{document}
 \title{Inhomogeneous field theory inside the arctic circle}

\author{Nicolas Allegra$^1$, J\'er\^ome Dubail$^1$, Jean-Marie St\'ephan$^2$ and Jacopo Viti$^2$}

\address{
$^1$ CNRS \& IJL-UMR 7198, Universit\'e de Lorraine, BP 70239, F-54506 Vand{\oe}uvre-l\`es-Nancy Cedex, France \\
$^2$Max Planck Institut f\"ur Physik Komplexer Systeme, N\"othnitzer Str. 38 D01187 Dresden, Germany}

 \eads{\mailto{jstephan@pks.mpg.de}}
 
\begin{abstract}
Motivated by quantum quenches in spin chains, a one-dimensional toy-model of fermionic particles evolving in imaginary-time from a domain-wall initial state is solved. The main interest of this toy-model is that it exhibits the {\it arctic circle} phenomenon, namely a spatial phase separation between a critically fluctuating region and a frozen region. Large-scale correlations inside the critical region are expressed in terms of correlators in a (euclidean) two-dimensional massless Dirac field theory. It is observed that this theory is inhomogenous: the metric is position-dependent, so it is in fact a Dirac theory in curved space. The technique used to solve the toy-model is then extended to deal with the transfer matrices of other models: dimers on the honeycomb and square lattice, and the six-vertex model at the free fermion point ($\Delta=0$). In all cases, explicit expressions are given for the long-range correlations in the critical region, as well as for the underlying Dirac action. Although the setup developed here is heavily based on fermionic observables, the results can be translated into the language of height configurations and of the gaussian free field, via bosonization. Correlations close to the phase boundary and the generic appearance of Airy processes in all these models are also briefly revisited in the appendix.
\end{abstract}
 \date{\today}
\maketitle

\tableofcontents
\newpage

\section{Introduction and overview}
\label{sec:intro}

The {\it arctic circle} is the name given by Jockush, Propp \& Shor to a phenomenon they discovered in 1995 \cite{jockusch1998random} while studying dimer coverings on the so-called Aztec diamond \cite{elkies1992alternating, elkies1992alternating2}. The phenomenon consists in the appearance, in the thermodynamic limit, of a curve that separates two macroscopic domains, one in which the dimers appear to be frozen, meaning that one knows their exact position with a probability that goes quickly to one in the thermodynamic limit, and another domain where the dimers fluctuate. The same phenomenon was observed in other models \cite{propp_boxed}, in particular in the six-vertex model with domain-wall boundary conditions \cite{korepin_zinnjustin,palamarchuk20106,colomo2008arctic,colomo2010arctic}, a system that already possessed a long history  connected to integrability and to the evaluation of the norm of Bethe states \cite{korepin1982calculation,izergin1987partition,izergin1992determinant}.  In fact, the phenomenon itself appears to have been known as early as 1977, in a different context though: the statistics of large Young diagrams \cite{kerov_vershik,logan1977variational}. It was also well studied in the early eighties in the context of the physics of crystal growth, see for instance \cite{rottman1984statistical, nienhuis1984triangular}.

Since the late nineties, there has been an intense activity in this area, which lies at the frontier between physics, mathematics and computer science. Among the many results that were collected are the limiting distribution of dimers around the origin in the thermodynamic limit \cite{cohn1996local,johansson2005}, the fluctuations of the boundary described by the Airy process \cite{johansson2005}, a general theory for dimer models \cite{kenyon2009lectures} and the calculation of the corresponding arctic curves in connection with algebraic geometry \cite{kenyon_okounkov}, calculations of various correlation functions \cite{bogoliubov2002boundary,OkounkovReshetikhin,okounkov2007random}, steps towards extension to interacting models \cite{korepin_zinnjustin,colomo2010arctic,colomo2005two} ({\it i.e.} models that cannot be mapped to free fermions; for example: the six-vertex model at $\Delta \neq 0$, or dimers with aligning interactions \cite{alet2005interacting}), and numerical investigations \cite{Zvonarev_arctic,allison2005numerical,Cugliandolo_arctic}. There are also connections with the physics of glassy systems \cite{perret2012super}.

\vspace{0.4cm}
Our motivation for revisiting this ancient problem comes from the physics of quantum quenches in spin chains, and in particular their description by two-dimensional conformal field theory (CFT). This topic, launched by Calabrese \& Cardy in 2005/2006 \cite{calabrese2005time,calabrese2006time}, received a lot of attention in recent years \cite{calabrese2007quantum,calabrese2007entanglement,sotiriadis2008inhomogeneous,sotiriadis2009quantum,sotiriadis2010quantum,gambassi2011quantum,dubail2011universal,stephan2011local,asplund2011evolution,StephanDubail2013,asplund2014mutual,kennes2014universal,kuns2014non,cardy2015quantum,BernardDoyon_hydro}. The basic idea is the following: the usual quantum/classical correspondence tells us that, modulo Wick rotation, the real-time evolution of a one-dimensional quantum model is equivalent to a two-dimensional classical statistical model. In that setup, the initial state of the quantum system becomes a boundary condition in the classical model. Then the game is to chose the quantum system and the initial state such that (i) the bulk is critical, which is achieved when the Hamiltonian of the spin chain is gapless and (ii) the boundary condition flows (in the RG sense) towards a conformal boundary condition (see appendix \ref{appx:real} for some details). The second property is crucial and underlies most of the published works in that area. Fortunately, conformal boundary conditions are the rule rather than the exception: it is usually argued that, as soon as the correlations in the initial state are short range, the initial state corresponds to a conformal boundary condition in the thermodynamic limit (perturbed by irrelevant operators, to complete the picture). Good discussions of these ideas exist in the literature, see for instance \cite{gambassi2011quantum}.

\vspace{0.4cm}

What is the relation between the arctic circle and quantum quenches in spin chains ? Hopefully,
this will become clear to the reader already in the next page. There, we introduce a quantum quench model which,
we argue, may be regarded as a toy-model for arctic circle problems.
The crucial ingredients are the ${\rm U}(1)$-symmetry, or conservation of the number of particles,
together with the {\it domain-wall initial state} (DWIS), see below and Fig. \ref{fig:param}. 
Because of particle conservation, the standard argument about the initial state renormalizing to a conformal
boundary condition {\it does not apply} (see also the discussion in \cite{stephan2014emptiness}).
This is in very sharp contrast with what happens, for instance, in the quantum Ising chain, which does not
possess a ${\rm U}(1)$ symmetry (the magnetization is {\it not} conserved), and where the boundary RG flow argument can be
safely exploited \cite{calabrese2008time,DV_interface}.

\vspace{0.4cm}

The question that motivates the rest of this introduction is: how does
one apply the CFT framework of \cite{calabrese2006time} to the quantum quench
in the XX chain studied previously in \cite{antal1999transport}?
We will soon see that the arctic circle shows up naturally in
the process of answering that question. The purpose of the rest of the paper is then to revisit other {\it arctic circle problems} (dimer models and the six-vertex model) from that perspective.

Finally, let us emphasize that the model we deal with in the rest of this introduction can be mapped exactly on the {\it PNG droplet model} that appeared previously in the literature \cite{prahofer2000universal, prahofer2002scale, spohn2006exact}---this was pointed out to us by H. Spohn while we were completing this paper---. We shall not use this terminology here, because our own motivation really comes from quantum quenches in spin chains, so we prefer to view the model as the XX spin chain (namely the XXZ chain at $\Delta=0$). Also, although there is some rather large overlap between this introduction and the results of \cite{prahofer2002scale}, we think that our goal and our point of view are sufficiently different from the ones of \cite{prahofer2000universal, prahofer2002scale, spohn2006exact} so that they are worth explaining here independently.

\subsection*{A toy-model for arctic-circle phenomena: the XX chain in imaginary-time}

The model is a translation-invariant chain of free fermions living on lattice sites $x \in \mathbb{Z}+ \frac{1}{2}$. The Hamiltonian is diagonal in $k$-space, with a single band and a dispersion relation $\varepsilon(k)$,
\begin{equation}
	\label{eq:Ham}
	H \, = \, \int_{-\pi}^\pi \frac{dk}{2\pi} \, \varepsilon(k) \, c^\dagger(k) c(k) \, .
\end{equation}
The fermion creation/annihilation modes obey the canonical anti-commutation relations $\{ c^\dagger(k) ,c(k')\} \, =\, 2\pi \, \delta(k-k')$ in $k$-space, or $\{ c^\dagger_x, c_{x'} \} = \delta_{x,x'}$ in real space. Our convention for Fourier transforms is $c^\dagger(k) \, = \, \sum_{x \in \mathbb{Z} + \frac{1}{2}} e^{i k x} c_x^\dagger$; notice that $c^\dagger(k+2\pi) = -c^\dagger (k)$. For simplicity, we specialize to the case of nearest-neighbor hopping only, such that the dispersion relation may be put in the form\footnote{possibly after a gauge transformation, in case of complex hopping amplitudes.}
\begin{equation*}
	\varepsilon(k) \, = \, - \cos k \,.
\end{equation*}
In other words, the Hamiltonian is the one of the XX chain, after a Jordan-Wigner transformation. We have normalized $\varepsilon(k)$ such that the maximal group velocity $v(k) \, = \, \frac{d}{dk} \varepsilon (k)$ is $1$ (at $k = \frac{\pi}{2}$).
\begin{figure}[!ht]
\center{
	\begin{tikzpicture}[scale=0.4]
		\draw (-8,2) node{{\bf Filled sites}};
		\draw (8,2) node{{\bf Empty sites}};
		\draw[thick] (-15,0) -- (15,0);
		\draw[dashed] (-17,0) -- (-15,0);
		\draw[dashed] (17,0) -- (15,0);
		\foreach \x in {-15,-13,...,-1} \filldraw (\x,0) circle (0.4cm);
		\foreach \x in {15,13,...,1} \filldraw[fill=white] (\x,0) circle (0.4cm);
		\draw[->] (-15,-1) -- (15,-1) node[below]{$x$};
		\draw (0,-0.7) -- ++(0,-0.6) ++(0,-0.2) node[below]{$0$};
		\draw (-1,-0.7) -- ++(0,-0.6) node[below]{$-\frac{1}{2}$};
		\draw (1,-0.7) -- ++(0,-0.6) node[below]{$\frac{1}{2}$};
		\draw (-3,-0.7) -- ++(0,-0.6) node[below]{$-\frac{3}{2}$};
		\draw (3,-0.7) -- ++(0,-0.6) node[below]{$\frac{3}{2}$};
		\draw (-5,-0.7) -- ++(0,-0.6) node[below]{$-\frac{5}{2}$};
		\draw (5,-0.7) -- ++(0,-0.6) node[below]{$\frac{5}{2}$};
	\end{tikzpicture}}
	\caption{The toy-model is a chain of free fermions hopping on the lattice $x \in \mathbb{Z} + \frac{1}{2}$, evolving from the DWIS, which is completely filled on the left, and completely empty on the right.}
	\label{fig:param}
\end{figure}
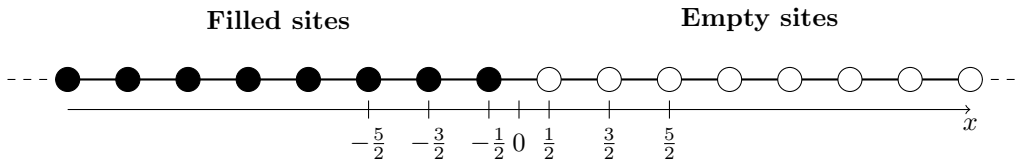%
We focus on the evolution of the fermions from a DWIS, see Fig. \ref{fig:param}. The DWIS $\ket{\Psi_0}$ is completely filled on the left (density equal to one), and completely empty on the right (density equal to zero):
\begin{equation}
	\label{eq:definition_DWIS}
	\rho_x \ket{\Psi_0} \, = \, c^\dagger_x c_x \ket{\Psi_0} \, = \, \Theta(-x) \ket{\Psi_0} \, .
\end{equation}
$\Theta$ is the Heaviside step function: $\Theta(-x) =1$ if $x<0$ and $\Theta(-x)=0$ if $x>0$. On top of this $\ket{\Psi_0}$ is normalized: $\left< \Psi_0 \left| \Psi_0 \right> \right. = 1$. This completely determines $\ket{\Psi_0}$, up to an irrelevant phase factor.
The main originality of the model, compared to previous work on the DWIS in the XX chain, is that we focus on imaginary-time evolution (the real time dynamics is discussed in \ref{appx:real}). We want to compute correlators of local observables $\mathcal{O}(x,y)$, defined as the (imaginary-)time-ordered expectation values 
\begin{eqnarray*}\fl
\nonumber	\left< \mathcal{O}_1(x_1,y_1) \dots \mathcal{O}_n(x_1,y_1)  \right> & \equiv & \frac{ \bra{\Psi_0} e^{-2R \, H} \, \mathcal{T} \left[ \left( e^{(R+y_1) H} \mathcal{O}_1 (x_1) e^{-(R+y_1)H}  \right) \dots  \left( e^{(R+y_n) H} \mathcal{O}_n (x_n) e^{-(R+y_n)H}  \right)  \right]\ket{\Psi_0} }{ Z }, 
\end{eqnarray*}
where $\mathcal{T}\left[ . \right]$ stands for time-ordering (the factors must appear from left to right with decreasing values of $y$). $Z$ is the partition function of the model, it ensures  $\braket{1}=1$. For our toy model it is given by the exact formula
\begin{equation}\label{eq:exactZ}
	Z \, = \, \bra{\Psi_0} e^{-2R \,H} \ket{\Psi_0} \,=e^{R^2/2} . 
\end{equation}
A proof of (\ref{eq:exactZ}) is given in \ref{sec:partitionsfunctions}. 

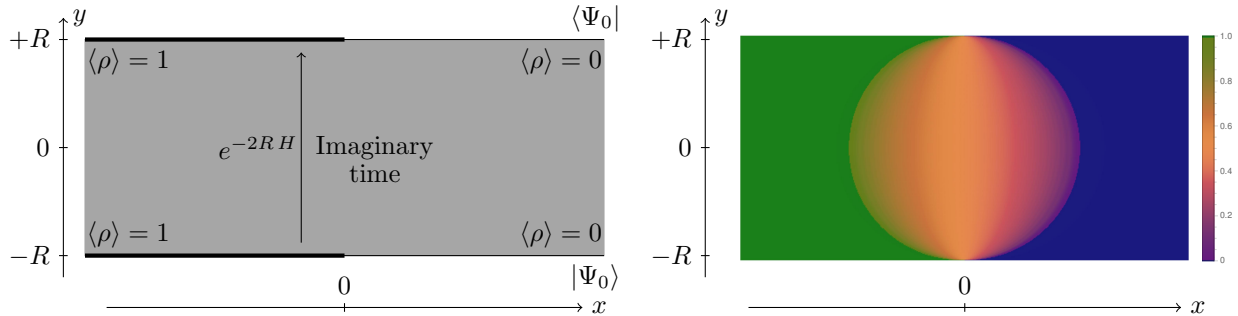
\begin{figure}[!ht]
\center{
\begin{tikzpicture}[scale=0.575]
	\filldraw[gray!70] (-6,-2.5) rectangle (6,2.5);
	\draw (-6,-2.5) -- (6,-2.5);
	\draw (-6,2.5) -- (6,2.5);
	\draw[ultra thick] (-6,2.5) -- (0,2.5);
	\draw[ultra thick] (-6,-2.5) -- (0,-2.5);
	\draw (5,-2) node {$\braket{\rho} = 0$};
	\draw (5,2) node {$\braket{\rho} = 0$};
	\draw (-5,2) node {$\braket{\rho} = 1$};
	\draw (-5,-2) node {$\braket{\rho} = 1$};
	\draw (5.8,3) node {$\bra{\Psi_0}$};
	\draw (5.8,-3) node {$\ket{\Psi_0}$};
	\draw[->] (-1,-2.2) -- (-1,2.2);
	\draw (0.7,0) node {Imaginary};
	\draw (0.7,-0.6) node {time};
	\draw (-2,0) node {$e^{-2R \, H}$};
	\draw[->] (-6.5,-3) -- (-6.5,3) node[right]{$y$};
	\draw (-6.6,2.5) node[left]{$+R$} -- (-6.4,2.5);
	\draw (-6.6,-2.5) node[left]{$-R$} -- (-6.4,-2.5);
	\draw (-6.6,0) node[left]{$0$} -- (-6.4,0);
	\draw[->] (-5.5,-3.7) -- (5.5,-3.7) node[right]{$x$};
	\draw (0,-3.8) -- (0,-3.6) node[above]{$0$};
\end{tikzpicture}
	\hfill
	\begin{tikzpicture}[scale=0.575]
\node at (0,0) {\includegraphics[height=3.1cm]{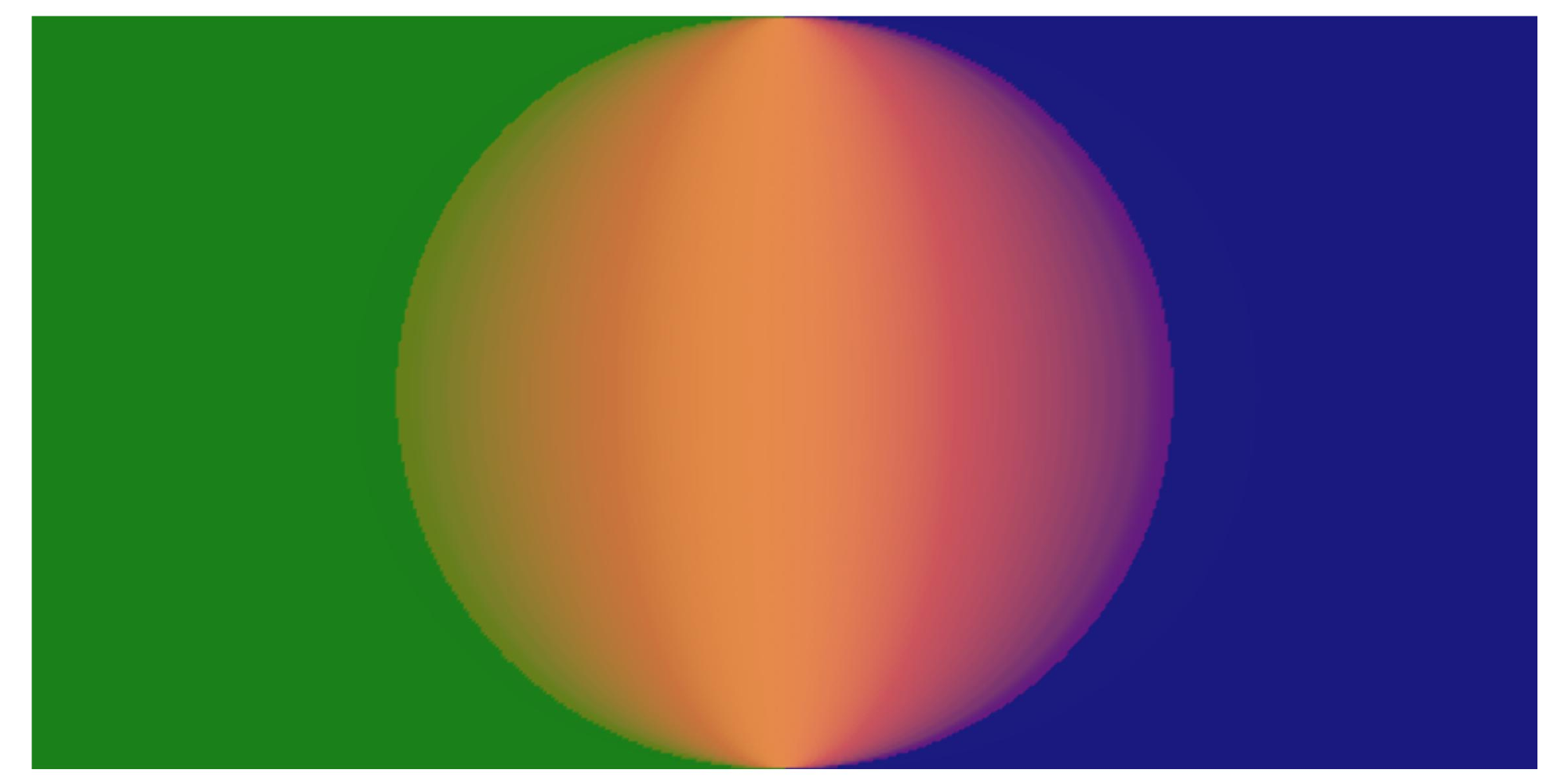}};
\node at (3.9*2-2,-0.19) {\includegraphics[height=3.35cm]{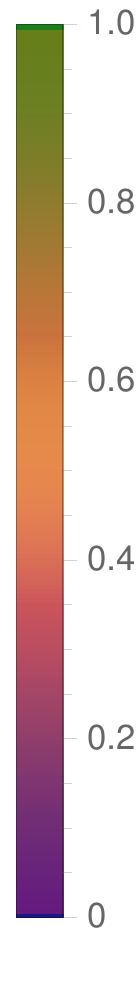}};
	\draw[->] (-6,-3) -- (-6,3) node[right]{$y$};
	\draw (-6.1,2.5) node[left]{$+R$} -- (-5.9,2.5);
	\draw (-6.1,-2.5) node[left]{$-R$} -- (-5.9,-2.5);
	\draw (-6.1,0) node[left]{$0$} -- (-5.9,0);
	\draw[->] (-5,-3.7) -- (5,-3.7) node[right]{$x$};
	\draw (0,-3.8) -- (0,-3.6) node[above]{$0$};
\end{tikzpicture}
}
	\caption{(Left) Cartoon of the toy-model: the system is a strip of width $2R$, with boundary conditions corresponding to the DWIS. (Right) Density profile in the toy-model, measured numerically for $R = 128$. 
	}
	\label{fig:param2}
\end{figure}

The constant $R>0$ is a parameter that sets the effective size of the system. Indeed, although the chain is infinite in the $x$-direction, it has a width $2R$ in the (imaginary-time) $y-$direction. We have the following cartoon in mind (see Fig. \ref{fig:param2}). One starts from the initial state $\ket{\Psi_0}$ at imaginary time $y = -R$ and one lets the system evolve freely up to imaginary time $y = +R$, where it is conditioned to come back to the initial state $\ket{\Psi_0}$. 
At imaginary time $y = - R$ the degrees of freedom are frozen for any $x$, because all the sites are occupied on the left and empty on the right, so all correlations are trivial. Then the particles are released and can hop between left and right. So, as soon as $y >-R$, a small region around $x=0$ must appear, where the average density is between $0$ and $1$. In that region, there are density fluctuations, with non-trivial correlations. At $y = +R$, the system is conditioned to come back to its initial state, so we expect the fluctuating region to shrink back to a point as $y$ increases and approaches $+R$. In fact, it is easy to see that the model is symmetric under $y \rightarrow -y$, so the fluctuating region is reflection-invariant across the horizontal axis. 

Our model possesses the same phenomenology as the arctic circle problem in its original formulation for the Aztec diamond dimer model \cite{jockusch1998random}. This is consistent with the scaling of (\ref{eq:exactZ}). 
Indeed, from general statistical mechanical arguments, the partition function is expected to scale as the exponential of the fluctuating area, which is proportional to $R^2$. The connection with dimers, and with other models, will be developed further in the main parts of the paper. In this introduction, our purpose is to use this toy-model as an illustration of the approach that will be followed later, in the main parts. As can be seen in Fig. \ref{fig:param2}, in the scaling limit $R \rightarrow \infty$, the exact shape of the fluctuating region is a disc of radius $R$ centered at the origin. Let us be more precise about what we mean by {\it scaling limit} here and in the rest of the paper. We are interested in expectation values and correlation of observables of the form $\left< \mathcal{O}_1(x_1,y_1) \dots \mathcal{O}_n(x_n,y_n) \right>$ in the {\it scaling limit}
\begin{equation}
	\label{eq:scaling_regime}
	x_j/R \quad {\rm and } \quad y_j/R \quad {\rm fixed \; for \; any \; }j \, , \qquad \quad{\rm and} \quad R \rightarrow \infty \, .
\end{equation}
In this limit, the density inside the arctic circle takes a very simple form, interpolating from $\left< \rho \right> = 1$ on the left to $\left< \rho \right> = 0$ on the right (this formula is derived below):
\begin{equation*}
	(x^2+y^2 < R^2)\qquad \quad \left< \rho_{x,y} \right> \, = \, \frac{1}{\pi} {\rm arccos} \frac{x}{\sqrt{R^2 - y^2}} \, .
\end{equation*}
{\bf The main focus of this paper, however, is neither on the shape of the critically fluctuating region, nor on the density profile inside it. Instead, what we are really eager to understand is the field theory inside the disc}. What do we mean by that ? Simply the fact that the critically fluctuating region is, by definition, the region where one finds observables with long-range (connected) correlations. Such correlations must be describable by a euclidean quantum field theory with a local action; moreover this theory should be massless. In fact, there presumably is a theory that describes both the interior and the exterior of the disc, with a non-zero mass term only outside the disc. But, in the scaling limit (\ref{eq:scaling_regime}), any such mass term with $m>0$  is automatically renormalized to $m = +\infty$, so the theory outside the disc is trivial in that regime. The theory inside the disc, on the contrary, is non-trivial in the scaling limit; our goal is to find out what it is.

\vspace{0.5cm}

In fact, it is rather easy to guess what the field theory should be. Since we are dealing with a quadratic Hamiltonian for fermionic particles, we are looking for a free fermion theory. Moreover, a key aspect of the model is the particle conservation, meaning that it is invariant under ${\rm U}(1)$ transformations $c^\dagger \rightarrow e^{i \varphi} c^\dagger$, $c \rightarrow e^{-i \varphi} c$. The field theory must inherit this symmetry from the lattice model. So, there is not much choice, it has to be a massless Dirac theory in two dimensions, or possibly a direct product of several of those (meaning a theory with more than one species of massless fermions). The basic object in any such theory is the two-component Dirac spinor, $\Psi = \left( \begin{array}{c}  \psi \\ \overline{\psi} \end{array} \right)$, $\Psi^\dagger = \left( \begin{array}{cc}  \psi^\dagger & \overline{\psi}^\dagger \end{array} \right)$. We pick our conventions for the (euclidean) gamma matrices as $\gamma^1 \, = \, -\sigma^2$ and $\gamma^2 \, = \, \sigma^1$ (the $\sigma^j$'s are the Pauli matrices), such that $\gamma^5 \, = \, i \gamma^2 \gamma^1 \, = \, \sigma^3$ is diagonal and refer to $\psi$ and $\overline{\psi}$ as the chiral and anti-chiral components. The latter are the local fields in the continuous theory. At a given point, one must be able to express the lattice degrees of freedom in terms of the continuous fields. Such relations hold in the sense that lattice correlators are equal, in the scaling limit, to field theory correlators of carefully chosen fields at the same point. There is always some freedom in the choice of the relation between the lattice and the continuous degrees of freedom, because local perturbations may be absorbed either in this relation, or in the action, in the form of perturbations by irrelevant operators. Here, what we want to do is to fix this relation once and for all, and then work out the action of the theory.
The lattice fermion $c^\dagger_{x,y}$ must be a linear combination of $\psi^\dagger$ and $\overline{\psi}^\dagger$, so let us fix
\begin{equation}
	\label{eq:lattice_continuous_XX}
	\left\{ \begin{array}{rcl} \displaystyle c^\dagger_{x,y} &=& \displaystyle \frac{1}{\sqrt{2\pi}}\, \psi^\dagger(x,y) \, + \, \frac{1}{\sqrt{2\pi}} \, \overline{\psi}^\dagger(x,y) \\ 
	\displaystyle c_{x,y} &=& \displaystyle \frac{1}{\sqrt{2\pi}} \, \psi(x,y) \, + \, \frac{1}{\sqrt{2\pi}} \, \overline{\psi}(x,y) \, . \end{array} \right. 
\end{equation}
The normalization factors $1/\sqrt{2\pi}$ are introduced for later convenience; global phase factors may always be absorbed in the definition of the continuous fields. The point is that we chose the expression of the lattice degree of freedom in terms of the continuous fields such that it is the same everywhere, there are no position-dependent phase or scale factors. The question is now: what is the action ? The system is clearly not homogeneous---because the density is position-dependent---so there is no reason for the field theory to be homogeneous. Therefore, we are looking for some massless Dirac action in which the parameters vary with position. But what parameter is there to vary in the Dirac action, if not the mass ?

\vspace{0.5cm}

The simplest possible answer is: the metric. After all, it makes sense that the metric of a field theory describing a non-homogeneous system could vary. Also, there might be some background gauge potentials. Thus, we are led to contemplate the Dirac action in fully covariant form, allowing the possibility of both vector (v) and axial (a) background gauge potentials
\begin{equation}
	\label{eq:Dirac}
	S \, = \, \frac{1}{2\pi} \int d^2 {\rm x}  \sqrt{ g}   \, e^{\mu}_{\phantom{a}a} \left[  \overline{\Psi} \gamma^a \left( \frac{1}{2} \overset{\leftrightarrow}{\partial_\mu} \, + \, i A^{({\rm v})}_\mu \, + \, i A^{({\rm a})}_\mu \gamma^5  \right)  \Psi \right] \, .
\end{equation}

Here, the 'Dirac adjoint' $\overline{\Psi}$ is $\Psi^\dagger \gamma^2$, $e^\mu_{\phantom{a}a}({\rm x})$ is the tetrad, and $ (d^2 {\rm x} \sqrt{ g}) =  (d^2{\rm x}\, |e|^{-1})$ is the volume element. The spin connection drops out of the two-dimensional Dirac action ({\it cf.} \cite{nakahara2003geometry} for details). This action (\ref{eq:Dirac}) is {\it a priori} the most general possibility for the field theory describing the long-range correlations inside the critically fluctuating region (apart from the possibility of having more species of massless fermions, as already mentioned). The goal is now to properly identify the tetrads and the background gauge potentials in our toy-model. Before we proceed to this identification, let us stress the important difference between the vector and the axial potentials: the ${\rm U}(1)$-symmetry of the lattice model ({\it i.e.} particle number conservation) is a true symmetry. These ${\rm U}(1)$ gauge transformations act on the vector part, not the axial one. Physical observables must be gauge-invariant. In fact, as we will see, both the vector and axial parts are pure gauge: $A_\mu^{({\rm v})} \, = \, \partial_\mu f$ for some function $f$ (and the same for $A_\mu^{({\rm a})}$ with a different function), so the vector part can simply be gauged away, and $A_\mu^{({\rm v})}$ can really be dropped. It can never appear in physical observables. The situation is different for the axial part, as {\it it does not correspond to a symmetry of the model}. 
There is no axial symmetry on the lattice; it happens to be a symmetry of the continuous Dirac Lagrangian, but it is well-known that the 'symmetry' is not preserved by any UV regularization procedure required to define the path integral: the 'symmetry' is anomalous. Stated in a more pedestrian way: the two gauge potentials $A_\mu^{({\rm v})}$ and  $A_\mu^{({\rm a})}$ are not on the same footing. Physical observables must not depend on $A_\mu^{({\rm v})}$, but they can (and they do) depend on $A_\mu^{({\rm a})}$.

\subsection*{Stationary phase and long-range correlations in the toy-model}

The parameters in the action would be easily obtained if we knew the long-range behavior of fermionic correlators $\left<   c^\dagger_{x_1,y_1} \dots c^\dagger_{x_n,y_n}  c_{x'_1,y'_1} \dots c_{x'_n,y'_n} \right>$ in the strip of width $2R$. As we shall see in part \ref{sec:ktr}, these correlators can be computed exactly. They turn out to be directly related to the correlators in the DWIS, through a rather simple linear transformation,
\begin{eqnarray}\fl
	\label{eq:correl_XX}
\nonumber&	 \left<   c^\dagger_{x_1,y_1} \dots c^\dagger_{x_n,y_n}  c_{x'_1,y'_1} \dots c_{x'_n,y'_n} \right>  \\\fl
	 &= \, \int  \prod_j \frac{dk_j}{2\pi} e^{-i k_j x_j + y_j \varepsilon(k_j) - i  R \tilde{\varepsilon}(k_j)}  \prod_p \frac{dk'_p}{2\pi} e^{i k'_p x'_p - y'_p \varepsilon(k'_p) + i  R \tilde{\varepsilon}(k'_p)}  \bra{\Psi_0} c^\dagger (k_1) \dots c^\dagger(k_n)  c(k'_1) \dots c(k'_n) \ket{\Psi_0} \, .
\end{eqnarray}
Here $\tilde{\varepsilon}(k)$ is the {\it Hilbert transform} of the dispersion relation $\varepsilon (k)$. For $\varepsilon(k) = -\cos k$, this is simply $\tilde{\varepsilon}(k) = - \sin k$. We will come back to the Hilbert transform at length in the main parts of the paper, so we do not insist on it in this introduction. Of course, since we are dealing with a free fermion model, Eq. (\ref{eq:correl_XX}) requires a proof only for $n=1$; the case $n > 1$ then follows immediately from applying Wick's theorem on both sides,
\begin{eqnarray*}
\fl
\det_{1 \leq i,j\leq n} \left(  \,  \left< c^\dagger_{x_i,y_i} c_{x'_j,y'_j}  \right>   \,  \right)  &=&  \det_{1\leq i,j \leq n} \left( \,  \int  \frac{dk}{2\pi} \frac{dk'}{2\pi}  e^{-i k x_i + y_i \varepsilon(k) - i  R \tilde{\varepsilon}(k)}  e^{i k' x'_j - y'_j \varepsilon(k') + i  R \tilde{\varepsilon}(k')} \bra{\Psi_0} c^\dagger (k)  c(k') \ket{\Psi_0}   \, \right)   .
\end{eqnarray*}
For later purposes, notice that the propagator in the DWIS appearing in the r.h.s is
\begin{equation}
	\label{eq:prop_DWIS}
	\bra{\Psi_0} c^\dagger(k) c(k') \ket{\Psi_0} \, = \,  \sum_{x \in \mathbb{Z} + \frac{1}{2}}  e^{ i (k-k') x} \Theta(-x) \, = \, \frac{1}{ 2 i \sin \left( \frac{k-k'}{2} - i 0 \right)}   \, .
\end{equation}
The formula (\ref{eq:correl_XX}) and the role played by $\tilde{\varepsilon}(k)$ are a key aspect of this paper, analyzed in great detail in part \ref{sec:ktr}. For now, let us take this formula for granted, and proceed to the analysis of the scaling limit of fermionic correlators. First we introduce some useful piece of notation. We write
\begin{equation}
	\label{eq:notation_doteq}
	\left\{   \begin{array}{rcl}   c^\dagger_{x,y} &   \doteq   & \displaystyle  \int_{-\pi}^\pi \frac{dk}{2\pi} e^{-i k x + y \varepsilon(k) - i  R \tilde{\varepsilon}(k)}   c^\dagger (k)    \\  \\
	 c_{x,y} &   \doteq   & \displaystyle  \int_{-\pi}^\pi \frac{dk}{2\pi} e^{i k x - y \varepsilon(k) + i  R \tilde{\varepsilon}(k)}   c (k)   , 
	 \end{array}   \right.
\end{equation}
where we use the symbol $\doteq$ to emphasize that the relation must be understood in the sense of Eq. (\ref{eq:correl_XX}). It is {\it not an equality between operators}. Instead, it holds inside correlators, the l.h.s being inserted inside the bracket $\left<  .\right>$, and the r.h.s between $\bra{\Psi_0}$ and $\ket{\Psi_0}$. Eq. (\ref{eq:notation_doteq}) is really nothing more than the equality (\ref{eq:correl_XX}), but it is written in a more compact way.

\begin{figure}[hbt]
	\begin{tikzpicture}
		\begin{scope}[scale=0.7]
			\filldraw[black] (-5,2) rectangle (0,-2);
			\draw (-5,-2) rectangle (5,2);
			\draw[thick] (-5,-2) -- (5,-2);
			\draw[thick] (-5,2) -- (5,2);
			\filldraw[thick,fill=gray] (0,0) circle (2);
			\draw[<->] (0.1,0) -- (1.9,0);
			\draw (1,0.4) node{$R$};
			\draw[->] (-5.5,-2.5) -- (-5.5,2.5) node[right]{$y$};
			\draw[->] (-5.5,-2.5) -- (5.5,-2.5) node[below]{$x$};
			\draw (-5.4,0) -- (-5.6,0) node[left]{$0$};
			\draw (-5.4,2) -- (-5.6,2) node[left]{$+R$};
			\draw (-5.4,-2) -- (-5.6,-2) node[left]{$-R$};
			\draw (0,-2.4) -- (0,-2.6) node[below]{$0$};
			\filldraw (-1,1.3) circle (0.6mm) node[right]{$(x,y)$};
		\end{scope}
		\draw[->] (4.5,0) -- (6,0);
		\draw (5.25,0.5) node{$(x,y) \, \mapsto \, z$};
		\begin{scope}[xshift=8.5cm, scale=0.6]
			\draw[->] (0,-5) -- (0,5) node[right]{${\rm x}^2 = {\rm Im } \, z$};
			\filldraw[fill=gray,gray] (0,-4) rectangle (2,4);
			\filldraw[fill=black] (2,-4) rectangle (2.12,4);			
			\filldraw[fill=white,draw=white] (-0.06,-4) rectangle (0.06,4);	
			\draw[->] (-1,0) -- (3,0) node[right]{${\rm x}^1 = {\rm Re}\, z $};
			\draw (-0.06,-4) rectangle (-0.06,4);	
			\draw (0.06,-4) rectangle (0.06,4);	
			\draw (-0.4,-0.4) node{$0$};
			\draw (2.4,-0.4) node{$\pi$};
			\filldraw (1.7,-1.3) circle (0.6mm) node[left]{$z$};
		\end{scope}		
	\end{tikzpicture}
	\caption{The critically fluctuating region $x^2+y^2 < R^2$ is mapped to an infinite strip of width $\pi$ by $z(x,y) \, = \, {\rm arccos}\frac{x}{\sqrt{R^2-y^2}} - i\, {\rm arcth} \frac{y}{R}$.}
	\label{fig:strip}
\end{figure}
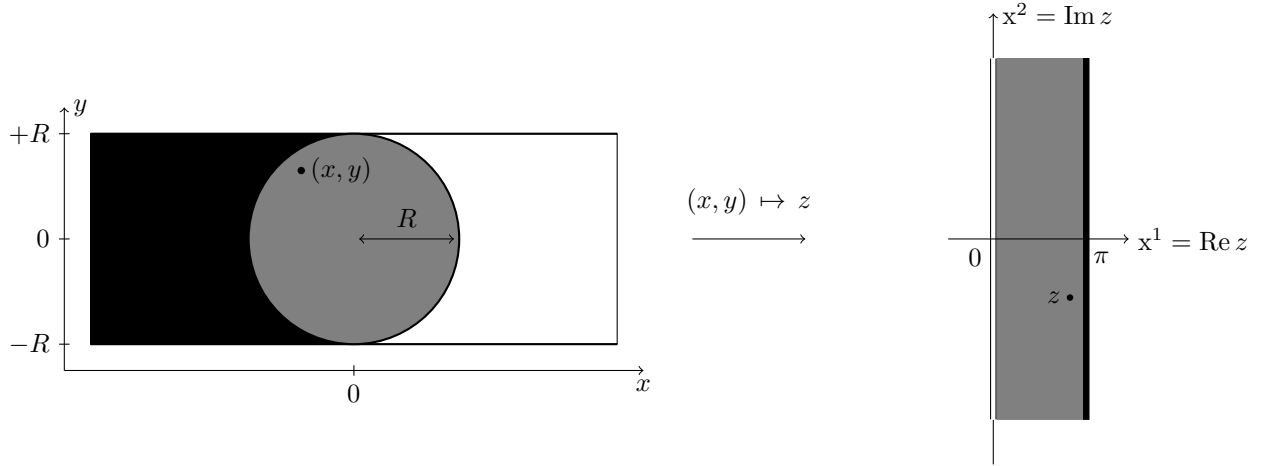
Next, we observe that the scaling limit (\ref{eq:scaling_regime}) of the correlators may be obtained with the steepest descent method. In the relations (\ref{eq:notation_doteq}), the contour of integration over $k$ may be deformed such that it passes through the points where the phase is stationary; these are the solutions of
\begin{equation*}
	\frac{x}{R} \, + \, i \frac{y}{R} \, \frac{d\varepsilon(k)}{dk} \, +  \, \frac{d \tilde{\varepsilon}(k)}{dk} \, = \, \frac{1}{R} \frac{d}{dk} \left[ k x + i y \varepsilon (k)  + R \tilde{\varepsilon}(k)  \right] \, = \, 0 \, .
\end{equation*} 
For $\varepsilon(k) \, = \, -\cos k$ and $\tilde{\varepsilon}(k) = - \sin k$, this is a quadratic equation in $e^{i k}$, which is solved straightforwardly. As long as $x^2+y^2 < R^2$, there are two simple roots: $k \, = \, z(x,y)$ and $k \, =\, -z^*(x,y)$, where $z^*$ is the complex conjugate of $z$, with
\begin{eqnarray*}
	z\, = \, z(x,y) \, \equiv \, {\rm arccos } \frac{x}{\sqrt{R^2-y^2}} \, - \, i \, {\rm arcth}\frac{y}{R} \, .
\end{eqnarray*}
The integral over $k$ is then evaluated by the stationary phase approximation around these points: we approximate the argument of the exponential by
\begin{equation*}
	\left\{  \begin{array}{rcll}
		e^{-i k x \, + \,  y  \varepsilon(k) \, - \, i R \tilde{\varepsilon}(k)}  & \simeq & e^{ - i \varphi \, - \, \frac{i}{2}(k-z)^2\, e^{\sigma}  }     & \qquad  {\rm around } \quad k = z  \\
		e^{-i k x \, + \,  y  \varepsilon(k) \, - \, i R \tilde{\varepsilon}(k)}  & \simeq & e^{  i \varphi^* \, + \, \frac{i}{2}(k+z^*)^2\, e^{\sigma}  }     & \qquad  {\rm around } \quad k = -z^* \, ,
	\end{array} \right.
\end{equation*}
where we have defined
\begin{eqnarray*}
	\varphi \, = \, \varphi(x,y) & \equiv & z(x,y)\, x \, + \, i y \, \varepsilon(z(x,y)) \, + \, R \, \tilde{\varepsilon}(z(x,y))  \, =  \,z(x,y)\, x \, - \, \sqrt{R^2-x^2-y^2}  \, ,\\ 
	e^{\sigma} \, = \, e^{\sigma(x,y)} & \equiv &  \, i y \, \frac{d^2 \varepsilon}{dk^2} (z(x,y)) \, + \, R \, \frac{d^2 \tilde{\varepsilon}}{dk^2}(z(x,y)) \, = \, \sqrt{R^2 - x^2 -y^2} \, .
\end{eqnarray*}
Finally, one performs the integral in (\ref{eq:notation_doteq}), which is now gaussian; it yields the asymptotic formulas
\begin{equation}
	\label{eq:notation_doteq_asympt}
	\left\{   \begin{array}{rcl}   c^\dagger_{x,y} &   \underset{ R \rightarrow \infty  }{\doteq}   & \displaystyle   \frac{ e^{-i \varphi(x,y)  - i \frac{\pi}{4} } }{\sqrt{2\pi}} e^{-\frac{1}{2} \sigma(x,y)}    c^\dagger (z)    \,  + \,  \frac{ e^{i \varphi^*(x,y) + i \frac{\pi}{4} } }{\sqrt{2\pi}} e^{-\frac{1}{2} \sigma(x,y)}    c^\dagger (-z^*)   \\  \\
	 c_{x,y} &   \underset{ R \rightarrow \infty  }{\doteq}   & \displaystyle   \frac{ e^{i \varphi(x,y)  + i \frac{\pi}{4} } }{\sqrt{2\pi}} e^{-\frac{1}{2} \sigma(x,y)}    c (z)    \,  + \,  \frac{ e^{-i \varphi^*(x,y) - i \frac{\pi}{4} } }{\sqrt{2\pi}} e^{-\frac{1}{2} \sigma(x,y)}    c (-z^*) \, . 
	 \end{array}   \right.
\end{equation}
What exactly do we mean by $c^\dagger(z)$, $c(z)$ here, when $z \notin \mathbb{R}$? Again, these relations must be understood in the sense of Eqs. (\ref{eq:correl_XX})-(\ref{eq:notation_doteq}): the correlator of fermions in the l.h.s of (\ref{eq:notation_doteq_asympt}) are equal to expressions that involve $\bra{\Psi_0} c^\dagger(z) c(z')\ket{\Psi_0} = 1/(2i \sin \frac{z-z'}{2} )$, the analytic continuation of (\ref{eq:prop_DWIS}).

\subsection*{Identifying the parameters in the Dirac action}

Now let us come back to Eqs. (\ref{eq:lattice_continuous_XX})-(\ref{eq:Dirac}) and to the discussion around them. The resemblance with Eq. (\ref{eq:notation_doteq_asympt}) is striking, and clearly suggests that the continuous field $\psi^\dagger$ is, roughly speaking, nothing but $e^{- i \varphi - i \frac{\pi}{4} } e^{-\frac{1}{2} \sigma } c^\dagger$ (and similar relations for $\psi$, $\overline{\psi}^\dagger$ and $\overline{\psi}$). It is then very easy to extract the parameters of the Dirac action in the toy-model. One simple way of doing this is to match the local behaviors of the lattice and continuous propagators. The behavior of the lattice propagator follows from Eqs. (\ref{eq:notation_doteq_asympt}), which give its scaling limit in the form of a sum of four terms. When $(x',y')$ approaches $(x,y)$, it is dominated by only two of the terms, each of which diverge as
\begin{equation}
	\label{eq:prop_short}
	\left< c_{x,y}^\dagger c_{x',y'}  \right> \, \simeq \, \frac{e^{-\frac{\sigma  + \sigma' }{2} }}{2\pi i} \left( \frac{e^{-i (\varphi - \varphi' ) }}{z-z'}  - \frac{e^{i (\varphi^* - \varphi'^* ) }}{z^*-z'^*}   \right) \, .
\end{equation}
Looking back at Eq. (\ref{eq:lattice_continuous_XX}), one sees that the propagator of the continuous fields must have the local behavior 
\begin{equation*}
	\left< \psi^\dagger(x,y) \psi(x',y') \right> \, \simeq \, e^{-\frac{\sigma  + \sigma' }{2} }  \frac{e^{-i (\varphi - \varphi' ) }}{i(z-z')} \qquad \qquad  \qquad \left< \overline{\psi}^\dagger(x,y) \overline{\psi}(x',y') \right> \, \sim \, e^{-\frac{\sigma  + \sigma' }{2} }  \frac{e^{i (\varphi^* - \varphi'^* ) }}{-i(z^*-z'^*)} \, ,
\end{equation*}
implying that they are Green's functions for the following operators
\begin{equation*}
	\left\{	\begin{array}{cll}
	e^\sigma \left( i \partial_{\bar{z}} +  \frac{i}{2} (\partial_{\bar{z}} \sigma)  -  (\partial_{\bar{z}} \varphi)  \right) \big\langle \psi^\dagger(z,\bar{z}) \psi(z',\bar{z}') \,\big\rangle  & = &  \pi \, \delta^{(2)} (z-z') \\
	e^\sigma \left( -i \partial_z - \frac{i}{2} (\partial_z \sigma)  - (\partial_z \varphi^*)  \right) \big\langle \,\overline{\psi}^\dagger(z,\bar{z}) \overline{\psi}(z,\bar{z}') \,\big\rangle  & = &  \pi \, \delta^{(2)} (z-z')  \, .
	\end{array} \right.
\end{equation*}
We view the latter as arising from the action
\begin{equation*}\fl
	\begin{array}{c}
 	\displaystyle S \, = \, \frac{1}{\pi} \int d^2 z   \, e^{\sigma}  \left[  \left( \begin{array}{cc} \psi^\dagger  & \overline{\psi}^\dagger  \end{array}  \right)    \left( \begin{array}{cc}  - \frac{i}{2} \overset{\leftrightarrow}{\partial}_{\bar{z}}    -  (\partial_{\bar{z}} \varphi)   & 0  \\  0  &   \frac{i}{2} \overset{\leftrightarrow}{\partial}_{z}    -  (\partial_{z} \varphi^*)      \end{array}\right)    \left(  \begin{array}{c}  \psi \\ \overline{\psi} \end{array}  \right)  \right] \\ \\
 \left( = \,  \frac{1}{\pi} \int d^2 z   \, e^{\sigma} \, \left[    \left( [ i \partial_{\bar{z}} +  \frac{i}{2} (\partial_{\bar{z}} \sigma)  -  (\partial_{\bar{z}} \varphi )]   \psi^\dagger   \right)  \psi +  \left( [ -i \partial_{z} -  \frac{i}{2} (\partial_{z} \sigma)  -  (\partial_{z} \varphi^*)  ]  \overline{\psi}^\dagger \right)   \overline{\psi}    \right]   \quad + \quad {\rm boundary \; terms} \right)  \, .
	\end{array}
\end{equation*}
This, as anticipated in the discussion above, is a particular case of the generic Dirac action (\ref{eq:Dirac}). To match the two expressions, one needs to write (\ref{eq:Dirac}) in the coordinate system provided by the map $(x,y) \mapsto z(x,y)$,
\begin{equation*}
\left\{	\begin{array}{rcl}  {\rm x}^1 \, + \, i \, {\rm x}^2 & = & z(x,y) \\
	   {\rm x}^1 \, - \, i \, {\rm x}^2 & = & z^*(x,y) \, , \end{array}  \right.
\end{equation*}
where $0 <{\rm x}^1 < \pi$ and ${\rm x}^2 \in \mathbb{R}$ (Fig. \ref{fig:strip}). The tetrad in (\ref{eq:Dirac}) must be taken as the diagonal matrix
\begin{equation}
	e^\mu_{\phantom{b}a} \, = \, e^{-\sigma} \delta^\mu_{\phantom{b}a} \, ,
\end{equation}
so the corresponding metric is $d s^2 = e^{2 \sigma} [(d {\rm x}^1)^2 + (d {\rm x}^2)^2]$. In other words, what the map $(x,y) \mapsto z(x,y)$ does for us is that it provides isothermal coordinates, for free. The background gauge potentials are readily identified as
\begin{equation}
	A^{({\rm v})}_\mu \, = \, -i \partial_\mu {\rm Im} \,\varphi    \qquad \qquad \qquad  A^{({\rm a})}_\mu \, = \, -\partial_\mu {\rm Re}\, \varphi \, .
\end{equation}
They are both pure gauge. It may look problematic that $A^{({\rm v})}_\mu$ is imaginary, but again,
it can never appear in physical observables, which must be gauge invariant.
[Notice that the propagator itself, as defined so far, {\it is not a physical observable}, as {\it it is not gauge invariant}.
To make it gauge invariant, one would have to define it with the insertion of a Wilson line.] The axial potential,
on the other hand, can (and does) appear in physical quantities. One example is the density:
it may be obtained from Eq. (\ref{eq:prop_short}) by taking $y'=y$ first,
and then $x'\rightarrow x$: $\left< \rho_{x,y} \right> \, = \,  -\frac{1}{\pi} \frac{\partial_x {\rm Re } \varphi }{e^{\sigma}
\partial_x z}$.
The formula for the density can be further simplified using $e^{\sigma} \partial_x z=-1$. This identity can be checked directly using the formulas for $z$ and $e^\sigma$ or
by observing that the stationary phase equation is of the form $F(x,y,z)=0$, where  $F(x,y,z)=x+iyv(z)+
R\tilde v(z)$ and  $v(z)=\frac{d\varepsilon}{dz}$, $\tilde{v}(z)=\frac{d\tilde{\varepsilon}}{dz}$. We also have 
$e^{\sigma}=\partial_z F$. By the chain rule and the fact that $dF=0$ when $z=z(x,y)$ we obtain
$\partial_x z(x,y)=-\partial_x F/\partial_z F$ and the result follows from $\partial_x F=1$. Coming back to the density
we obtain 
\begin{eqnarray*}
	\left< \rho_{x,y} \right> \, =   \,-\frac{1}{\pi} A^{({\rm a})}_{x} \, ,
\end{eqnarray*}
and plugging the explicit formula for
$A^{({\rm a})}_{x} $,
one recovers $\left< \rho_{x,y} \right> = \frac{1}{\pi} {\rm arccos} \frac{x}{\sqrt{R^2-y^2}} $ as claimed above.

\subsection*{Discussion}

Let us pause and take a look back at the logic of the previous paragraphs. We defined a two-dimensional toy-model that displays an arctic circle; it is discrete in one direction (space coordinate $x$) and continuous in the other direction (imaginary-time coordinate $y$). It was argued, on general grounds (fermion statistics, ${\rm U}(1)$ symmetry, locality), that the long-range correlations inside the fluctuating region should be described by a massless Dirac theory. Then:
\begin{itemize}
	\item independently of these field theory considerations, we exhibited an exact formula for the lattice fermion propagator $\left< c_{x,y}^\dagger c_{x',y'} \right>$, to be proved later (part \ref{sec:ktr}).
	\item the {\it scaling limit} of this propagator was analyzed with the stationary phase approximation. We found that the stationary points are the solution of a quadratic equation. If one solution is called $z$, the other is $-z^*$. The mapping $(x,y) \mapsto z$ is a diffeomorphism that sends the interior of the arctic circle to a certain subset of $\mathbb{C}$ (here, an infinite strip), so it can be used as a coordinate system. Moreover, in the analysis of the stationary points, two other functions $\sigma$ and $\varphi$ show up naturally.
	\item trying to match the local asymptotic behavior of the lattice fermion propagator with the one coming from the generic Dirac action (\ref{eq:Dirac}), we were able to fix the free parameters in this action, namely the tetrad and the background gauge potentials. It turns out that the coordinate system provided by $(x,y) \mapsto z$ is isothermal, with a conformal factor given by $e^\sigma$, and that the gauge potentials are given by the real- and imaginary-part of $\partial_\mu \varphi$.
\end{itemize}
Some important aspects were neglected in the present discussion, for instance the boundary conditions that should be imposed on the continuous fields (we will focus on the boundary in part \ref{sec:dirac_honey}). But, once again, the point we wish to emphasize in this introduction is that, if one is willing to describe the fluctuating region by a field theory, then the theory must be {\it inhomogeneous}, in the sense that, once a relation between the lattice and the continuous degrees of freedom is fixed (such as Eq. (\ref{eq:lattice_continuous_XX}) above), then the parameters in the action vary with position. Since the components of the metric themselves are such parameters, it is natural that one ends up with a theory in curved euclidean space. 
[In the isothermal coordinate system above, $\sigma \, = \, \log ( R \,\sin ({\rm x}^1) / \cosh ({\rm x}^2) )$, which gives the non-zero scalar curvature $K  \, = \, - e^{2 \sigma} \Delta \sigma  \, = \, \left(  \sin^2 ({\rm x}^1) + \cosh^2({\rm x}^2) \right) / (R^2 \sin^4 ({\rm x}^1)) $ , so it is really a curved space.]

\vspace{0.3cm}

The reader might be puzzled by our logic: usually, when one discusses field theory in relation with lattice models, the goal is to use arguments or calculation tools that are simpler in the continuous setting to make interesting predictions about lattice observables. Here, even though we initially promised a field theory investigation, technically, we have proceeded backwards: we first provided the solution of the lattice model with formula (\ref{eq:correl_XX}), and then imposed some field theory interpretation on top of it. 
It is our opinion that the observation that 'arctic circle problems' are related to field theories in curved space deserves to be highlighted, and that the precise derivation of the action deserves a proper exposition. 
We also believe that some of the ideas exposed here will be useful to tackle other problems involving inhomogeneous systems, including inhomogeneous quantum quenches, particles trapped by inhomogeneous potentials as well as closely related random matrix questions; this will be discussed elsewhere. 
\subsection*{Organization of the paper}

The goal of the rest of the paper is to extend the approach we just presented to other free-fermion models that have appeared previously in the arctic circle context: dimer models on the honeycomb and square lattice, and the six-vertex model at $\Delta = 0$.
We also need to complete the derivation of the key formula (\ref{eq:correl_XX}), which was left aside in this introduction. This is achieved in part \ref{sec:ktr}: we obtain the propagator for the toy-model, in the form of a closed, simple, exact formula. We also extend the calculation to the case of two-band models, needed later to solve the six-vertex model and dimers on the square lattice. The most important aspect of part \ref{sec:ktr} is the appearance of the Hilbert transform of the dispersion relation; it is instrumental throughout the paper. Dimers on the honeycomb lattice are studied in Part \ref{sec:honeycomb}. Because it is a one-band model, the honeycomb model is the simplest possible extension of the toy-model; the results are similar to the ones we sketched in this introduction. In parts \ref{sec:square} and \ref{part:6v}, we repeat the analysis, for the dimer model on the square lattice, and for the six-vertex model. Essentially, these are nothing but the toy-model on steroids. Intermediate calculations become increasingly cumbersome, while all main conclusions remain qualitatively  identical: the large-scale correlations inside the fluctuating region are those of a Dirac theory, and the parameters in the action (\ref{eq:Dirac}) are identified from a pair of stationary points. In part \ref{part:GFF} we translate our results into the language of discrete height configurations and their continuum limit, the gaussian free field. While the approach taken in this paper is really tied to free fermions, we think that presenting the results in the alternative (bosonic) way should be useful to facilitate connections with other references, for example \cite{kenyon2009lectures}. 

Some additional information is gathered in five appendices. In \ref{appx:real}, we explain how the results of the toy-model may be analytically continued to real time.  \ref{sec:airy} deals with the behavior of the correlations close to the arctic curve, which were neglected in the main text. We provide a simple heuristic explanation for the genereric appearance of the Airy kernel as well as the Tracy-Widom distribution in that case. In \ref{appx:hilb} we compute the various Hilbert transforms needed throughout the paper, while all the technical steps needed to obtain exact finite-size formulae for the two-band models are gathered appendix.~\ref{sec:moredetails}. Finally, in \ref{sec:partitionsfunctions}, we explain how the well known partition functions for all the models studied here may be recovered from bosonization arguments.

\newpage

\newpage

\section{Propagator in imaginary-time and appearance of the Hilbert transform}
\label{sec:ktr}

In this part, we prove the formula (\ref{eq:correl_XX}) of the introduction. More precisely, we prove the $n=1$ case: we compute the propagator inside the strip.  The result involves the Hilbert transform of the dispersion relation $\varepsilon (k)$, which comes from the DWIS. Indeed, the DWIS acts as a projector
onto the negative real axis in real-space, which, when Fourier transformed, gives the Hilbert transform. We derive this key formula in full details in the case of a single band, and then sketch the derivation for the two-band case, which differs from the former case only by relatively minor details. Not surprisingly, the main technical tool in this section is Wick's theorem; we use it in a relatively non-standard bosonized form though, which considerably simplifies the calculations. This bosonization trick will not be needed in the rest of the paper; only the results will be, and the latter will be recalled when needed. Therefore, the impatient reader may want to skip this technical part and go directly to section \ref{sec:honeycomb}.

\subsection{Imaginary time propagator for one-band models}

\noindent This subsection is devoted to the calculation of the propagator in $k$-space for the one-band Hamiltonian (\ref{eq:Ham}). Let us recall that this propagator is defined as
\begin{equation*}
	\left< c^\dagger(k,y) c(k',y') \right>  \, \equiv \, \left\{ \begin{array}{c} \displaystyle \frac{\bra{\Psi_0} e^{ -(R-y)  H} c^\dagger(k) e^{-(y -y') H} c(k') e^{- (R+ y')  H} \ket{\Psi_0}}{\bra{\Psi_0} e^{-2 R\, H} \ket{\Psi_0}}   \qquad {\rm if} \;(y>y')   \\ \\
\displaystyle - \,\frac{\bra{\Psi_0} e^{ -(R-y') \,  H} c(k) e^{-(y' -y) H} c^\dagger(k') e^{- (R+ y) \, H} \ket{\Psi_0}}{\bra{\Psi_0} e^{- 2 R \, H} \ket{\Psi_0}}   \qquad {\rm if} \;(y<y')  \,.\end{array}  \right.
\end{equation*}
We are going to show that
\begin{equation}
	\label{eq:ktr}
	\left< c^\dagger(k,y) c(k',y') \right> \, = \, \frac{e^{y \varepsilon(k)  -i R  \tilde{\varepsilon}(k)}  e^{-y' \varepsilon(k')  + i R \tilde{\varepsilon}(k') } }{ 2i \sin \left( \frac{k-k'}{2} - i 0 \right)} \, ,
\end{equation}
where $\tilde{\varepsilon}(k)$ is the Hilbert transform of the dispersion relation $\varepsilon(k)$:
\begin{equation}
	\label{eq:Hilbert}
	\tilde{\varepsilon}(k) \, = \, {\rm p.v.} \int_{-\pi}^\pi \frac{dk'}{2\pi} \varepsilon(k') ~ {\rm cot} \left( \frac{k-k'}{2}\right).
\end{equation}
The formula (\ref{eq:ktr}) is quite remarkable, and it completely solves the toy-model, as we explained in the introduction. It also solves the dimer model on the honeycomb lattice, as we explain in part \ref{sec:honeycomb}. To prove this formula, we proceed as follows. First, we note that, since
\begin{equation*}
\left\{ 	\begin{array}{ccccc}
		c^\dagger(k,y) & = & e^{y H} c^\dagger (k) e^{-y H} & = & e^{ y \varepsilon(k)} c^\dagger (k)\\
		c(k',y') & = & e^{y' H} c (k') e^{-y' H} & = & e^{- y' \varepsilon(k)} c (k')\, ,
	\end{array} \right.
\end{equation*}
the $y$- and $y'$-dependence of (\ref{eq:ktr}) is trivial. Instead, what we really need to show is
\begin{equation}
	\label{eq:ktr0}
	\left< c^\dagger(k,0) c(k',0) \right> \, = \, \frac{e^{-i R \left[ \tilde{\varepsilon}(k) - \tilde{\varepsilon}(k') \right]} }{ 2i \sin \left( \frac{k-k'}{2} - i 0 \right)} \, .
\end{equation}
We introduce the normal order $:.:$ adapted to our initial state $\ket{\Psi_0}$, namely
\begin{equation*}
	: c^\dagger_x c_x : \;\equiv  \; \left\{ \begin{array}{rcl}  c^\dagger_x c_x && {\rm if} \quad x>0  \\   - c_x c_x^\dagger  && {\rm if} \quad x<0 \, ,   \end{array} \right.
\end{equation*}
such that
\begin{eqnarray*}
\nonumber	:c^\dagger (k) c(k'): & = & c^\dagger(k) c(k') \, - \, \bra{\Psi_0}  c^\dagger (k) c(k') \ket{\Psi_0}  \\ 
	& = &  c^\dagger(k) c(k') \, - \,  \frac{1}{ 2i \sin \left( \frac{k-k'}{2} -i 0\right)} \, .
\end{eqnarray*}
It is probably possible to get to formula (\ref{eq:ktr0}) using Wick's theorem for fermions directly, but we haven't been able to do so; the reader may try and convince themselves that the computation is rather cumbersome. So, instead, we chose to use standard bosonization formulas that make the various computational steps much lighter. For this purpose, we introduce a (real) free bosonic field $\varphi(k)$, and we bosonize the fermion creation/annihilation operators, as well as the Hamiltonian, according to the rules
\begin{equation}
	\label{eq:bosonization}
({\rm bosonization}) \qquad	\left\{ \begin{array}{rcl} c^\dagger(k) & \rightarrow & : e^{i \varphi(k)}: \\ 
	c(k) & \rightarrow & : e^{-i \varphi(k)}: \\  
	\int \frac{dk}{2\pi} \varepsilon(k) \, c^\dagger(k) c(k)	& \rightarrow &  \int \frac{dk}{2\pi} \varepsilon(k) \,  \partial \varphi(k) \, .  \end{array} \right.
\end{equation}
The propagator of the bosonic field is chosen as
\begin{equation}
	\label{eq:bosprop}
	\bra{\Psi_0}  \varphi(k - i \epsilon ) \varphi(k')  \ket{\Psi_0} \, = \, - \log  \left[  2 i \sin \left( \frac{k- i \epsilon -k'}{2} \right) \right]
\end{equation}
such that $\bra{\Psi_0} : e^{i \varphi(k- i \epsilon)} :\, :e^{-i \varphi(k')} : \ket{\Psi_0} \, = \, \bra{\Psi_0}  c^\dagger(k) c(k')  \ket{\Psi_0} $ when $\epsilon \rightarrow 0^+$. When computing correlators or commutators, one needs to carefully regularize the bosonic expressions. This is done by adding some small imaginary part to the argument $k$, which is larger, or smaller, depending on the order of appearance of the operators in the different expressions. An instructive example consists in recovering the commutator $\left[ H , c^\dagger(k) \right] \, = \, \varepsilon(k) \, c^\dagger(k)$ in bosonized form:
\begin{eqnarray}
\nonumber	\left[ H , c^\dagger(k) \right] & \rightarrow & \lim_{\epsilon  \rightarrow 0^+} \int \frac{dq}{2\pi}  \varepsilon(q) \left[  \partial \varphi(q- i \epsilon ) \,: e^{i \varphi(k)} :\;  -\;  :e^{i \varphi(k)}:  \, \partial \varphi(q+ i \epsilon ) \right] \\
\nonumber & = & \lim_{\epsilon  \rightarrow 0^+} \int \frac{dq}{2\pi}  \varepsilon(q) \left[  i \partial_{q}  \bra{\Psi_0}  \varphi(q- i \epsilon ) \varphi(k) \ket{\Psi_0} \;  -\;  i\partial_{q}  \bra{\Psi_0}  \varphi(k ) \varphi(q+i \epsilon) \ket{\Psi_0} \,  \right] \,  : e^{i \varphi(k)} :  \\
\nonumber &=&   \lim_{\epsilon  \rightarrow 0^+} \int \frac{dq}{2\pi}  \varepsilon(q) \left[  -\frac{i}{2}\, {\rm cot} \left(\frac{q - i \epsilon - k}{2} \right) \;  +\;  \frac{i}{2}\, {\rm cot} \left(\frac{q + i \epsilon - k}{2} \right) \,  \right] \,  : e^{i \varphi(k)} :  \\
\nonumber &=&  \int \frac{dq}{2\pi}  \varepsilon(q)  \, 2\pi\delta(q-k) \,  : e^{i \varphi(k)} :  \;\, = \; \varepsilon(k) \, :e^{i \varphi(k)}: \, ,
\end{eqnarray}
which is the expected result. In the first line, we used the bosonization formulas (\ref{eq:bosonization}), introducing the regulator $\varepsilon > 0$ according to the respective position of the two operators; in the second line we applied Wick's theorem; in the third line we used the bosonic propagator (\ref{eq:bosprop}); in the last line we used
\begin{equation}
	\label{eq:delta_pv}
	\lim_{\epsilon \rightarrow 0^+} {\rm cot} \left(\frac{k  - k' \pm i \epsilon }{2} \right) \; = \;  \mp \,  2\pi i\, \delta(k-k') \, + \, {\rm p.v.} \left[ {\rm cot} \left(\frac{k - k'}{2} \right) \right] \, .
\end{equation}

With these bosonization tricks at hand, it is a relatively pleasant exercise to prove formula (\ref{eq:ktr0}). Indeed, the quantity that we need to evaluate becomes
\begin{eqnarray}
	\label{eq:interm0}
\nonumber	&& \frac{\bra{\Psi_0}  e^{-R H} c^\dagger(k) c(k')   e^{-R H} \ket{\Psi_0}}{ \bra{\Psi_0}  e^{-2 R H} \ket{\Psi_0}  }  \\
	  &&  \rightarrow \; \frac{  \bra{\Psi_0}: e^{ -R  \int \frac{dq}{2\pi} \varepsilon(q) \, \partial \varphi(q - 2i \epsilon)  } :  \;:e^{i \varphi(k- i \epsilon)}: \; :e^{-i \varphi(k')}: \;  : e^{ -R  \int \frac{dq'}{2\pi} \varepsilon(q') \,\partial \varphi(q'+ i \epsilon)  } :\ket{\Psi_0}    }{  \bra{\Psi_0} :e^{ - R  \int \frac{dq}{2\pi} \varepsilon(q) \, \partial \varphi(q)  }: \; :e^{ - R  \int \frac{dq'}{2\pi} \varepsilon(q') \, \partial \varphi(q')  }:  \ket{\Psi_0} }\; ,
\end{eqnarray}
and we may compute this by applying the celebrated (exact) fusion formula for vertex operators
\begin{equation*}
	: e^{\alpha \varphi(k)} : \, : e^{  \beta \varphi(k')}: \;\, = \;  e^{\alpha \beta \bra{\Psi_0}  \varphi(k) \varphi(k') \ket{\Psi_0} } \,  :e^{ \alpha \varphi(k) +  \beta \varphi(k') }:
\end{equation*}
which follows from Wick's theorem. In (\ref{eq:interm0}), we start by fusing the two central vertex operators that come from $c^\dagger(k)$ and $c(k')$. This gives
\begin{equation*}\fl
 \bra{\Psi_0}  c^\dagger(k) c(k')  \ket{\Psi_0}  \, \times \,	\frac{  \bra{\Psi_0}: e^{ -R  \int \frac{dq}{2\pi} \varepsilon(q) \, \partial \varphi(q - 2i \epsilon)  } :  \;:e^{i \varphi(k- i \epsilon) -i \varphi(k')}: \;  : e^{ -R  \int \frac{dq'}{2\pi} \varepsilon(q') \,\partial \varphi(q'+ i \epsilon)  } :\ket{\Psi_0}    }{  \bra{\Psi_0} :e^{ - R \int \frac{dq}{2\pi} \varepsilon(q) \, \partial \varphi(q)  }: \; :e^{ - R  \int \frac{dq'}{2\pi} \varepsilon(q') \,\partial \varphi(q')  }:  \ket{\Psi_0} } .
\end{equation*}
Next, we fuse the vertex operator involving the integral over $q$ with the one involving the integral over $q'$, both in the numerator and in the denominator. Clearly, the scalar factor that comes out of the fusion is the same in both cases, so it cancels when we take the ratio, and we do not need to evaluate it. The denominator is then the expectation value of a single normal-ordered exponential, which is one by definition. Thus, we are left with
\begin{equation*}
 \bra{\Psi_0}  c^\dagger(k) c(k')  \ket{\Psi_0}  \, \times \,  \bra{\Psi_0}: e^{ -R  \int \frac{dq}{2\pi} \varepsilon(q) \, \partial \varphi(q - 2i \epsilon)  \, -\, R  \int \frac{dq'}{2\pi} \varepsilon(q') \,\partial \varphi(q'+ i \epsilon)  } : \;:e^{i \varphi(k- i \epsilon) -i \varphi(k')}: \ket{\Psi_0}   .
\end{equation*}
We use the fusion formula one last time to get
\begin{eqnarray*}
\nonumber \bra{\Psi_0}  c^\dagger(k) c(k')  \ket{\Psi_0}  \, \times \, e^{ -R  \int \frac{dq}{2\pi} \varepsilon(q) \, i \partial_{q} \bra{\Psi_0} \varphi(q - 2i \epsilon)  \varphi(k- i \epsilon) \ket{\Psi_0} \, -\,R  \int \frac{dq'}{2\pi} \varepsilon(q') \,i \partial_{q'} \bra{\Psi_0 }\varphi(q'+ i \epsilon)  \varphi(k- i \epsilon)  \ket{\Psi_0} }  \\
\times \;  e^{ R  \int \frac{dq}{2\pi} \varepsilon(q) \, i \partial_{q} \bra{\Psi_0} \varphi(q - 2i \epsilon)  \varphi(k') \ket{\Psi_0} \, +\,R  \int \frac{dq'}{2\pi} \varepsilon(q') \,i \partial_{q'} \bra{\Psi_0 }\varphi(q'+ i \epsilon)  \varphi(k')  \ket{\Psi_0} }  \,   .
\end{eqnarray*}
Finally, we use the explicit form of the bosonic propagator (\ref{eq:bosprop}),
\begin{eqnarray*}\fl
\nonumber \bra{\Psi_0}  c^\dagger(k) c(k')  \ket{\Psi_0}  \, \times \, e^{ i R  \int \frac{dq}{2\pi} \varepsilon(q)  \left[  \frac{1}{2} {\rm cot} \left( \frac{q  - k -  i \epsilon}{2} \right) \, +\, \frac{1}{2}{\rm cot} \left( \frac{q - k + 2 i \epsilon }{2} \right)  \right] }  \;  e^{ - i  R  \int \frac{dq}{2\pi} \varepsilon(q)  \left[ \frac{1}{2} {\rm cot} \left( \frac{q  - k' - 2 i \epsilon}{2} \right) \, +\, \frac{1}{2} {\rm cot} \left( \frac{q - k' + i \epsilon }{2} \right)  \right] } \,   ,
\end{eqnarray*}
and we take the limit $\epsilon \rightarrow 0^+$, using (\ref{eq:delta_pv}). The integrals in the exponentials become the principal value that appears in the definition of the Hilbert transform (\ref{eq:Hilbert}). We arrive at
\begin{equation*}
	\frac{\bra{\Psi_0}  e^{-R H} c^\dagger(k) c(k')   e^{-R H} \ket{\Psi_0}}{ \bra{\Psi_0}  e^{-2 R H} \ket{\Psi_0}  }   \; = \;   e^{- i R\left[ \tilde{\varepsilon}(k) - \tilde{\varepsilon}(k')   \right]}  \; \bra{\Psi_0}  c^\dagger(k) c(k')  \ket{\Psi_0},  
\end{equation*}
as claimed above.

\subsection{Imaginary time propagator for two-band models}

In parts \ref{sec:square} and \ref{part:6v}, we will see that, in our setup, the dimer model on the square lattice and the six-vertex model are both models with {\it two} fermionic degrees of freedom per unit cell. Therefore, their transfer matrices $T$---as well as their 'Hamiltonians' $H \, =\, -\log T$---have two bands, instead of one. So, in order to prepare for these models, we need to generalize the previous calculation to the two-band case. We have not presented the six-vertex and square-lattice dimer models yet, and actually, for the purposes of this section, we do not need them. Instead, we introduce only the few features that are strictly necessary for now; the details that are more specific to either model will be given later, in parts \ref{sec:square} and \ref{part:6v}. \\

First, we are still dealing with fermionic particles on the lattice $\mathbb{Z} + \frac{1}{2}$, and with the DWIS $\ket{\Psi_0}$. The transfer matrix $T$, however, is not invariant under translations of {\it one site}, but only of {\it two sites}. Therefore, in $k$-space, the transfer matrix can be put in the form (possibly up to an unimportant global numerical factor)
\begin{equation*}
	T \, = \,  \exp \left[ -  \int_{-\frac{\pi}{2}}^{\frac{\pi}{2}} \frac{dk}{2\pi} \left( \begin{array}{cc}  c^\dagger(k)  &  c^\dagger(k+\pi)  \end{array} \right)    h(k)  \left( \begin{array}{c}  c(k) \\  c(k+\pi)  \end{array} \right)  \right] \, ,
\end{equation*}
where $h(k)$ is some $2 \times 2$ matrix whose elements depend on the details of the model. Now let us make the following assumptions. They are all  satisfied by the models tackled later; these assumptions allow
to express the results in a more compact way, which will be most useful in parts \ref{sec:square} and \ref{part:6v}. We assume that:
\begin{enumerate}[label=(\roman*)]
	\item \label{item:hermitian}the transfer matrix $T$ is a hermitian operator; equivalently $h(k)$ is a hermitian $2 \times 2$ matrix
	\item \label{item:eigenvalues}the two (real) eigenvalues of $h(k)$ are opposite to each other, namely there is a unitary $2\times 2$ matrix $U(k)$ and a real-valued function $\varepsilon (k) $ such that
		\begin{equation*}
			h(k) \, = \, U^t(k)  \left(  \begin{array}{cc}   \varepsilon(k)  & 0 \\  0 & - \varepsilon (k)    \end{array}  \right) U^*(k)
		\end{equation*}
	\item \label{item:continuation}the function $\varepsilon (k)$, which is initially defined on $[ - \frac{\pi}{2} , \frac{\pi}{2} ]$, may be analytically continued to the real axis $\mathbb{R}$. Moreover, this function $\varepsilon : \mathbb{R} \rightarrow \mathbb{R}$ is {\it anti-periodic} with period $\pi$: $\varepsilon(k + \pi )  = - \varepsilon (k)$. Similarly, the matrix-valued function $U(k)$ can be chosen such that it can be analytically continued to $\mathbb{R}$, with the property
	\begin{equation}
		\label{eq:periodU}
		U(k + \pi) \, = \, \left(  \begin{array}{cc}  0 & 1 \\ -1 & 0  \end{array} \right) U(k) \left(  \begin{array}{cc}  0 & -1 \\ 1 & 0  \end{array} \right) \, .
	\end{equation}
\end{enumerate}
The second and third assumptions are not really crucial to the expansion of the formalism of the previous section to two-band models, but they hold for the two models of interest in this paper, and they will considerably simplify the calculations in parts \ref{sec:square} and \ref{part:6v}. Therefore, we chose to rely quite heavily on these properties already in the present section, in order to set up the proper notations for later parts; but let us stress that, if we were interested in more general two-band models in which these two properties do not hold, it would still be possible to generalize the calculation of the previous section. In fact, these additional assumptions reflect the very special structure of both the dimer model on the square lattice and the six-vertex model, compared to more generic two-band models. These properties come from the fact these two models are somehow one-band models in disguise. This is rather clear for the six-vertex model, because it is well-known that, when formulated in the standard language of the row-to-row transfer matrix---as opposed to the diagonal-to-diagonal transfer matrix which we use in this paper, see part \ref{part:6v}---the six-vertex model at $\Delta = 0$ is a genuine one-band model. Our setup, however, based on the DWIS, makes it more natural for us to work with the diagonal-to-diagonal transfer matrix, and therefore, with a two-band model. Presumably, using the row-to-row transfer matrix instead would lead us to a framework that should resemble the one of Okounkov and Reshetikhin \cite{OkounkovReshetikhin}. But let us not pursue this direction here, and come back to our purpose, which is to get the imaginary-time propagator of a model that satisfies the three assumptions above.

\vspace{0.3cm}

The canonical transformation 
\begin{equation*}
	\left( \begin{array}{c}  c^\dagger_+(k)    \\   c^\dagger_-(k)  \end{array} \right) \, =  \, U(k)  \left( \begin{array}{c}  c^\dagger(k)    \\   c^\dagger (k+\pi)  \end{array} \right) 
\end{equation*}
diagonalizes the transfer matrix,
\begin{equation*}
	T \, = \,  \exp \left[ -  \int_{-\frac{\pi}{2}}^{\frac{\pi}{2}} \frac{dk}{2\pi}  \varepsilon(k) \left(  c^\dagger_+(k) c_+(k) - c^\dagger_-(k) c_-(k)   \right)  \right]   \, .
\end{equation*}
Moreover, thanks to property (\ref{eq:periodU}), we have
\begin{equation*}
	\left\{  \begin{array}{rcl}  c^\dagger_+(k+\pi)    &  = &  c^\dagger_-(k) \\ 
	  c^\dagger_-(k+\pi)  &=& -c^\dagger_+(k) \, , 
	  \end{array} \right.
\end{equation*}
and this, together with the anti-periodicity of $\varepsilon(k)$, allows us to rewrite the transfer matrix as
\begin{equation*}
	T \, = \,  \exp \left[ -  \int_{-\pi}^{\pi} \frac{dk}{2\pi}  \varepsilon(k)   c^\dagger_+(k) c_+(k)  \right]   \, .
\end{equation*}
It is then obvious how to adapt the bosonization trick learned above. Essentially, we just need to replace the modes $c^\dagger(k)$ and $c(k)$ by $c^\dagger_+(k)$ and $c_+(k)$. We introduce a free real boson $\varphi_+$, with the propagator
\begin{equation*}
	\left< \varphi_+ (k-i \varepsilon )  \varphi_+ (k') \right> \, = \, \log \left[  \bra{\Psi_0} c^\dagger_+(k) c_+(k')  \ket{\Psi_0} \right] \, .
\end{equation*}
The r.h.s needs to be computed. This, in fact, is the only important difference with the previous section: instead of being simply $1/(2 i \sin ( \frac{k-k'}{2} - i0 ) )$, the correlator in the r.h.s depends on the change of basis $U(k)$. 
More precisely, we have
\begin{equation*}\fl
	\left(   \begin{array}{cc}  \bra{\Psi_0}  c_+^\dagger(k) c_+(k')  \ket{\Psi_0}  & \bra{\Psi_0}  c_+^\dagger(k) c_-(k')  \ket{\Psi_0}   \\
	\bra{\Psi_0}  c_-^\dagger(k) c_+(k')  \ket{\Psi_0}  & \bra{\Psi_0}  c_-^\dagger(k) c_-(k')  \ket{\Psi_0}   \end{array}    \right) \, = \,     U(k)  \left(   \begin{array}{cc}  \bra{\Psi_0}  c^\dagger(k) c(k')  \ket{\Psi_0}  & \bra{\Psi_0} c^\dagger(k) c(k'+\pi)  \ket{\Psi_0}   \\
	\bra{\Psi_0}  c^\dagger(k+\pi) c(k')  \ket{\Psi_0}  & \bra{\Psi_0}  c^\dagger(k+\pi) c(k'+\pi)  \ket{\Psi_0}   \end{array}    \right)   U^\dagger(k') \, ,
\end{equation*}
where the matrix in the middle in the r.h.s is $ \left(   \begin{array}{cc} 1/( 2 i \sin  \frac{k-k' }{2}   )   &  -1 /( 2 i \cos  \frac{k-k' }{2}  )  \\
	1/( 2 i \cos \frac{k-k' }{2}   )  & 1 /(2 i \sin  \frac{k-k' }{2}   )  \end{array}    \right) $. 
So we find that the propagator in the DWIS is given by the $(1,1)$ matrix element
\begin{equation}
	\label{eq:fU}
	\bra{\Psi_0} c^\dagger_+(k) c_+(k')  \ket{\Psi_0} \, = \,  \left[  U(k)  \left(   \begin{array}{cc} \frac{1}{ 2 i \sin \left( \frac{k-k' }{2}  \right) }   & \frac{-1}{ 2 i \cos \left( \frac{k-k' }{2}  \right) }  \\
	\frac{1}{ 2 i \cos \left( \frac{k-k' }{2}  \right) }  & \frac{1}{ 2 i \sin \left( \frac{k-k' }{2}  \right) }  \end{array}    \right)  U^\dagger(k')    \right]_{11} \, .
\end{equation}
The rest of the calculation is straightforward. We apply the bosonization rules
\begin{equation*}
	\left\{ \begin{array}{ccc}
		c^\dagger_+ (k) & \rightarrow & e^{ i \varphi_+(k)} \\
		c_+ (k) & \rightarrow & e^{ -i \varphi_+(k)} \\
		T & \rightarrow & \exp \left( - \int_{-\pi}^\pi \frac{dk}{2\pi} \varepsilon (k) \, \partial \varphi_+(k)  \right) \, ,
	\end{array} \right.
\end{equation*}
and we compute the imaginary-time propagator as in the previous section; the computational steps are exactly the same, so we do not repeat them here. In the end, we find that the propagator at $y=y'=0$ is:
\begin{eqnarray*}
\nonumber	\left< c^\dagger_+ (k,0) c_{+} (k',0) \right> & \equiv & \frac{  \bra{\Psi_0} T^N  c^\dagger_+ (k) c_{+}(k')  T^N  \ket{\Psi_0}  }{\bra{\Psi_0} T^{2N} \ket{\Psi_0} } \\
\nonumber	&=& e^{- i N \int \frac{dq}{2\pi} \varepsilon (q)  \,  \left[ \partial_{q}  \log \bra{\Psi_0} c^\dagger_+(k-i \epsilon) c_+(q-2i \epsilon)  \ket{\Psi_0}   +  \partial_{q}   \log \bra{\Psi_0} c^\dagger_+(k-i \epsilon) c_+(q+i \epsilon)  \ket{\Psi_0}  \right] } \\
\nonumber	&& \quad \times \; e^{i N \int \frac{dq}{2\pi} \varepsilon (q)  \,  \left[ \partial_{q}      \bra{\Psi_0} c^\dagger_+(q-2 i \epsilon) c_+(k')  \ket{\Psi_0}     + \partial_{q}  \log    \bra{\Psi_0} c^\dagger_+(q+ i \epsilon) c_+(k')  \ket{\Psi_0} \right] }  \\
 &&  \quad \quad  \times \;  \bra{\Psi_0} c^\dagger_+(k) c_+(k')  \ket{\Psi_0} \, .
\end{eqnarray*}
Finally, since the matrix elements of $U(k)$ are analytic, the only possible singularities in $\bra{\Psi_0} c^\dagger_+(k) c_+(k')  \ket{\Psi_0} $ come from the entries $1/(2i \sin  ( \frac{k-k'}{2} ))$ in (\ref{eq:fU}); therefore the kernel $\partial_{q} \log \bra{\Psi_0} c^\dagger_+(q) c_+(k)  \ket{\Psi_0} $ must have a single poles at $k = k' + 2\pi \mathbb{Z}$. Hence, the expressions in the two exponentials are again principal values. Bringing back the (trivial) dependence on $y$ and $y'$, we end up with the formula
\begin{equation}
	\label{eq:prop_interm_2}
	\left< c^\dagger_+ (k,y) c_{+} (k',y') \right> \, =\,    e^{ y  \varepsilon(k) - i N  \tilde{\varepsilon}_{U} (k)} \, e^{ -y'  \varepsilon(k') + i N  \tilde{\varepsilon}_{U} (k')}    \, \bra{\Psi_0} c^\dagger_+(k) c_+(k')  \ket{\Psi_0}  \, ,
\end{equation}
where the function $\tilde{\varepsilon}_{U}(k)$ is defined as
\begin{equation}
	\tilde{\varepsilon}_{U}(k) \, \equiv \,    {\rm p.v.} \int_{-\pi}^\pi \frac{dq}{2\pi} \varepsilon (q)  \; 2 \partial_{q} \log \bra{\Psi_0} c^\dagger_+(q) c_+(k)  \ket{\Psi_0}  \, .
\end{equation}
It does not seem possible to go further without being more specific about the matrix elements of $U(k)$. In the concrete models of parts \ref{sec:square} and \ref{part:6v}, we will compute these matrix elements, and the associated kernel $ 2 \partial_{q}\log \bra{\Psi_0} c^\dagger_+(q) c_+(k)  \ket{\Psi_0}  $. It will turn out that, both for the dimer model on the square lattice and for the six-vertex model, this kernel takes a particularly simple form. We will see that, in both cases, $\tilde{\varepsilon}_{U}(k)$ is the Hilbert transform of $\varepsilon (k)$, modulo a conjugation by a change of variables.
Let us mention that a quite similar bosonization procedure can be used to compute the partition functions of these models (see \ref{sec:partitionsfunctions} for details).

\newpage
\newcommand{\Te}{T_{\rm e}}
\newcommand{\To}{T_{\rm o}}

\section{Dimers on the honeycomb lattice}
\label{sec:honeycomb}
In the introduction, we focused on a simple toy model, the XX chain in imaginary time. A particularly neat feature of that model is the simple dispersion relation $\varepsilon(k)=-\cos k$, which implies that the Hilbert transform is also very simple: $\tilde{\varepsilon}(k)=-\sin k$. The correlations were then analyzed using the steepest descent method. Crucial in the whole analysis were of course the stationary points: we saw that they satisfy the equation $x+i y v(k)+R\tilde{v}(k)=0$, where $ v(k)=\frac{d\varepsilon}{dk}$ is the group velocity, and $\tilde{v}(k)=\frac{d\tilde{\varepsilon}}{dk}$ is its Hilbert transform. Solving this equation is elementary, as it is quadratic in $e^{ik}$. In particular the arctic curve is the set $(x,y)$ where the two solutions are degenerate; it turns out to be the circle $x^2+y^2=R^2$. Of course, this circle can be turned into an ellipse by a trivial rescaling of the dispersion relation.

\begin{figure}[htbp]
 \includegraphics[width=\textwidth]{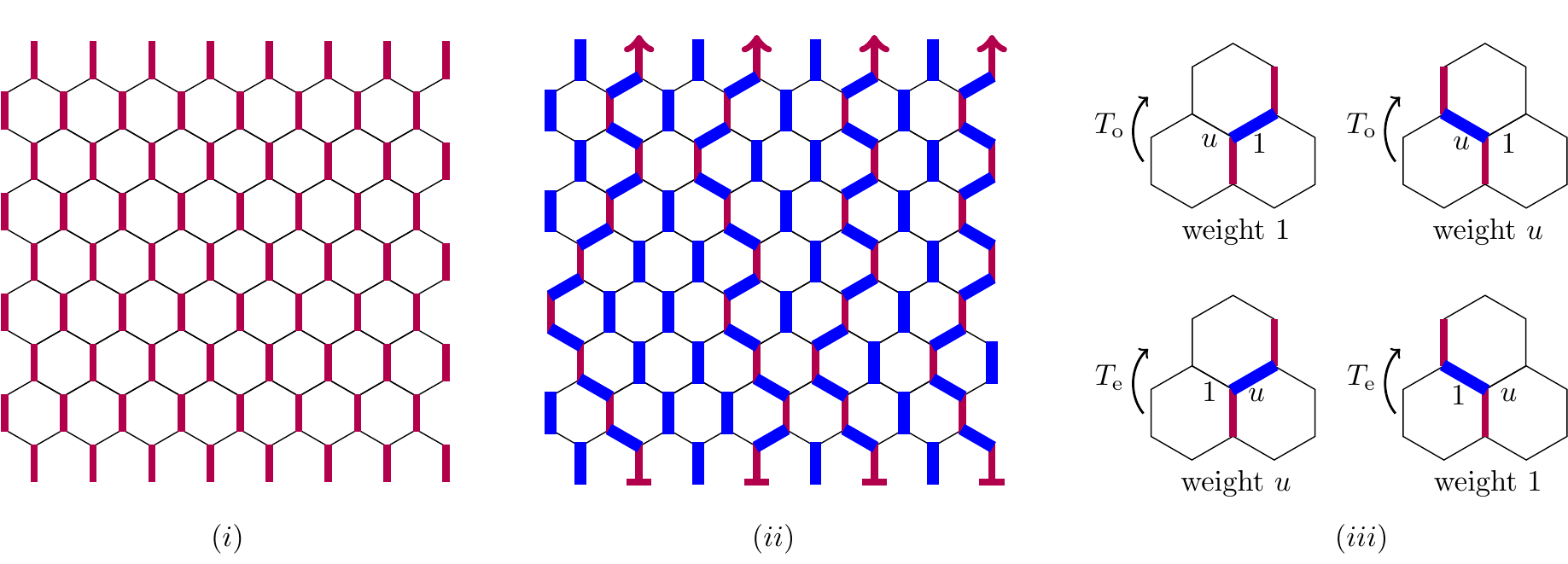}
 \caption{Mapping onto a system of non-intersecting lattice paths (particle trajectories). \emph{(i)}: Reference configuration with staggered dimers (shown in red). \emph{(ii)}: A configuration of dimers (shown in thick blue) is superimposed on the reference configuration. This generates four lattice paths which propagate from bottom to top. \emph{(iii)}: Rules for the trajectories. A particle is defined as a vertical link not occupied by a real (thick blue) dimer. Note the distinction between even and odd rows.}
 \label{fig:fermions_honey}
\end{figure}

There is, in fact, a whole (one-parameter) family of dispersion relations that lead to a quadratic equation for the stationary phase. It is obtained by considering a group velocity of the form
\begin{equation}\label{eq:vk}
 v(k)=\frac{i/2}{1+u e^{ik}}-\frac{i/2}{1+u e^{-ik}}=\frac{u \sin k}{1+u^2+2u \cos k},
\end{equation}
which is a sum of two simple poles in $e^{ik}$ and $e^{-ik}$. Here $u$ is some real parameter in the interval $(0,1)$; the XX case is recovered by considering $v(k)/u$ and by taking the limit $u\to 0$, or, equivalently, by rescaling the vertical coordinates as $y\to y/u$, $R\to R/u$ and then by taking the limit $u\to 0$. It turns out that there are a few models that realize the more general dispersion relation corresponding to (\ref{eq:vk}), and that are therefore essentially the same model, as they can be mapped onto each other. The first representative is the {\it multilayer PNG droplet} \cite{prahofer2002scale}. Another representative, on which we focus in this part, is the honeycomb dimer model on a strip, with boundary conditions at the top and bottom corresponding to the DWIS. In this model, the parameter $u$ is the fugacity of the dimers on a subset of the lattice edges (Fig. \ref{fig:fermions_honey}).

The dimer model can be mapped onto a system of free fermions as follows. We consider the ordered configuration of staggered dimers that is displayed in Fig.~\ref{fig:fermions_honey} (i); we refer to it as the \emph{reference configuration}. Then, any other dimer configuration may be superimposed on the reference configuration, as shown in Fig.~\ref{fig:fermions_honey} (ii). This procedure generates a set of lattice paths, all going upwards. It is convenient to think of these as trajectories of particles, where a particle is defined as a vertical link not occupied by a real dimer. The particle trajectories have the following important properties:
\begin{itemize}
 \item There is a one-to-one correspondence between the particle trajectories and the dimer coverings.
 \item The number of particles is conserved from one row to the next.
 \item The trajectories do not cross.
\end{itemize}
The last property implies that the statistics of the particles does not matter. We are thus free to consider them as fermions, and look at the transfer matrix as an operator acting on the fermion Fock space. The rules for the fermion trajectories are summarized in Fig.~\ref{fig:fermions_honey} (iii). Each fermion can go either to the nearest left or to the nearest right site from one row to the next. However, the two jumps are not symmetric: for even rows we assign a weight $u$ to a right jump, and $1$ to the left jump. For odd rows the rule is reversed, a weight $u$ is assigned to the left jump and a weight $1$ to the right jump.

\begin{figure}[htbp]
 \includegraphics[width=\textwidth]{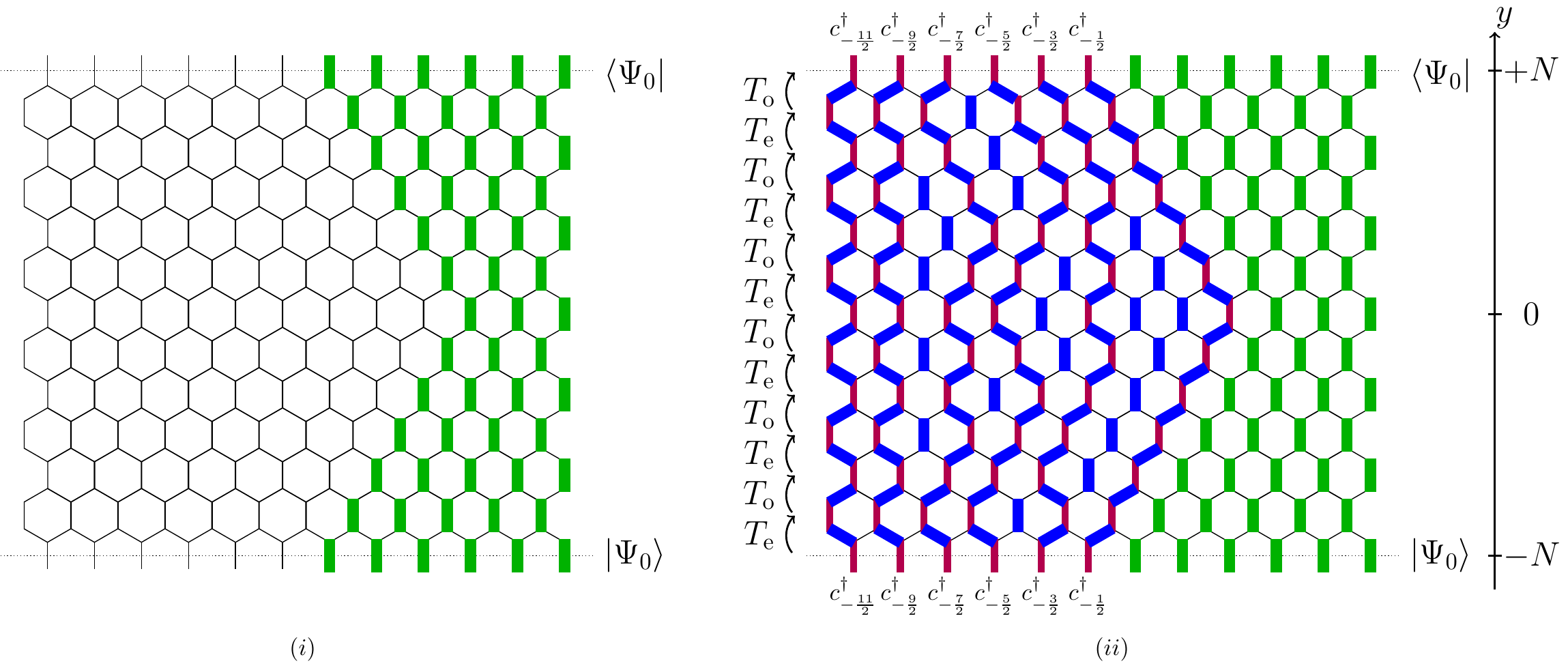}
 \caption{(\emph{i}) Geometry obtained by enforcing the DWIS as bottom and top boundary conditions. The corresponding dimers cross either of the dotted lines, and are shown in green. Other dimer occupancies are also set to one as a result; these are also shown in green. (\emph{ii}) Example of a possible set of particle trajectories, from bottom to top.}
 \label{fig:arctic_honey}
\end{figure}

Before we turn to the construction of the transfer matrix, let us explain the geometry induced by the DWIS boundary conditions. In dimer language, the DWIS $\ket{\Psi_0}=\prod_{x<0} c_x^\dag \ket{0}$ corresponds to all vertical links on the right \emph{occupied} by dimers and all links on the left \emph{empty}. As a result, many other dimer occupancies are automatically set to one, see Fig.~\ref{fig:arctic_honey}(i). An example of corresponding fermions trajectories is pictured in Fig.~\ref{fig:arctic_honey} (ii). 

\subsection{The transfer matrix}

\label{sec:honey_tm}
The transfer matrix acts on the fermion Fock space that represents the configurations of particles along a given row, namely on linear combinations of states of the form $\prod_{i} c_{x_i}^\dag \ket{0}$. It is however more elegant to look at the action of the transfer matrix on a single fermion creation operator. Given the rules shown in Fig.~\ref{fig:fermions_honey} (iii), a transfer matrix satisfying the rules
\begin{eqnarray*}
 T_{\rm e} c_x^\dag T_{\rm e}^{-1}&=& c_x^\dag+u c_{x+1}^\dag,\\
  T_{\rm o} c_x^\dag T_{\rm o}^{-1}&=&u c_{x-1}^\dag + c_{x}^\dag,
\end{eqnarray*}
generates all possible dimer configurations, with the correct weights. The action on any many-fermion configuration may be reconstructed using  $T_{\rm e}\prod_{i} c_{x_i}^\dag \ket{0} =\left(\prod_{i} T_{\rm e} c_{x_i}^\dag T_{\rm e}^{-1} \right)T_{\rm e}\ket{0}$ and the fact that $T_{\rm e}\ket{0}=\ket{0}$ (the same goes for $T_{\rm o}$). In particular, the partition function is simply $Z=\braket{\Psi_0|\left(T_{\rm o} T_{\rm e}\right)^{N}|\Psi_0}$. Notice that $T_{\rm e}$ and $T_{\rm o}$ are {\it not} Hermitian; however $T^2=T_{\rm o} T_{\rm e}$ is. Therefore, $T^2$ is the natural hermitian transfer matrix we focus on, and to which one can readily apply the formalism of Part~\ref{sec:ktr}. The action of $\Te$, $\To$, $T^2$ on the Fourier modes of the fermions (recall that they are defined as $ c^\dag(k)=\sum_{x\in \mathbb{Z}+1/2} e^{ikx} c_x^\dag$) is diagonal:
\begin{eqnarray*}
 \Te \,c^\dag(k)\,\Te^{-1}&=&\left(1+u e^{ik}\right) c^\dag(k),\\
 \To \,c^\dag(k)\,\To^{-1}&=& \left(1+ue^{-ik}\right) c^\dag(k),\\
 T^2 \,c^\dag(k)\, T^{-2}&=&\left(1+u^2 +2u \cos k\right) c^\dag(k).
\end{eqnarray*}
More explicitely, $T^2$ reads
\begin{equation*}
 T^2=\exp\left(-2 \int \frac{dk}{2\pi} \varepsilon(k) c^\dag(k)c(k)\right) \;, \qquad \quad {\rm with}\qquad \varepsilon(k)=-\frac{1}{2}\log \left(1+u^2+2u \cos k\right).
\end{equation*}
Notice that the derivative of the dispersion relation $\varepsilon(k)$ is exactly Eq.~(\ref{eq:vk}).

\subsection{Fermionic correlators: exact finite-size formulas}

In the rest of this part, we assume for simplicity that $N$ is \emph{even}. The vertical axis is labelled by \emph{integers} $-N\leq y\leq N$, while the horizontal axis is still labelled by half integers $x\in \mathbb{Z}+1/2$. 
The expectation value of an observable $\mathcal{O}$ at position $x$ and $y$ is
\begin{eqnarray}
	\label{eq:hex_eveny}
\braket{\mathcal{O}_{x,y}}=\left\{
\begin{array}{ccccc}
 \displaystyle{\frac{  \bra{\Psi_0} (T^2)^{\frac{N-y}{2}} \mathcal{O}_x (T^2)^{\frac{N+y}{2}}  \ket{\Psi_0}  }{ \bra{\Psi_0} T^{2N} \ket{\Psi_0}    } }&&,&& y \,{\rm even}\\\\
 \displaystyle{\frac{  \bra{\Psi_0} (T^2)^{\frac{N-y-1}{2}} T_{\rm o}\mathcal{O}_x T_{\rm o}^{-1} (T^2)^{\frac{N+y+1}{2}} \ket{\Psi_0} }{ \bra{\Psi_0} T^{2N} \ket{\Psi_0} }} &&,&& y {\rm odd} \, .
\end{array}
\right.
\end{eqnarray}
Similar formulae hold for more than one local observable. In particular, the fermion propagator is, using the result of part \ref{sec:ktr},
\begin{equation}\fl
	\label{eq:prop_honey}
	\left<  c_{x,y}^\dagger c_{x',y'} \right> \, = \,  \int \frac{dk}{2\pi} e^{- i k x + y \varepsilon(k) - i N \tilde{\varepsilon}(k)} r_y(k) \int \frac{dk'}{2\pi} e^{ i k' x' - y' \varepsilon(k) + i N \tilde{\varepsilon}(k')}    s_{y'}(k') \, \bra{\Psi_0} c^\dagger(k) c(k') \ket{\Psi_0} \, ,
\end{equation}
where the propagator in the DWIS is still $\bra{\Psi_0} c^\dagger(k) c(k') \ket{\Psi_0}  = 1/(2 i \sin \frac{k-k'}{2})$. The factors $r_y(k)$ and $s_y(k)$ are
\begin{equation*}\fl
	r_y(k) = s_y(k) = 1 \qquad {\rm if} \quad y \quad {\rm is \;even} \, , \qquad \qquad \qquad \quad
		r_y(k) \,=\, \frac{1}{s_y(k)} \,=\, \left( \frac{1+ u e^{- ik}}{ 1+ u e^{i k}} \right)^{\frac{1}{2}}  \quad {\rm if} \quad y \quad {\rm is \;odd} \, .
\end{equation*}
These coefficients follow from the exponentials in (\ref{eq:hex_eveny}): for $y$ odd, one uses the relations $\To c^\dag (k)\To^{-1}=(1+u e^{-ik})c^\dag (k)$ and $\To c(k)\To^{-1} =(1+u e^{-ik})^{-1}c(k)$, and one takes into account the small shift $y\rightarrow y+1$ in (\ref{eq:hex_eveny}) compared to the formulae in part \ref{sec:ktr}. This gives $r_y(k) = 1/s_y(k) = e^{\varepsilon(k)} ( 1+u e^{-i k})$, as claimed. 
In what follows, it is convenient to use again the notation from the introduction, see Eq.~(\ref{eq:notation_doteq}). We rewrite (\ref{eq:prop_honey}) as
\begin{equation}
\label{eq:ident_hex}
	\left\{  \begin{array}{rcl} c^\dagger_{x,y}   & \doteq &  \displaystyle \int_{-\pi}^\pi \frac{dk}{2\pi} e^{-i k x + y \varepsilon (\kappa)  - i N \tilde{\varepsilon}(\kappa) }  \,  r_{y} (k)  \, c^\dagger(k) \\ \\
	c_{x,y}   & \doteq & \displaystyle  \int_{-\pi}^\pi \frac{dk}{2\pi} e^{i k x-y \varepsilon (\kappa)  + i N \tilde{\varepsilon}(\kappa) }  \,  s_{y} (k)  \, c(k) \, , \end{array} \right.
\end{equation}
where the dot means that this is an equality that holds only when the l.h.s is evaluated within the bracket $\left< . \right>$, and the r.h.s is the expectation value in the DWIS $\bra{\Psi_0} . \ket{\Psi_0}$. Finally, we note that, here, the Hilbert transform of the dispersion relation is (see the Appendix \ref{appx:hilb})
 \begin{equation*}
  \varepsilon(k)=-\frac{1}{2}\log \left(1+u^2+2u \cos k\right) \qquad\Rightarrow\qquad \tilde{\varepsilon}(k)=\frac{i}{2}\log \left(\frac{1+u e^{ik}}{1+u e^{-ik}}\right) \, .
 \end{equation*}

 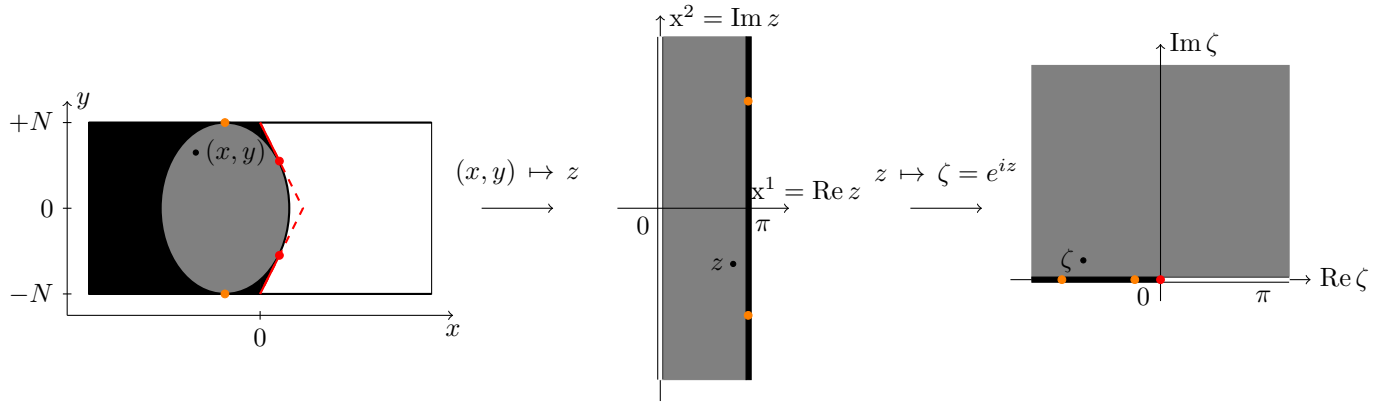
\begin{figure}[hbt]
	\begin{tikzpicture}[scale=0.95]
		\begin{scope}[scale=0.6]
			\filldraw[black] (-5,2) rectangle (-1,-2);
			\filldraw[black] (-1,2) -- (0,0) -- (-1,-2);
			\filldraw[white] (-0.5,2) rectangle (2,-2);
			\draw (-5,-2) rectangle (3,2);
			\draw[thick] (-5,-2) -- (3,-2);
			\draw[thick] (-5,2) -- (3,2);
			\filldraw[thick,fill=gray] (-1.82,0) ellipse (1.5cm and 2cm);
			\draw[red,dashed,thick] (-1,2) -- (0,0) -- (-1,-2);
			\draw[red,thick] (-1,2) -- (-0.55,1.1);
			\filldraw[red] (-0.55,1.1) circle (0.9mm);
			\draw[red,thick] (-1,-2) -- (-0.55,-1.1);
			\filldraw[red] (-0.55,-1.1) circle (0.9mm);
			\filldraw[orange] (-1.82,2) circle (0.9mm);
			\filldraw[orange] (-1.82,-2) circle (0.9mm);
			\draw[->] (-5.5,-2.5) -- (-5.5,2.5) node[right]{$y$};
			\draw[->] (-5.5,-2.5) -- (3.5,-2.5) node[below]{$x$};
			\draw (-5.4,0) -- (-5.6,0) node[left]{$0$};
			\draw (-5.4,2) -- (-5.6,2) node[left]{$+N$};
			\draw (-5.4,-2) -- (-5.6,-2) node[left]{$-N$};
			\draw (-1,-2.4) -- (-1,-2.6) node[below]{$0$};
			\filldraw (-2.5,1.3) circle (0.6mm) node[right]{$(x,y)$};
		\end{scope}
		\draw[->] (2.5,0) -- (3.5,0);
		\draw (3,0.5) node{$(x,y) \, \mapsto \, z$};
		\begin{scope}[xshift=5cm, scale=0.6]
			\draw[->] (0,-4.5) -- (0,4.5) node[right]{${\rm x}^2 = {\rm Im } \, z$};
			\filldraw[fill=gray,gray] (0,-4) rectangle (2,4);
			\filldraw[fill=black] (2,-4) rectangle (2.12,4);			
			\filldraw[fill=white,draw=white] (-0.06,-4) rectangle (0.06,4);	
			\draw[->] (-1,0) -- (3,0);
			\draw (3.4,0) node[above]{${\rm x}^1 = {\rm Re}\, z $};
			\draw (-0.06,-4) rectangle (-0.06,4);	
			\draw (0.06,-4) rectangle (0.06,4);	
			\draw (-0.4,-0.4) node{$0$};
			\draw (2.4,-0.4) node{$\pi$};
			\filldraw (1.7,-1.3) circle (0.6mm) node[left]{$z$};
			\filldraw[orange] (2.05,-2.5) circle (0.9mm);
			\filldraw[orange] (2.05,2.5) circle (0.9mm);
		\end{scope}
		\draw[->] (8.5,0) -- (9.5,0);
		\draw (9,0.5) node{$z \, \mapsto \, \zeta = e^{i z}$};
		\begin{scope}[xshift=12cm, yshift=-1cm, scale=0.6]
			\draw[->] (-3.5,0) -- (3.5,0) node[right]{${\rm Re}\, \zeta $};
			\filldraw[fill=gray,gray] (-3,0) rectangle (3,5);
			\filldraw[fill=black] (0,-0.06) rectangle (-3,0.06);			
			\filldraw[fill=white,draw=white] (0,-0.06) rectangle (3,0.06);
			\draw[->] (0,-0.5) -- (0,5.5) node[right]{${\rm Im } \, \zeta$};
			\draw (3,-0.06) -- (0,-0.06);	
			\draw (3,0.06) -- (0,0.06);	
			\draw (-0.4,-0.4) node{$0$};
			\draw (2.4,-0.4) node{$\pi$};
			\filldraw (-1.8,0.45) circle (0.6mm) node[left]{$\zeta$};
			\filldraw[orange] (-0.6,0) circle (0.9mm);
			\filldraw[orange] (-2.3,0) circle (0.9mm);
			\filldraw[red] (0,0) circle (0.9mm);
		\end{scope}				
	\end{tikzpicture}
	\caption{The critically fluctuating region $X^2+y^2 < N^2$ is mapped to an infinite strip of width $\pi$ by $( x,y) \mapsto z(x,y)$, see Eq. (\ref{eq:zhoney}). The red curve is $x = \frac{N-|y|}{2}$; it is tangent to the ellipse at the two red points. Those two points are mapped to ${\rm x}^2 = \pm  \infty$ on the strip. The two points $(x,y) = (-\frac{u^2}{1-u^2}N,\pm N)$ in orange are mapped to $({\rm x}^1,{\rm x}^2) = (\pi, \mp {\rm arcth}\left( \frac{1-u^2}{1+u^2}\right))$. The strip itself can be mapped conformally to the upper half-plane, $\zeta = e^{i z}$. Because this last mapping is conformal, it makes no difference to use the coordinate $z$ or $\zeta$ as an isothermal coordinate system. Here we chose to use $z$.}
	\label{fig:strip_honey}
\end{figure}
 
 \subsection{Fermionic correlators: scaling limit}
 
Exactly like in the introduction, we apply the steepest descent method to obtain the large scale correlations in our honeycomb dimer model. The stationary points in (\ref{eq:ident_hex}) are the solutions to ($v(k) = \frac{d}{dk} \varepsilon(k)$ and $\tilde{v}(k) = \frac{d}{dk} \tilde{\varepsilon}(k)$):
\begin{equation*}
 x+iy v(k)+N\tilde{v}(k)=0 \, .
\end{equation*}
This equation possesses two roots, $k=z(x,y)$ and $k=-z^*(x,y)$, with
 \begin{equation}\label{eq:zhoney}
 z(x,y)={\rm arccos} \left(\frac{(1+u^2)x-N u^2}{u\sqrt{(N-2x)^2-y^2}}\right)-i\, {\rm arctanh}\left( \frac{y}{N-2x}\right) \, ,
\end{equation}
or equivalently $\zeta (x,y) \, = \, e^{i z(x,y)} \, = \, \frac{ \frac{1+u^2}{u} x - N u  + i\sqrt{R^2 - X^2 - y^2}  }{N-2 x- y}$,
where $X=\frac{1-u^2}{u}x+N u$. The two solutions are usually distinct; they are degenerate only if 
 \begin{equation*}
  X^2+y^2 \, = \, N^2   \, .
 \end{equation*}
This curve, an ellipse, is the arctic curve of the model. Inside this ellipse the long-range correlations are critical. Around the points where the phase is stationary, the argument in the exponential in (\ref{eq:ident_hex}) is approximated by
\begin{equation*}
	\left\{ \begin{array}{ccl}
		 e^{-ikx+y\varepsilon(k)-i N \tilde{\varepsilon}(k)} & \simeq & e^{-i \varphi-\frac{i}{2}(k-z)^2 e^{\sigma+i\theta}} \qquad \,{\rm around}\quad k=z \\
 		e^{-ikx+y\varepsilon(k)-i N\tilde{\varepsilon}(k)}& \simeq & e^{i \varphi^*+\frac{i}{2}(k+z^*)^2 e^{\sigma-i\theta}}\qquad {\rm around}\quad k=-z^* \, ,
	\end{array} \right.
\end{equation*}
where
\begin{equation}
	\label{eq:phi_honey}
 	\left\{ \begin{array}{rcl}
		\varphi  & \equiv &\displaystyle z(x,y)x+iy \varepsilon(z(x,y))+N\tilde{\varepsilon}(z(x,y)) \\ \\
		 e^{\sigma +i \theta} &\equiv & \displaystyle i  y\, \frac{d^2\varepsilon}{dk^2}(z(x,y)) \, + \, N\,\frac{d^2\tilde{\varepsilon}}{dk^2}(z(x,y)) \\
 		&=&\frac{u\sqrt{(N-2x)^2-y^2}\sqrt{N^2-X^2-y^2}}{(1-u^2)\sqrt{N^2-y^2}}\times\exp\left[-i{\rm arccos}\left(\frac{N(N-2x)-\frac{1+u^2}{1-u^2}y^2}{\sqrt{N^2-y^2}\sqrt{(N-2x)^2-y^2}}\right)\right]  \, .
		\end{array} \right.
\end{equation}
Performing the gaussian integration, we find that the scaling limit of (\ref{eq:ident_hex}) is
\begin{equation}
	\label{eq:scaling_ident_hex}
	\left\{\begin{array}{rcl}
		c^\dagger_{x,y} &\doteq& \displaystyle \frac{e^{-i \frac{\pi}{4}}}{\sqrt{2\pi}} e^{- \frac{\sigma + i \theta}{2}} e^{- i \varphi} r_y(z) c^\dagger(z) \, + \, \frac{e^{i \frac{\pi}{4}}}{\sqrt{2\pi}} e^{- \frac{\sigma - i \theta}{2}} e^{i \varphi^*} r_y(-z^*) c^\dagger(-z^*)  \\ \\
		c_{x,y} &\doteq& \displaystyle \frac{e^{i \frac{\pi}{4}}}{\sqrt{2\pi}} e^{- \frac{\sigma + i \theta}{2}}   e^{ i \varphi} s_y(z) c(z) \, + \, \frac{e^{-i \frac{\pi}{4}}}{\sqrt{2\pi}} e^{- \frac{\sigma - i \theta}{2}} e^{-i \varphi^*} s_y(-z^*) c(-z^*)  \, .
	\end{array} \right.
\end{equation}

\subsection{Identification of the Dirac theory inside the arctic ellipse}\label{sec:dirac_honey}

Again, we proceed by repeating the procedure of the introduction. There are two important differences though. The first is in the relation between the lattice degrees of freedom and the continuous fields. Here we chose
\begin{equation}\fl
\begin{array}{lcr}
	\left\{ \begin{array}{rcl}
		c^\dagger_{x,y} &=& \displaystyle \frac{1}{\sqrt{2\pi}} \left[ \psi^\dagger(x,y) \, + \,   \overline{\psi}^\dagger(x,y) \right] \\ 
		c_{x,y} &=& \displaystyle \frac{1}{\sqrt{2\pi}} \left[ \psi(x,y) \, + \,  \overline{\psi}(x,y) \right]
	\end{array}\right. &\qquad& (y \quad{\rm even})  \\ \\
	\left\{ \begin{array}{rcl}
		c^\dagger_{x,y} &=& \displaystyle \frac{1}{\sqrt{2\pi}} \left[ \psi^\dagger(x,y+1) + u \, \psi^\dagger(x+1,y+1) \, + \, \overline{\psi}^\dagger(x,y+1) + u \, \overline{\psi}^\dagger(x+1,y+1)\right] \\ 
		c_{x,y} &=& \displaystyle \frac{1}{\sqrt{2\pi}} \left[ \psi(x,y-1)+ u \, \psi(x+1,y-1)  \, + \,   \overline{\psi}(x,y-1) + u  \, \overline{\psi}(x+1,y-1) \right] 
	\end{array}\right. &\qquad& (y \quad{\rm odd}) \, .
\end{array}
\end{equation}
We make this choice so as to naturally incorporate the coefficients $r_y(k)$ and $s_y(k)$ above; the field $\psi^\dagger$ is then essentially $e^{-i\frac{\pi}{4}} e^{-\frac{\sigma + i\theta}{2}} e^{-i \varphi} c^\dagger(z)$, and similar relations for $\overline{\psi}^\dagger$, $\psi$ and $\overline{\psi}$, see Eq. (\ref{eq:scaling_ident_hex}). It follows that the short-distance behavior of the propagator of the continuous fields is
\begin{equation*}\fl
	\left<  \psi^\dagger(x,y) \psi(x',y') \right>  \, \sim \,  e^{-\frac{\sigma + i \theta}{2} -\frac{\sigma' + i \theta'}{2} }  \frac{e^{-i (\varphi - \varphi')}}{i (z-z')} \quad , \qquad  \qquad \left<  \overline{\psi}^\dagger(x,y) \overline{\psi}(x',y') \right>  \, \sim \,  e^{-\frac{\sigma - i \theta}{2} -\frac{\sigma' - i \theta'}{2} }  \frac{e^{i (\varphi^* - \varphi'^*)}}{i (z^*-z'^*)} \, .
\end{equation*}
Here we see the second main difference with the case treated in the introduction: the appearance of the angle $\theta = \theta(x,y)$, which was absent from the propagator in the XX chain, see Eq. (\ref{eq:prop_short}). This angle, however, does not change the fact that $\left< \psi^\dagger(z) \psi (z') \right>$ is the propagator of a euclidean Dirac theory with action (\ref{eq:Dirac}). The only difference with the introduction is that, now, the action written in $z$-coordinates must be
\begin{equation*}
 	\displaystyle S \, = \, \frac{1}{\pi} \int d^2 z   \, e^{\sigma}  \left[  
 	\left( \begin{array}{cc} \psi^\dagger  & \overline{\psi}^\dagger  \end{array}  \right)    \left( \begin{array}{cc}  -e^{i\theta} [\frac{i}{2} \overset{\leftrightarrow}{\partial}_{\bar{z}}    +  (\partial_{\bar{z}} \varphi)]   & 0  \\  0  &   e^{-i\theta}[\frac{i}{2} \overset{\leftrightarrow}{\partial}_{z}    -  (\partial_{z} \varphi^*)]      \end{array}\right)    \left(  \begin{array}{c}  \psi \\ \overline{\psi} \end{array}  \right)  \right] \, ,
\end{equation*}
while in the introduction there was no phase $e^{i \theta}$. But, comparing this to the covariant expression (\ref{eq:Dirac}), we see that this phase factor can be absorbed in the choice of the tetrad,
\begin{equation}
	\label{eq:tetrad_honey}
	e^{\mu}_{\phantom{a}a} \, = \, e^{- \sigma} \left( \cos \theta \, \delta^\mu_{\phantom{a}a}  \, - \, \sin \theta\,  \varepsilon^\mu_{\phantom{a}a}  \right)
\end{equation}
where $\varepsilon^{1}_{\phantom{1}2} = -\varepsilon^{2}_{\phantom{1}1} = 1$ and $\varepsilon^{1}_{\phantom{1}1} = \varepsilon^{2}_{\phantom{1}2} =0$. In complex coordinates (using complex coordinates both in the tangent space and in the moving frame) the components of the tetrad are $e^z_{\phantom{a}z}  =  e^{-\sigma + i \theta}$, $e^{\bar{z}}_{\phantom{a}\bar{z}} = e^{-\sigma - i \theta}$ and $e^z_{\phantom{a}\bar{z}} = e^{\bar{z}}_{\phantom{a}z}=0$. Multiplied by the Jacobian from the volume element $\sqrt{g} d^2 {\rm x} = |e|^{-1} d^2 {\rm x}$, this gives $|e|^{-1} e^{z}_{\phantom{z}z} = e^{\sigma + i\theta}$ in the action above. Thus, we end up with exactly the same conclusion as in the introduction: the long-range correlations inside the arctic ellipse follow from the Dirac action (\ref{eq:Dirac}), with the tetrad (\ref{eq:tetrad_honey}) and with background gauge potentials that are again derivatives of the function $\varphi(x,y)$,
\begin{equation*}
	A^{({\rm v})}_\mu \, =\, - i \partial_{\mu}  {\rm Im}\, \varphi \qquad {\rm and}\qquad  A^{({\rm a})}_\mu \, =\, - \partial_{\mu}  {\rm Re} \,\varphi \, . 
\end{equation*}
Finally, there is one point that we left aside in the introduction, that we wish to discuss now: the boundary conditions obeyed by the continuous fields. We use the coordinate system $({\rm x}^1, \,{\rm x}^2) \, = \, ({\rm Re} z , \, {\rm Im} z)$; the boundaries are then at ${\rm x}^1 = 0$ and ${\rm x}^1=\pi$. We claim that the boundary conditions are
\begin{equation*}
	\begin{array}{lll}
	{\rm x}^1 \, = \, 0 \,: &\quad & \left\{ \begin{array}{rcl}
		e^{i\frac{\pi}{4} + i {\rm Re}\, \varphi} \psi^\dagger & = & e^{-i \frac{\pi}{4} } e^{-i  {\rm Re}\,  \varphi}  \overline{\psi}^\dagger \\
		e^{-i\frac{\pi}{4} - i {\rm Re}\, \varphi} \psi & = & e^{i \frac{\pi}{4} } e^{i  {\rm Re}  \,\varphi}  \overline{\psi} \\
	\end{array} \right.  \\ \\
	{\rm x}^1 \, = \, \pi \quad {\rm and } \quad |{\rm x}^2|< {\rm arcth}\left( \frac{1-u^2}{1+u^2} \right) : &\quad & \left\{ \begin{array}{rcl}
		e^{i\frac{\pi}{4} + i {\rm Re} \,\varphi} \psi^\dagger & = & -\, e^{-i \frac{\pi}{4} } e^{-i  {\rm Re} \, \varphi}  \overline{\psi}^\dagger \\
		e^{-i\frac{\pi}{4} - i {\rm Re}\, \varphi} \psi & = & \, -\, e^{i \frac{\pi}{4} } e^{i  {\rm Re} \, \varphi}  \overline{\psi}
	\end{array} \right.  \\ \\
	{\rm x}^1 \, = \, \pi \quad {\rm and } \quad |{\rm x}^2|> {\rm arcth}\left( \frac{1-u^2}{1+u^2} \right) : &\quad & \left\{ \begin{array}{rcl}
		e^{i\frac{\pi}{4} + i {\rm Re}\, \varphi} \psi^\dagger & = & e^{-i \frac{\pi}{4} } e^{-i  {\rm Re}\,  \varphi}  \overline{\psi}^\dagger \\
		e^{-i\frac{\pi}{4} - i {\rm Re}\, \varphi} \psi & = & e^{i \frac{\pi}{4} } e^{i  {\rm Re} \, \varphi}  \overline{\psi} \, .
	\end{array} \right. 
	\end{array}
\end{equation*}
To see that, one can focus on the propagator in the DWIS, $\bra{\Psi_0} c^\dagger(z) c(z') \ket{\Psi_0}$, which is equal to $1/(2i \sin \frac{z-z'}{2})$. It is clear from that formula that $c^\dagger (z) = c^\dagger (-z^*)$ if ${\rm x}^1 = {\rm Re} \, z = 0$. Then it follows from (\ref{eq:scaling_ident_hex}) that $e^{i \frac{\pi}{4}} e^{\frac{\sigma + i \theta}{2}} e^{i \varphi} \psi^\dagger \, = \, e^{-i \frac{\pi}{4}} e^{\frac{\sigma-i \theta}{2}} e^{-i \varphi^*} \overline{\psi}^\dagger$. Moreover, one can check that $\theta = 0$ if ${\rm x}^1 = 0$; this gives the boundary condition relating $\psi^\dagger$ and $\overline{\psi}^\dagger$ above. The other cases are similar; we use the fact that $c^\dagger (z) = -c^\dagger (-z^*)$ if ${\rm x}^1 = \pi$, and that $\theta =0$ or $\pi$ if ${\rm x}^1 = \pi$ and $|{\rm x}^2 | < {\rm arcth} \left( \frac{1-u^2}{1+u^2} \right)$ or $|{\rm x}^2 | > {\rm arcth} \left( \frac{1-u^2}{1+u^2}\right)$ respectively (see Fig. \ref{fig:strip_honey}). Notice that these boundary conditions look complicated only because we want to keep the dependence on ${\rm Re} \varphi$; clearly, if we took the liberty of making an axial gauge transformation $\psi \rightarrow e^{ i {\rm Re}\, \varphi} \psi$, $\overline{\psi} \rightarrow e^{- i {\rm Re}\, \varphi} \overline{\psi}$, then the phases above would drop. The remaining factors $e^{\pm i \frac{\pi}{4}}$ can be traced back to the fact that, in the $z$-coordinate system, the boundary is vertical. Using the $\zeta$-coordinates instead (see Fig. \ref{fig:strip_honey}), the boundary would simply be the real axis, and these phase factors would drop as well. Therefore, the only meaningful information in the boundary conditions above is the sign.

\subsection{Density profile}
\label{sec:density_honey}

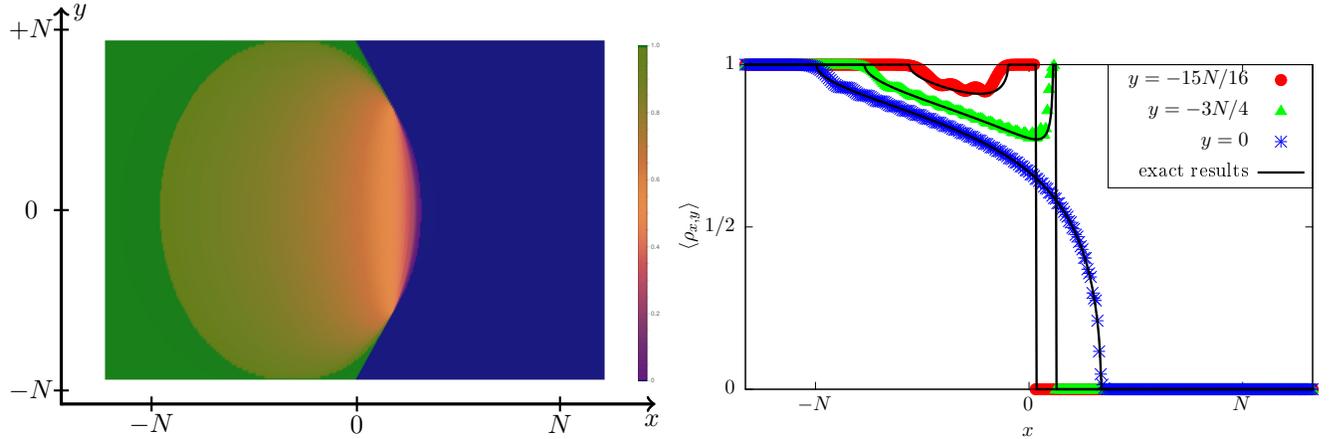
\begin{figure}[htbp]
\begin{tikzpicture}
\node at (0,0) {\includegraphics[height=4.7cm]{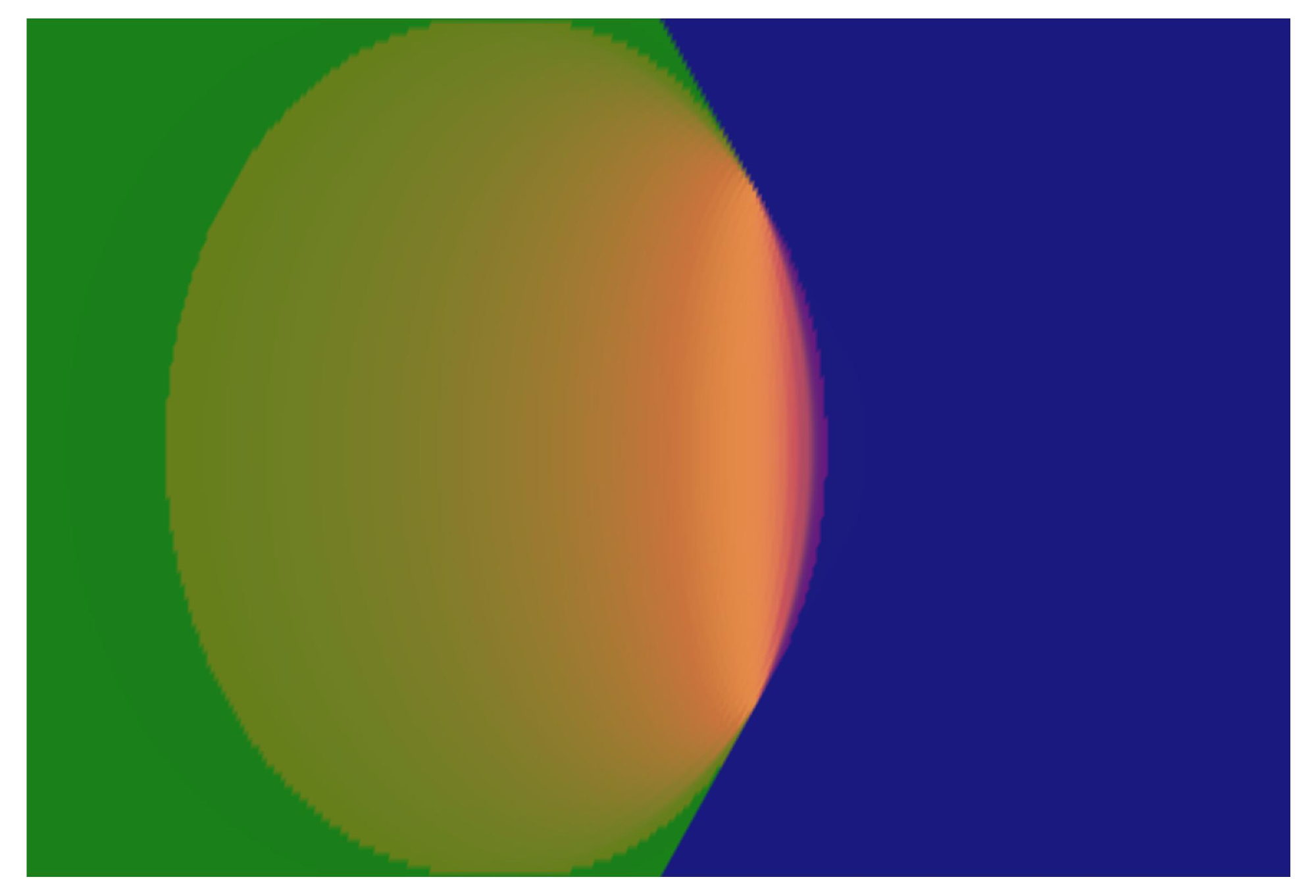}};
\node at (3.9,-0.115) {\includegraphics[height=4.8cm]{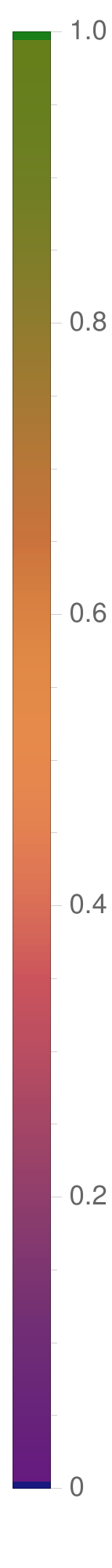}};
\draw[very thick,->] (-3.9,-2.6) -- (-3.9,2.7);\draw (-3.65,2.63) node {$y$};
\draw[thick] (-4,-2.4)-- (-3.8,-2.4);\draw (-4.3,-2.4) node {$-N$};
\draw[thick] (-4,0)-- (-3.8,0);\draw (-4.3,0) node {$0$};
\draw[thick] (-4,2.4)-- (-3.8,2.4);\draw (-4.3,2.4) node {$+N$};
\draw[very thick,->] (-3.9,-2.58) -- (4,-2.58);\draw (3.95,-2.8) node {$x$};
\draw[thick] (0.03,-2.68) -- (0.03,-2.48);\draw (0.03,-2.85) node {$0$};
\draw[thick] (-2.7,-2.68) -- (-2.7,-2.48);\draw (-2.7,-2.85) node {$-N$};
\draw[thick] (2.73,-2.68) -- (2.73,-2.48);\draw (2.73,-2.85) node {$N$};
\end{tikzpicture}
\hfill
 \includegraphics[height=5.3cm]{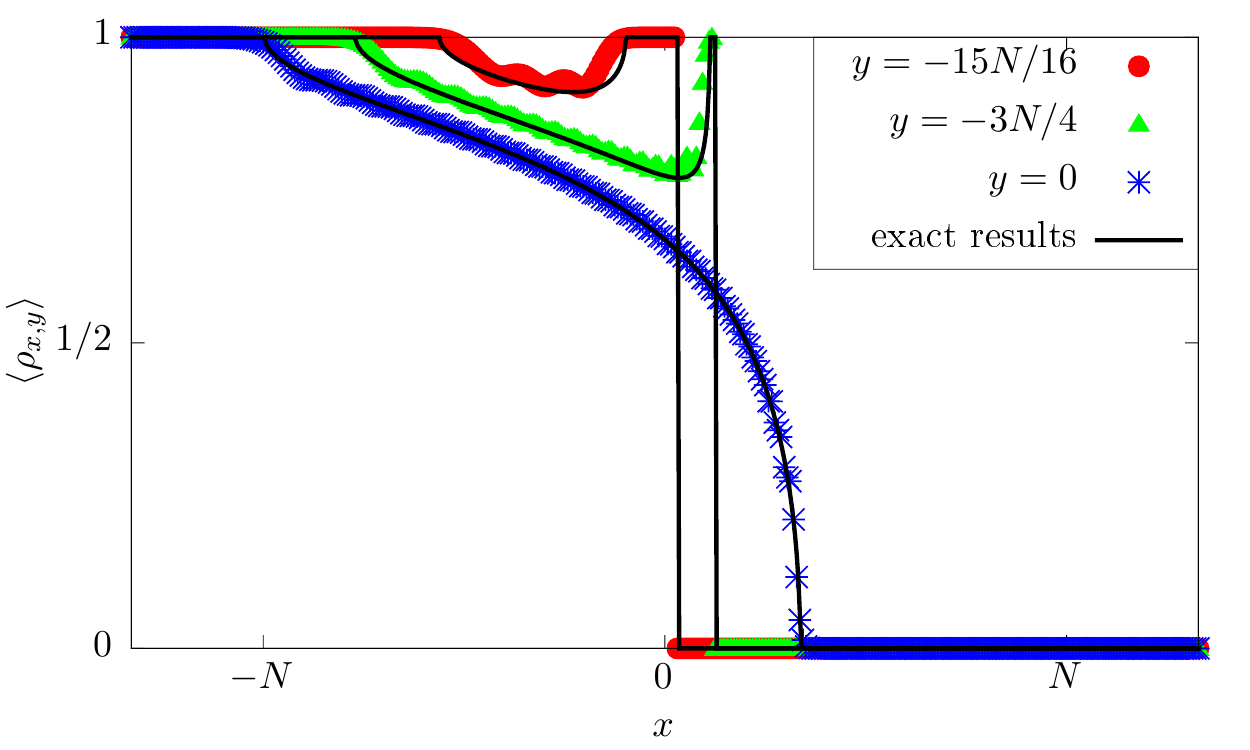}
 \caption{Left: Numerical density profile in a system of height $N=128$, for an anisotropy $u=1/2$. The simulation were performed on a finite system of size $L=512$. The color convention is the following: Green is $\rho=1$, Blue is $\rho=0$, and intermediate values are shown with a mixture of green, yellow and red. The fluctuating region is in perfect qualitative agreement with Eq.~(\ref{eq:densityhexa}), which gives the inside of an ellipse $(N/2+3x/2)^2+y^2<N^2$. Right: Quantitative comparison between a numerical simulation for the same system size and the exact asymptotic solution (\ref{eq:densityhexa}) for $y=0$ (blue stars), $y=-N+N/4=-3N/4$ (green triangles), $y=-N+N/16=-15N/16$ (red circles). The black lines are the analytical solution. The agreement is excellent.}
 \label{fig:densityplots}
\end{figure}

What is the density profile inside the ellipse ? Like in the introduction, one way of obtaining the density is to take the limit $x' \rightarrow x$ of the propagator $\left< c^\dagger_{x,y} c_{x',y} \right>$. From the previous paragraphs, we know that the short-distance behavior of the propagator is
\begin{equation*}
	\left< c^\dagger_{x,y} c_{x',y} \right> \, \simeq \, \frac{e^{-\frac{\sigma+\sigma'}{2}}}{2\pi i} \left[   \frac{e^{-i \frac{\theta+\theta'}{2}}  e^{-i (\varphi - \varphi')}   }{z-z'}    -    \frac{e^{i \frac{\theta+\theta'}{2}}  e^{i (\varphi^* - \varphi'^*)}  }{z^*-z'^*}   \right] \, .
\end{equation*}
Expanding the $\varphi$-exponentials to the first order, one finds
\begin{equation*}
	\left< c^\dagger_{x,y} c_{x',y} \right> \, \underset{x' \rightarrow x}{ \simeq } \, \frac{e^{-\sigma}}{\pi } \left[  {\rm Im} \left( \frac{e^{-i \theta}}{z-z'} \right)   -   {\rm Re} \left( \frac{e^{-i \theta} (\varphi - \varphi')}{z-z'} \right)  \, + \, \dots \right] \, .
\end{equation*}
It looks like this expression diverges; however it does not: in fact, one can check from the above explicit formulas that $e^{\sigma + i \theta} \partial_x z = -1$ (see also the Introduction). Using this, we see that ${\rm Im} \left( \frac{e^{-\sigma -i \theta}}{z-z'} \right) \, \simeq \, {\rm Im} \left( \frac{e^{-\sigma-i \theta}}{\partial_x z} \frac{1}{x-x'} \right) \, = \, 0$. Then, exactly like in the introduction, we are left with
\begin{eqnarray*}
	\left< \rho_{x,y} \right> &=& \underset{x' \rightarrow x}{\rm lim} \left< c^\dagger_{x,y} c_{x',y} \right> \, = \, - \frac{1}{\pi }   {\rm Re} \left( \frac{ \partial_x \varphi }{e^{\sigma+i \theta} \partial_x z} \right)   \\
	&=& -\frac{1}{\pi} A_x^{({\rm a})} \, .
\end{eqnarray*}
We then evaluate $\varphi(x,y)$ explicitly and find that, inside the ellipse,
\begin{eqnarray}
	\label{eq:Rephi_honeycomb}
\nonumber	\fl{\rm Re} \, \varphi(x,y) = x \, {\rm arccos}\left( \frac{(1+u^2)x - N u^2}{u \sqrt{(N-2x)^2-y^2} } \right) \, + \, \frac{|y|}{2}\, {\rm arccos} \left( \frac{ N(N-2x) - \frac{1+u^2}{1-u^2} y^2 }{ \sqrt{ (N^2-y^2) ((N-2x)^2 -y^2) } } \right) \\
	 \qquad \qquad - \, \frac{N}{2} \,{\rm arccos} \left( \frac{  (1-2u^2) (N-x)^2  + x^2-y^2   }{ \sqrt{ (N^2-y^2) ((N-2x)^2 -y^2) } } \right)  \,  .
\end{eqnarray}
Differentiating with respect to $x$, one gets $A_x^{({\rm a})} = - \partial_x {\rm Re} \varphi$. The formula (\ref{eq:Rephi_honeycomb}) holds inside the ellipse only, but similar tricks work also outside the ellipse. The final result is
\begin{equation}\fl\label{eq:densityhexa}
\left< \rho_{x,y} \right> \;=\;
 \left\{\begin{array}{ccl}
          \frac{1}{\pi}{\rm arccos} \left(\frac{(1+u^2)x-N u^2}{u\sqrt{(N-2x)^2-y^2}}\right)&&{\rm if} \quad X^2+y^2<N^2 \, , \\\\
  	1&&{\rm if} \quad X\leq -\sqrt{N^2-y^2}  \, , \\\\
          1&& {\rm if} \quad X\geq \sqrt{N^2-y^2} \quad {\rm  and } \quad  2x<N-|y|  \quad {\rm and } \quad \frac{y}{N} \geq \frac{1-u^2}{1+u^2} \, , \\\\
          0 && {\rm otherwise} \,.
        \end{array}
 \right.
\end{equation}
Numerical checks of the formula are shown in Fig.~\ref{fig:densityplots}, and show excellent agreement with (\ref{eq:densityhexa}).

\vspace{0.6cm}

{\bf Alternative derivation of the density profile.} 
The reader may be unsatisfied with the way we obtained the density profile, by a limiting procedure which may look suspicious. Of course, other derivations of the result (\ref{eq:densityhexa}) exist; let us quickly sketch another one, that perhaps looks more mathematically sound. From the previous paragraphs, one gets that the average density $\left< \rho(x,y) \right> = \left<  c_{x,y}^\dagger c_{x,y}\right>$ is given by the integral
\begin{equation*}
	\left< \rho_{x,y} \right>  \, =  \,  \int \frac{dk}{2\pi } \int \frac{dk'}{2\pi}  e^{- i (k-k') x + y (\varepsilon(k)-\varepsilon(k')) - i N (\tilde{\varepsilon}(k)-\tilde{\varepsilon}(k')) }   \frac{1}{2 i  \sin\left( \frac{k-k'}{2} \right)}  \, .
\end{equation*}
Trying to analyze this integral by stationary phase, one faces the issue that the two stationary points (for $k$ and $k'$) are identical, so the denominator is singular. To circumvent this issue, one can differentiate with respect to $y$ under the integral (notice that $y$, although discrete in our model, may be treated as a continuous parameter in the integral in the r.h.s). This differentiation removes the singularity in the integrand:  for a point $(x,y)$ inside the ellipse, we find 
\begin{eqnarray*}
	\frac{\partial \left< \rho_{x,y} \right> }{\partial y} & = &  \int \frac{dk}{2\pi}  \int \frac{dk'}{2\pi} e^{- i (k-k') x + y (\varepsilon(k)-\varepsilon(k')) - i N (\tilde{\varepsilon}(k)-\tilde{\varepsilon}(k')) }   \frac{\varepsilon(k)-\varepsilon(k')}{2 i  \sin\left( \frac{k-k'}{2} \right)}  \\
	& \underset{N \rightarrow \infty}{\simeq} & \frac{e^{-\sigma-i \theta}}{2\pi i}  \,\frac{d \varepsilon}{dk} (z)\, + \,  \frac{e^{-\sigma+i\theta}}{2\pi i}  \, \frac{d \varepsilon}{dk} (-z^*) \\
	& =  & \frac{1}{\pi} \, {\rm Im} \left[  e^{-\sigma-i\theta}\,\frac{d \varepsilon}{dk} (z) \right] \\
	&=& \frac{1}{\sqrt{R^2-X^2-y^2}}\times\frac{y(x-u^2(N-x))}{\pi u((N-2x)^2-y^2)}\, .
\end{eqnarray*}
Integrating back, we get the result (\ref{eq:densityhexa}) inside the ellipse, up to an undetermined additive constant, that may depend on $x$.
This constant can be fixed by setting $y=0$ and using a continuity argument. Outside the ellipse, one may repeat similar arguments; the derivative vanishes everywhere, except on the segments $N-2x=\pm y$, see Fig. \ref{fig:strip_honey}.

\newpage

\pagebreak

\section{Dimers on the square lattice}
\label{sec:square}
\begin{figure}[htbp]
 \includegraphics[width=\textwidth]{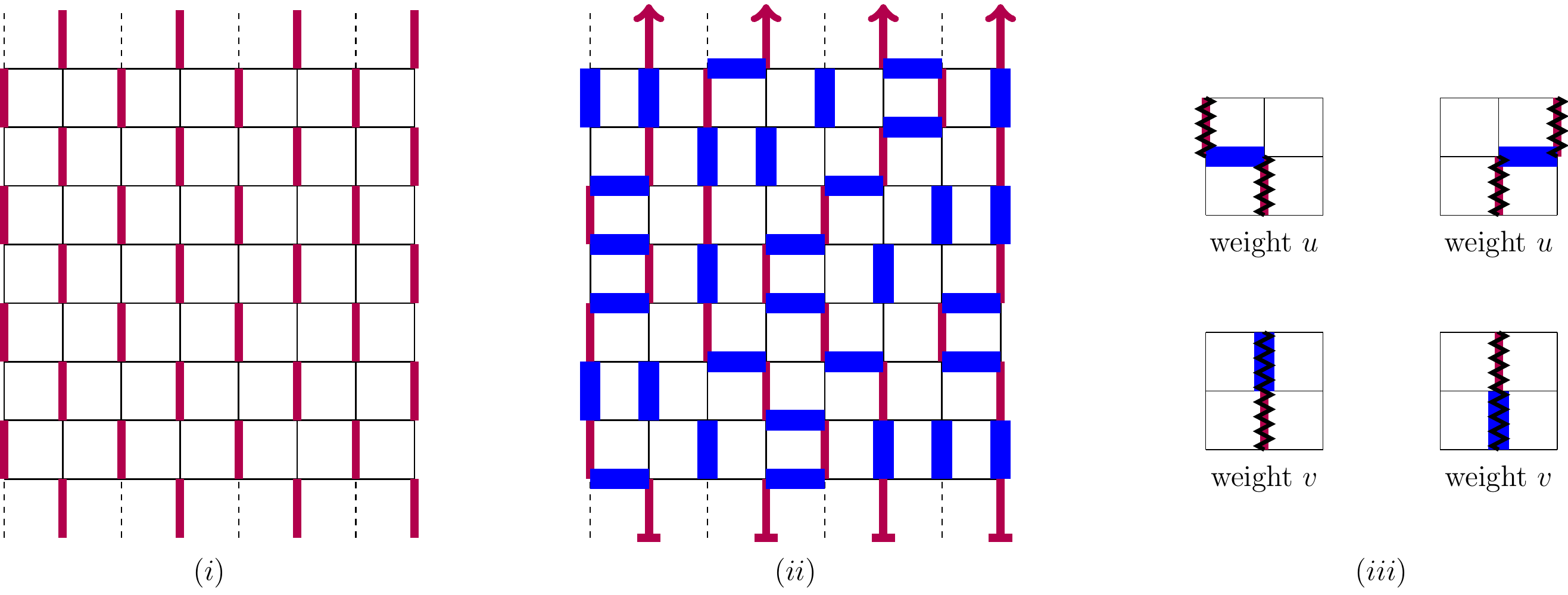}
 \caption{Mapping onto a system of non intersecting lattice paths (or particle trajectories). \emph{(i)}: Reference configuration with staggered dimers (shown in red). \emph{(ii)}: A configuration of dimers (shown in thick blue) is superimposed on the reference configuration. This generates four lattice paths which propagate from bottom to top. \emph{(iii)}: Rules for the trajectories. Recall a particle is defined as a vertical link occupied by a reference dimer only, or a vertical link occupied by a real dimer only. In the latter case the particle, represented by a black zigzag line, has to go straight ahead (bottom right). In the former there are three possibilities, the particle can either go straight ahead (bottom left), to the left (upper left) or to the right (upper right). A weight $u$ (resp. $v$) is assigned to horizontal (resp. vertical) dimers.}
 \label{fig:sqmapping}
\end{figure}

In this section, the method introduced in section \ref{sec:ktr} is applied to the dimer model on the Aztec diamond, where the arctic circle phenomenon has been originally discovered \cite{jockusch1998random}. As we shall see, it is possible to recover this exact geometry using our formalism. This case is slightly more intricate than the dimers on honeycomb, as the corresponding fermionic problem has two bands. We will consider a slight generalization where all dimers do not have the same weights. This is done by assigning a fugacity $u$ to all horizontal dimers, and a fugacity $v$ to all vertical dimers, so that the partition function reads
\begin{equation}
 Z=\sum_{\{c\}} u^{\# {\rm\, horizontal \;dimers} } v^{\# {\rm \,vertical\; dimers} }.
\end{equation}
The sum runs over all dimer coverings of the square lattice. Mapping the dimer model onto a system of free fermions is standard, see e.g. Refs.~\cite{lieb1967solution,alet2006classical,stephan2009shannon}, and can be done as follows.
 First, we consider a particular configuration, where all dimers are arranged in a staggered fashion. This is shown in Fig.~\ref{fig:sqmapping}(i); in the following we call it \emph{Reference configuration}. Then, any dimer configuration will be compared to this reference configuration, by superimposition of the two. For each dimer configuration, this procedure generates a set of lattice paths, an example being shown in Fig.~\ref{fig:sqmapping}(ii). It is convenient to think of these lattice paths as trajectories of particles propagating (say) upwards, where a particle is defined as a vertical link occupied by a reference dimer only, or by a ``real'' dimer only. While moving from one row to the next, the particles have to follow the rules shown in Fig.~\ref{fig:sqmapping}(iii) (Each particle is represented as a black zigzag line in the Figure).

The particle trajectories have the following important properties:
\begin{itemize}
 \item There is a one to one correspondence between the valid particle trajectories and the valid dimer coverings.
 \item The number of particle is conserved.
 \item The trajectories do not cross.
\end{itemize}
The later property means that we need not be careful about the statistics of the particles. As for dimers on the honeycomb lattice, we are allowed to represent them using fermionic operators, and use a transfer matrix acting on those operators. From now one we will call the particle fermions. We explicitly construct the transfer matrix in the following subsection. Before doing so, let us explain how the celebrated Aztec diamond may be recovered using our formalism, and the DWIS $\ket{\Psi_0}=\prod_{x<0}c_x^\dag \ket{0}$. In terms of dimer occupancies, the DWIS corresponds to an alternation $\ldots 101010$ for sites with index $x<0$ and $010101\ldots$ for sites with index $x>0$. This is shown in Fig.~\ref{fig:Aztecdimers}(i). The (imaginary-)time evolution is generated by several application of the transfer matrix, and the system is projected back onto the initial state $\ket{\Psi_0}$. Due to the staggered nature of the DWIS, many other dimer occupancies are automatically set to one, by construction of the dimer model. They are shown in thick green in Fig.~\ref{fig:Aztecdimers}(i). As can be seen, the remaining edges that are left free define precisely an Aztec diamond. In the following, we use this simple observation to compute all the large scale correlations, and derive the effective low energy theory. The phenomenology here is strikingly different from the homogenous case (see e. g. \cite{stephan2014emptiness,allegra2015exact}) where the standard machinery of boundary CFT can be safely used. Let us mention also that a slightly different transfer matrix approach has been used in \cite{boutillier2015dimers} to deal with more generic lattices. 
\subsection{The transfer matrix}
\begin{figure}[htbp]
 \includegraphics[width=\textwidth]{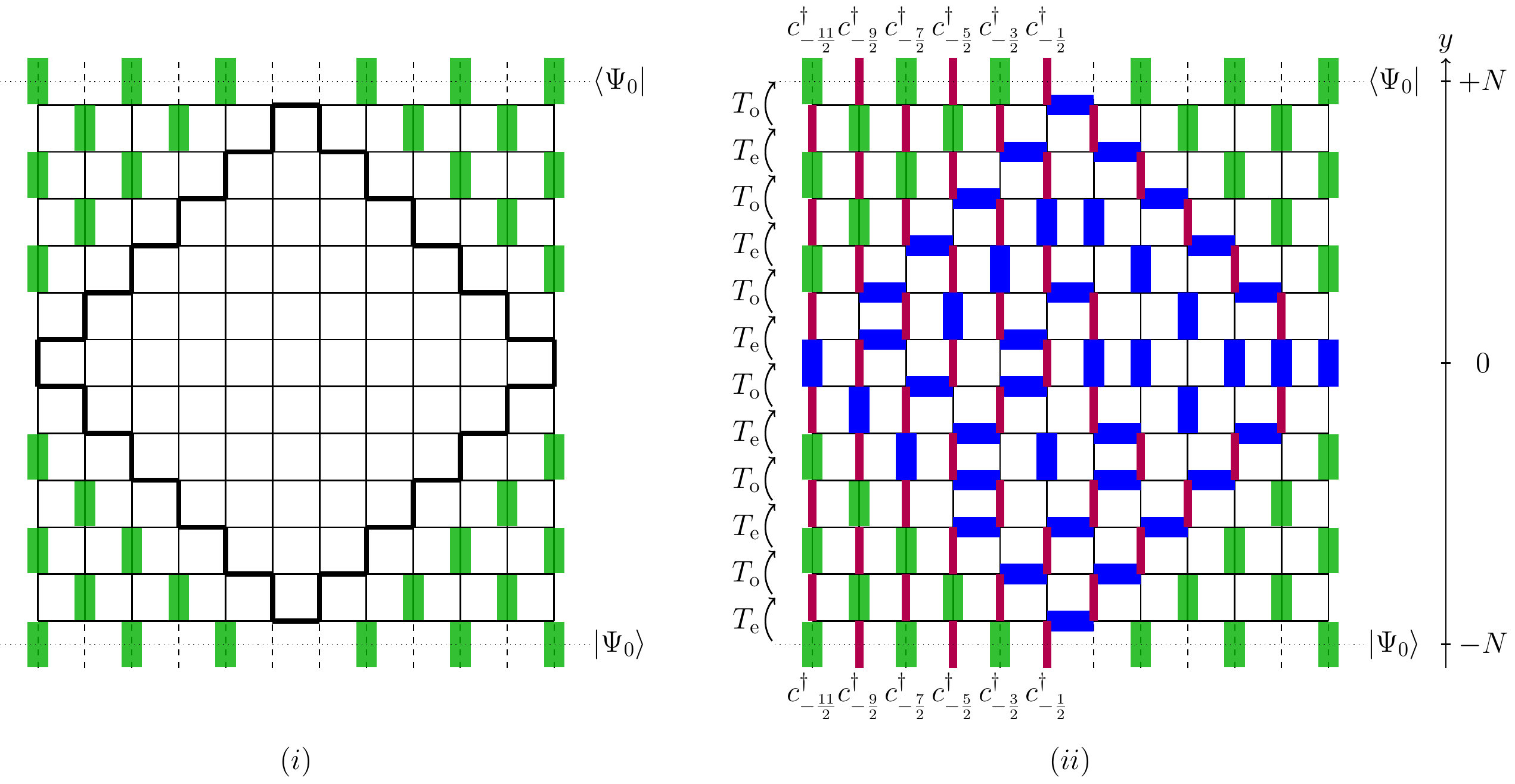}
 \caption{\emph{(i)}: Aztec diamond obtained when the DWIS $\ket{\Psi_0}$ is set as bottom and top boundary condition. The corresponding dimers cross either of the two dotted line, and are shown in thick green. Other dimers occupancies are set to one as a result; they are also shown in green. The remaining edges define the inside of an Aztec diamond, which is highlighted in thick black. \emph{(ii)}: Example of fermion trajectories generated by successive application of the transfer matrix on the DWIS.}
 \label{fig:Aztecdimers}
\end{figure}

Let us now write the transfer matrix in explicit form. First, we are still dealing with fermionic particles on the lattice $\mathbb{Z} + \frac{1}{2}$, and with the DWIS $\ket{\Psi_0}$. The transfer matrix, however, is not invariant under translations of {\it one site}, but only of {\it two sites}. To have a consistent labeling of the edges, it is necessary to distinguish between even and rows, so that there are in fact two transfer matrices $T_{\rm e}$ and $T_{\rm o}$. Their respective action on the many-fermion states follows from the rules shown in Fig.~\ref{fig:sqmapping}(iii). We have
\begin{eqnarray}\label{eq:even_action}
&&T_{\rm e}\ c^\dagger_{2j+1/2}\ T_{\rm e}^{-1}=v \,c^\dagger_{2j+1/2},\nonumber \\
&&T_{\rm e}\ c^\dagger_{2j-1/2}\ T_{\rm e}^{-1}=u \,c^\dagger_{2j-3/2}+v \,c^\dagger_{2j-1/2}+u\, c^\dagger_{2j+1/2},
\end{eqnarray}
and 
\begin{eqnarray}\label{eq:odd_action}
&&T_{\rm o}\ c^\dagger_{2j+1/2}\ T_{\rm o}^{-1}=u \,c^\dagger_{2j+3/2}+v\, c^\dagger_{2j+1/2}+u \,c^\dagger_{2j-1/2},\nonumber \\
&&T_{\rm o}\ c^\dagger_{2j-1/2}\ T_{\rm o}^{-1}=v \,c^\dagger_{2j-1/2}.
\end{eqnarray}
The action of the transfer matrix on the DWIS is illustrated in Fig.~\ref{fig:Aztecdimers}(ii), where an example of a possible trajectory for the fermions is shown. 
To diagonalize the transfer matrix, it is convenient to go to momentum space. Recall our convention for the Fourier transform is
\begin{equation}\label{eq:fouriertransform}
 c^\dag(k)=\sum_{x\in \mathbb{Z}+1/2}e^{ikx} c_x^\dag \qquad,\qquad c^\dag(k+\pi)=\sum_{x\in \mathbb{Z}+1/2} e^{i\pi x}e^{ikx}c_x^\dag.
\end{equation}
The action of the transfer matrix on those modes reads
\begin{equation}
 T_{\rm e} \ccvec T_{\rm e}^{-1}=\twomat{v+u\cos k}{-iu \cos k}{-iu \cos k}{v-u\cos k} \ccvec
\end{equation}
and
\begin{equation}
  T_{\rm o} \ccvec T_{\rm o}^{-1}=\twomat{v+u\cos k}{iu \cos k}{iu \cos k}{v-u\cos k} \ccvec.
\end{equation}
$T_{\rm e}$ and $T_{\rm o}$ do not satisfy the assumption \ref{item:hermitian} of section~\ref{sec:ktr}, because they are not hermitian, however $T^2=T_{\rm o}T_{\rm e}$ is. Its action on the Fourier modes is given by
\begin{equation}\fl
 T^2 \ccvec T^{-2}=M(k)\ccvec\quad,\quad M(k)=\twomat{u^2\cos^2 k+(v+u \cos k)^2}{2i u^2 \cos ^2 k}{-2iu^2 \cos^2 k}{u^2\cos^2 k+(v-u \cos k)^2}.
\end{equation}
$M(k)$ is hermitian, so may be diagonalized by unitary matrix: 
\begin{equation}\label{eq:umat}
 U(k)=\twomat{\cos \alpha(k)}{i\sin \alpha(k)}{i\sin\alpha(k)}{\cos \alpha(k)}.
\end{equation}
We parametrize the angle $\alpha$ as
\begin{equation}
 \alpha(k)=\frac{1}{2}\arcsin \left(\frac{u}{w}\cos \kappa\right)\qquad,\qquad w=\sqrt{u^2+v^2}.
\end{equation}
This choice of parametrization might appear somewhat artificial. However, as we shall see, it makes many subsequent computations much lighter. It is easy to check that $U(k)M(k)U^\dag(k)$ is diagonal provided the condition
\begin{equation}\label{eq:param}
 \tan \kappa =\tan \kappa(k)=\frac{v}{w}\tan k
\end{equation}
is fulfilled. 
The previous equation defines a map $k\mapsto \kappa(k)$ which is a bijection of $[-\pi/2,\pi/2]$ onto itself. $\kappa$ will be the natural variable to perform the steepest descent method. 
Using this change of basis we have
\begin{equation}\fl
 U(k)M(k)U^\dag(k)=\twomat{\lambda(k)}{0}{0}{1/\lambda(k)}\qquad,\qquad \lambda(k)=\left(u \cos k+\sqrt{v^2+u^2 \cos^2 k}\right)^2=
v^2\frac{w+u \cos \kappa}{w-u \cos \kappa}.
 \end{equation}
Introducing a new set of fermions through
\begin{equation}
 \cpcmvec=U(k) \ccvec
\end{equation}
allows to rewrite the transfer matrix as
\begin{eqnarray}\label{eq:tmsquare}
 T^2&=&\exp\left[-\int_{-\pi/2}^{\pi/2}\frac{dk}{2\pi} 2\varepsilon(k)\left(c_{+}^\dag(k)c_+(k)-c_{-}^\dag(k)c_-(k)\right)\right],\\\label{eq:square_energy}
 \varepsilon(k)&=&-\frac{1}{2}\log \lambda(k)=
-\frac{1}{2}\log \left(v^2\frac{w+u \cos \kappa}{w-u \cos \kappa}\right).
 \end{eqnarray}
The form (\ref{eq:tmsquare}) obviously satisfies the condition \ref{item:eigenvalues}. The last requirement, \ref{item:continuation}, may be fulfilled by extending the map $k\mapsto \kappa(k)$  to $\mathbb{R}$ in the following way:
\begin{eqnarray}
 \kappa(k+ p\pi)&=&\kappa(k)+p\pi\qquad\,,\qquad \forall p\in \mathbb{Z}
\end{eqnarray}
In particular, this implies $c^\dag_+(k+\pi)=c_-^\dag(k)$ and $\varepsilon(k+\pi)=-\varepsilon(k)$, so that the transfer matrix may be rewritten as
\begin{equation}
 T^2=\exp\left[-\int_{-\pi}^\pi \frac{dk}{2\pi}2\varepsilon(k)c_+^\dag(k)c_+(k)\right].
\end{equation}

\subsection{The Hilbert transform}
Now that all the assumptions \ref{item:hermitian},\ref{item:eigenvalues},\ref{item:continuation} are satisfied, we are ready to apply the formalism developed in section \ref{sec:ktr}. First, we need the propagator in the DWIS. We find, after a long calculation,
\begin{eqnarray}\label{eq:propdimers}\nonumber
 \braket{\Psi_0|c_+^\dag(k)c_{+}(k')|\Psi_0}&=&e^{i\frac{\beta(\kappa)-\beta(\kappa')}{2}} \frac{\cos \frac{\kappa-\kappa'}{2}}{i \sin (k-k')}\\\label{eq:correl_cp_square}
 &=&\left(\frac{d\kappa}{dk}\right)^{1/2}\left(\frac{d\kappa'}{dk'}\right)^{1/2} \frac{e^{i\frac{\beta(\kappa)-\beta(\kappa')}{2}}}{2i \sin \frac{\kappa-\kappa'}{2}}.
\end{eqnarray}
where $\kappa=\kappa(k)$, $\kappa'=\kappa(k')$, $\beta=\beta(k)$, and $\beta'=\beta(k')$. The ancillary angle $\beta$ is given by $$\beta(k)=\arctan\left(\frac{u}{v} \sin \kappa(k)\right).$$ The transform needed to obtain the full propagator $\braket{c_+^\dag(k,y)c_{+}(k',y')}$ is 
\begin{equation}
 \tilde{\varepsilon}(k)={\rm p.v.}\int \frac{dk'}{2\pi}\varepsilon(k')2\partial_{k'} \log \braket{\Psi_0|c_+^\dag(k')c_{+}(k)|\Psi_0}.
\end{equation}
We first remark that the phase in (\ref{eq:propdimers}) only gives a constant -- independent on $k$ -- contribution to $\tilde{\varepsilon}(k)$. Since only the difference $\tilde{\varepsilon}(k)-\tilde{\varepsilon}(k')$ enters any correlation function, this contribution may simply be ignored. Therefore, we set $\beta,\beta'=0$ in the remainder of the present subsection. 
Using Eq.~(\ref{eq:correl_cp_square}), we find that the appropriate transform is a Hilbert transform, modulo conjugation by the change of variables $k\mapsto\kappa(k)$:
\begin{eqnarray}
\nonumber	\tilde{\varepsilon}(k) & = & {\rm p.v.} \int_{-\pi}^\pi \frac{dk'}{2\pi} \varepsilon(k')  2\partial_{k'} \log  \bra{\Psi_0} c^\dagger_+(k) c_+(k') \ket{\Psi_0}  \\
\nonumber				&=& {\rm p.v.} \int_{-\pi}^\pi \frac{d\kappa'}{2\pi} \varepsilon(k')  2\partial_{\kappa'}  \log \left[ \frac{1}{2 i \sin \left( \frac{\kappa - \kappa'}{2} \right)} \right] \, + \,  {\rm p.v.} \int_{-\pi}^\pi \frac{d k'}{2\pi} \varepsilon(k')  \partial_{k'}  \log \left[  \frac{d\kappa'}{d k'}  \right]    \\
\nonumber				&=& {\rm p.v.} \int_{-\pi}^\pi \frac{d\kappa'}{2\pi} \varepsilon(\kappa')  \cot \left( \frac{\kappa - \kappa'}{2} \right) \quad + \quad 0\, .		
\end{eqnarray}
In the second line, we have split the function $\log  \bra{\Psi_0} c^\dagger_+(k) c_+(k') \ket{\Psi_0} $ into a sum of three terms, one of which trivially vanishes; then in the first term we changed variables $k' \rightarrow \kappa'$, using the fact that the Jacobian $\frac{dk'}{d\kappa'}$ in the integration measure cancels the one coming from the chain rule for the derivative $\partial_{k'} = \frac{d\kappa'}{dk'} \partial_{\kappa'}$. In this term, we also use the notation $\varepsilon (\kappa') $ for $\varepsilon(k')$. In the third line, the second term is zero because of the {\it anti-periodicity} of $\varepsilon(k)$, $\varepsilon(k+\pi) = -\varepsilon (k)$. Indeed, the derivative $\frac{d\kappa}{dk}$ is {\it periodic} (with period $\pi$), therefore the integrand $\varepsilon(\kappa')   \partial_{k'}  \log \left[  \frac{d\kappa'}{d k'}  \right]$ is {\it anti-periodic}, and its integral from $-\pi$ to $0$ cancels the one from $0$ to $+\pi$.

We call $\tilde{\varepsilon}(\kappa)$ the Hilbert transform of the function (of $\kappa$) $\varepsilon(\kappa)$. For the function of interest here, we have the result
\begin{equation}\label{eq:hilbertsquare}\fl
 \varepsilon(\kappa)=-\frac{1}{2}\log \left(v^2\frac{w+u \cos \kappa}{w-u \cos \kappa}\right)
 \qquad  \Leftrightarrow \qquad
 \tilde{\varepsilon}(\kappa)=\frac{i}{2}\log\left(\frac{v+i u \sin \kappa}{v-i u \sin \kappa}\right)-\log v=-\arctan \left(\frac{u}{v} \sin \kappa\right)-\log v,
\end{equation}
which is derived in the appendix \ref{appx:hilb}. We note that the function $\tilde{\varepsilon}(\kappa)$ is automatically anti-periodic with period $\pi$, a property it inherits from $\varepsilon(\kappa)$. 

\subsection{Fermionic correlators: exact finite-size formulas}

Here $N$ is assumed to be \emph{even}. The vertical axis is labelled by \emph{integers} $-N\leq y\leq N$, while the horizontal axis is still labelled by half integers $x\in \mathbb{Z}+1/2$. This way the left/right symmetry is $x\to -x$ and the up/down symmetry is $y\to -y$. 
For an observable $\mathcal{O}_{x,y}$ at position $x$ and $y$, we have
\begin{eqnarray}
\braket{\mathcal{O}_{x,y}}=\left\{
\begin{array}{ccccc}
 \displaystyle{\frac{\bra{\Psi_0}(T^2)^{\frac{N-y}{2}} \mathcal{O}_x (T^2)^{\frac{N+y}{2}}\ket{\Psi_0} }{\Braket{\Psi_0|T^{2N}|\Psi_0}}}&&,&& y \,{\rm  even}\\\\
 \displaystyle{\frac{\bra{\Psi_0}(T^2)^{\frac{N-y-1}{2}} T_{\rm o}\mathcal{O}_x T_{\rm o}^{-1} (T^2)^{\frac{N+y+1}{2}}\ket{\Psi_0}}{\Braket{\Psi_0|T^{2N}|\Psi_0}}} &&,&& y \,{\rm  odd}
\end{array}
\right.
\end{eqnarray}
We will now use the notation introduced in the introduction (see Eq.~(\ref{eq:notation_doteq})), and also used in Sec.~\ref{sec:honeycomb}.  We write
\begin{equation}
\label{eq:ident_square}
	\left\{  \begin{array}{rcl} c^\dagger_{x,y}   & \doteq &  \displaystyle \int_{-\pi}^\pi \frac{dk}{2\pi} e^{-i k x + y \varepsilon (\kappa)  - i N \tilde{\varepsilon}(\kappa) }  \,  r_{x,y} (k)  \, c^\dagger_+(k) \\ \\
	c_{x,y}   & \doteq & \displaystyle  \int_{-\pi}^\pi \frac{dk}{2\pi} e^{i k x-y \varepsilon (\kappa)  + i N \tilde{\varepsilon}(\kappa) }  \,  s_{x,y} (k)  \, c_+(k) \, , \end{array} \right.
\end{equation}
where the dot means that this is an equality that holds only when taking expectation values in the DWIS. The main difference with the toy model is the emergence of the extra factors $r_{x,y}(k)$, and $s_{x,y}(k)$. These coefficients are here because we are dealing with a two band problem; they only depend on the parity of $x$ and $y$. The aim of this subsection is to compute them. 

We start with the case $y\in 2\mathbb{Z}$, which is easiest. Using Eqs.~(\ref{eq:fouriertransform}), (\ref{eq:umat}) and extending the Brillouin zone to $[-\pi,\pi]$ we obtain
\begin{equation}
 c^\dag_{2j\pm 1/2}=\int_{-\pi}^{\pi}\frac{dk}{2\pi} e^{-i k (2j\pm 1/2)} \sqrt{1\mp (u/w)\cos \kappa}\; c_{+}^\dag(k).
\end{equation}
Now due to (\ref{eq:prop_interm_2}), it is easy to see that the coefficients we are looking for are given by
\begin{equation}
 r_{x,y}(k)\;=\;s_{x,y}(k)\;=\; \sqrt{1\mp \frac{u}{w}\cos \kappa},
\end{equation}
where the upper sign corresponds to $x\in 2\mathbb{Z}+1/2$ and the lower sign corresponds to $x\in 2\mathbb{Z}-1/2$. Note that the equality between $r_{x,y}(k)$ and $s_{x,y}(k)$ simply follows from hermiticity.

The case $y\in 2\mathbb{Z}+1$ is more tricky. Indeed, one has to shift $y$ by one, and conjugate with one application of $T_{\rm o}$, which is not Hermitian. For $x\in 2\mathbb{Z}+1/2$ we have -- see Eq.~(\ref{eq:odd_action}) -- $T_{\rm o} c_{x}^\dag T_{\rm o}^{-1}=v c_{x}^\dag +u(c_{x-1}^\dag +c_{x+1}^\dag)$, so that 
\begin{equation}
 c_{x,y}^\dag  \;\doteq \; v \,c_{x,y+1}^\dag +u\left[c_{x-1,y+1}^\dag+c_{x+1,y+1}^\dag\right],
\end{equation}
which yields
\begin{eqnarray}
 r_{x,y}(k)&=&e^{\varepsilon(\kappa)} \left[v \,r_{x,y+1}+2 u \cos k \,r_{x+1,y+1}\right]\\
 &=& \sqrt{1+ \frac{u}{w}\cos \kappa}.
\end{eqnarray}
For $x\in 2\mathbb{Z}-1/2$ we have $T_{\rm o} c_{x}^\dag T_{\rm o}^{-1}=v c_{x}^\dag$, and we find
\begin{equation}
 r_{x,y}(k)=\sqrt{1- \frac{u}{w}\cos \kappa}.
\end{equation}
Finally, it remains to perform a similar identification for $s_{x,y}(k)$. To do so, we first need to determine the action of the odd transfer matrix on the \emph{annihilation} operators. This is explained in Appendix.~\ref{sec:moredetails_dimers}. We find
\begin{eqnarray}\label{eq:odd_action_2}
&&T_{\rm o}\ c_{2j+1/2}\ T_{\rm o}^{-1}=\frac{1}{v}\, c_{2j+1/2},\nonumber \\
&&T_{\rm o}\ c_{2j-1/2}\ T_{\rm o}^{-1}=\frac{1}{v^2}\left(-u c_{2j-3/2}+ v c_{2j-1/2}-u c_{2j+1/2}\right),
\end{eqnarray}
and a similar calculation gives $r_{x,y}(k)=s_{x,y}(k)$ for $y$ odd. In total, all possible cases may be summarized by the following simple formula
\begin{equation}\label{eq:rsk_result}
 r_{x,y}(k)\,=\,s_{x,y}(k)\,=\, 
 \sqrt{1-\sigma \frac{u}{w}\cos \kappa(k)}\qquad,\qquad \sigma=(-1)^{x-y-1/2},
\end{equation}
which depends only on the choice of sublattice. In the following, we will refer to the case $x-y\in 2\mathbb{Z}+1/2$, for which $\sigma=+1$ as the \emph{even} sublattice. The \emph{odd} sublattice, $x-y\in 2\mathbb{Z}-1/2$, has $\sigma=-1$. 
\subsection{Fermionic correlators: scaling limit}

Once again, we are interested in the scaling limit of the correlators. These are obtained by the steepest descent method, exactly like in the introduction. We focus on the argument of the exponentials in Eq.~(\ref{eq:ident_square}). The stationary points are the solutions of
\begin{equation*}
	 \frac{x}{u} \, +\,  i \frac{y}{v} \sin \kappa  - \frac{N}{w} \cos \kappa     \, = \, \frac{1}{u}\,\frac{d}{dk} \left[  k x  \,+\,   i y   \varepsilon(\kappa) +  N   \tilde{\varepsilon}(\kappa)     \right] \, = \, 0 \, .
\end{equation*}
Here we have used the chain rule $\frac{d}{d k} = \frac{d\kappa}{dk} \frac{d}{d\kappa}$ and the formulas (see Eq.~(\ref{eq:hilbertsquare}))
\begin{equation*}
	\frac{d \varepsilon(\kappa)}{d \kappa} \, = \, \frac{ u w \, \sin \kappa }{w^2 - u^2  \cos^2 \kappa } \, , \qquad
	\frac{d \tilde{\varepsilon}(\kappa)}{d \kappa} \, = \, \frac{- u v \, \cos \kappa }{w^2 -  u^2   \cos^2 \kappa }  \, ,\qquad 	\frac{d k}{d \kappa} \, =\, \frac{v w}{ w^2 - u^2 \cos^2 \kappa } \, .
\end{equation*}
Generically, there are two non-degenerate stationary points
\begin{equation*}
	\kappa(x,y) \, =\,  z(x,y)   \qquad {\rm and}  \qquad \kappa(x,y) \, = \, -z^*(x,y)   \, ,
\end{equation*}
where
\begin{equation}
	\label{eq:z_square}
	z(x,y) \, = \, {\rm arccos} \frac{x/u}{\sqrt{ (\frac{N}{w} )^2 - (\frac{y}{v} )^2 }} \,-\, i \,{\rm arctanh} \left( \frac{y/v}{N/w} \right) \, .
\end{equation}
By 'generically', we mean 'unless $\frac{x^2}{u^2} + \frac{y^2}{v^2} = \frac{N^2}{w^2}$'. This curve is, of course, the arctic curve for the dimer model on the Aztec diamond. Now let us assume that $(x,y)$ is not on the arctic curve; then around the two stationary points, the exponential is approximated by
\begin{equation*}\fl
	\left\{  \begin{array}{rcl}
		e^{ -i N \left[  k \frac{x}{N} +   i \frac{y}{N} \varepsilon(k) +  \tilde{\varepsilon}(k)    \right]  } &  \simeq  &  e^{ -i  \varphi(x,y )  - i \frac{1}{2} (k - k(z) )^2 \, e^{\sigma (x,y) } \left(\frac{dk}{d\kappa}\right)^{-1}    }  \qquad {\rm around} \quad k = k(z)  \\ \\
		e^{ -i N \left[  k \frac{x}{2N} +   i \frac{y}{N} \varepsilon(k) +  \tilde{\varepsilon}(k)    \right]  } &  \simeq  &  e^{ i  \varphi^*(x,y )  - i \frac{1}{2} (k - k(-z^*) )^2 \, e^{\sigma (x,y)}  \left(\frac{dk^*}{d\kappa^*}\right)^{-1}     }     \qquad {\rm around} \quad  k = k(-z^*) ,
	\end{array} \right.
\end{equation*}
where
\begin{equation}\fl
	\label{eq:phi_sigma_square}
	\left\{  \begin{array}{rcl}
		\varphi \,= \, \varphi(x,y) & \equiv &   k(z(x,y)) x  +   i  y \varepsilon(z(x,y)) +  N  \tilde{\varepsilon}(z(x,y))      \\  \\
		e^{\sigma} \, = \, e^{\sigma(x,y)} &\equiv& \left(\frac{dk}{d\kappa}\right)  \frac{d^2}{dk^2} \left[  i  y  \varepsilon(k) +  N \tilde{\varepsilon}(k)   \right] \, = \,  u  \, \frac{d}{d\kappa}  \left[  -i  \frac{y}{v} \, \sin \kappa  +  \frac{N}{w}  \, \cos \kappa  \right] \, =\,   u \, \sqrt{ \left(\frac{N}{w}\right)^2 - \left(\frac{x}{u}\right)^2 - \left(\frac{y}{v}\right)^2  } \, .
	\end{array}  \right. 
\end{equation}
Performing the gaussian integration, we find that the identification (\ref{eq:ident_square}) becomes
\begin{equation}\fl
\label{eq:ident_square_2}
	\left\{  \begin{array}{rcl} c^\dagger_{x,y}   & \underset{N \rightarrow \infty}{\doteq} &  \displaystyle  \frac{ e^{-i \frac{\pi}{4}} }{\sqrt{2\pi}}  e^{ -i  \varphi}   e^{-\frac{\sigma }{2}  }   \,  r_{x,y} (z)  \,    \left( \frac{dk}{d \kappa} \right)^{\frac{1}{2}} c^\dagger_+(z)  \;  + \;   \frac{ e^{i \frac{\pi}{4}} }{\sqrt{2\pi}}  e^{ i  \varphi^*}   e^{-\frac{\sigma}{2}  }   \,  r_{x,y} (-z^*)  \,    \left( \frac{dk^*}{d \kappa^*} \right)^{\frac{1}{2}} c^\dagger_+(-z^*)   \\ \\
	c_{x,y}   & \underset{N \rightarrow \infty}{\doteq}  &  \displaystyle  \frac{ e^{i \frac{\pi}{4}} }{\sqrt{2\pi}}  e^{ i  \varphi}   e^{-\frac{\sigma }{2}  }   \,  s_{x,y} (z)  \,    \left( \frac{dk}{d \kappa} \right)^{\frac{1}{2}} c_+(z)  \;  + \;   \frac{ e^{-i \frac{\pi}{4}} }{\sqrt{2\pi}}  e^{ -i  \varphi^*}   e^{-\frac{\sigma }{2}  }   \,  s_{x,y} (-z^*)  \,    \left( \frac{dk^*}{d \kappa^*} \right)^{\frac{1}{2}} c_+(-z^*)  \, . \end{array} \right.
\end{equation}
Notice that the appearance of $ \left( \frac{dk}{d \kappa} \right)^{\frac{1}{2}} c^\dagger_+(z) $ and $ \left( \frac{dk}{d \kappa} \right)^{\frac{1}{2}} c_+(z) $ is particularly convenient, since the roots of the Jacobians are exactly what is needed to cancel the ones in (\ref{eq:correl_cp_square}). Their propagator is 
\begin{equation}\label{eq:someprop}
	\bra{\Psi_0} \left( \frac{dk}{d \kappa} \right)^{\frac{1}{2}} c^\dagger_+(z) \left( \frac{dk'}{d \kappa'} \right)^{\frac{1}{2}} c_+(z')  \ket{\Psi_0} \, =\, \frac{e^{i\frac{\beta(z)-\beta(z')}{2}}}{2 i \sin \left( \frac{z-z'}{2}\right)} \, . 
\end{equation}
We note that it is also possible to get rid of the the phase factor $e^{i\frac{\beta-\beta'}{2}}=e^{-i\frac{\tilde{\varepsilon}-\tilde{\varepsilon}'}{2}}$ in (\ref{eq:someprop}), and absorb it in the phase $\varphi(x,y)$. We will make use of this observation in the next subsection.

\subsection{Identification of the Dirac theory inside the arctic circle}

At this point, it is quite clear that the Dirac action in the dimer model is going to be similar to the one discussed in the introduction. 
There are however a couple differences. First, the relation between the lattice and continuous degrees of freedom is more complicated than in the case of the XX chain or honeycomb dimers, see Eq. (\ref{eq:lattice_continuous_XX}).
\begin{equation}
	\label{eq:square_c_psi}
	\left\{ \begin{array}{rcl}
		c_{x,y}^\dagger &=& \displaystyle  \sum_{n \in \mathbb{Z}}   \gamma_{x,y,n}  \left[  \frac{1}{\sqrt{2\pi}} \psi^\dagger(x+n,y) +  \frac{1}{\sqrt{2\pi}} \overline{\psi}^\dagger(x+n,y)  \right]    \\
		c_{x,y} &=& \displaystyle  \sum_{n \in \mathbb{Z}}   \gamma_{x,y,n}  \left[  \frac{1}{\sqrt{2\pi}} \psi(x+n,y) +  \frac{1}{\sqrt{2\pi}} \overline{\psi}(x+n,y)  \right]  
	\end{array} \right. 
\end{equation}
where the $\gamma_{x,y,n}$ are the $n$-th Fourier coefficients of $r_{x,y}$, {\it i.e.} $\gamma_{x,y,n}=\int_{-\pi}^{\pi} \frac{dk}{2\pi}e^{-ikn}r_{x,y}$. Since the $r_{x,y}$ are analytic in $k$, the coefficients decay exponentially fast with $n$. In that sense the identification (\ref{eq:square_c_psi}), despite the infinite number of terms, is still \emph{local}. 
The propagators of $\psi^\dagger$, $\psi$ and $\overline{\psi}^\dagger$, $\overline{\psi}$ are
\begin{eqnarray*}\fl
	\left<  \psi^\dagger(x,y) \psi(x',y')  \right> & = & e^{-i (\varphi - \varphi')} e^{-\frac{\sigma+\sigma'}{2}}  \bra{\Psi_0} \left( \frac{dk}{d \kappa} \right)^{\frac{1}{2}} c^\dagger_+(z) \left( \frac{dk'}{d \kappa'} \right)^{\frac{1}{2}} c_+(z')  \ket{\Psi_0} \, = \,  \frac{e^{-i (\varphi+\frac{\tilde{\varepsilon}}{2} - \varphi'-\frac{\tilde{\varepsilon}'}{2})} e^{-\frac{\sigma+\sigma'}{2}}}{2i \sin \left( \frac{z-z'}{2} \right)}  \\\fl
	\left<  \overline{\psi}^\dagger(x,y) \overline{\psi}(x',y')  \right> & = & e^{-i (\varphi - \varphi')} e^{-\frac{\sigma+\sigma'}{2}}  \bra{\Psi_0} \left( \frac{dk^*}{d \kappa^*} \right)^{\frac{1}{2}} c^\dagger_+(-z^*) \left( \frac{dk'^*}{d \kappa'^*} \right)^{\frac{1}{2}} c_+(-z'^*)  \ket{\Psi_0} \, = \, \frac{ e^{i (\varphi^*+\frac{\tilde{\varepsilon}^*}{2} - \varphi'^*-\frac{\varepsilon'^*}{2})} e^{-\frac{\sigma+\sigma'}{2}}}{2i \sin \left( \frac{z^*-z'^*}{2} \right)} \, .
\end{eqnarray*}
where $\tilde{\varepsilon}=\tilde{\varepsilon}(z)$.
These are exactly the same propagators as in the introduction, with functions $\varphi$ and $\sigma$ that are now given by (\ref{eq:phi_sigma_square}), and $\varphi$ replaced by $\varphi+\tilde{\varepsilon}/2$. We thus find that the field theory inside the arctic circle is again the generic Dirac action (\ref{eq:Dirac}), with a diagonal tetrad $e^\mu_a \, = \, e^{-\sigma} \delta_{a}^\mu$, and with gauge fields $A^{({\rm a})}_\mu \, = \, -\partial_\mu {\rm Re} [\varphi+\tilde{\varepsilon}/2]$ and $A^{({\rm v})}_\mu \, = \,- i\partial_\mu {\rm Im} [\varphi+\tilde{\varepsilon}/2]$. Let us finally comment on the boundary conditions for the fields $\psi,\psi^\dag, \overline{\psi}, \overline{\psi}^\dag$. As is explained in detail in Sec.~\ref{sec:dirac_honey}, they follows from the relation $1/\sin \left(\frac{z-z'}{2}\right)=\pm 1/\sin\left(\frac{z^*-z'}{2}\right)$, where the upper sign corresponds to the boundary ${\rm Re}\, z=0$, and the lower sign corresponds to the boundary ${\rm Re}\, z=\pi$.

\subsection{Density profile}
For the sake of completeness, we also show the density profile $\Braket{\rho_{x,y}}=\braket{c_{x,y}^\dag c_{x,y}}$ inside the arctic ellipse. A possible way to derive it is to take the derivative with respect to $y$:
\begin{eqnarray*}\fl
	\frac{\partial \Braket{\rho_{x,y}}}{\partial y} & = &  \int \frac{dkdk'}{(2\pi)^2}  e^{- i (k-k') x + y (\varepsilon(\kappa)-\varepsilon(\kappa')) - i N (\tilde{\varepsilon}(\kappa)-\tilde{\varepsilon}(\kappa')) }  r_{x,y} (k) s_{x,y} (k')  \left( \frac{d\kappa}{dk}  \right)^{\frac{1}{2}}  \left( \frac{d\kappa'}{dk'}  \right)^{\frac{1}{2}} \frac{(\varepsilon(\kappa)-\varepsilon(\kappa'))e^{i \frac{\beta(\kappa)-\beta(\kappa')}{2}}}{2 i  \sin\left( \frac{\kappa-\kappa'}{2} \right)}  \\\fl
	& \underset{N \rightarrow \infty}{\simeq} & \frac{e^{-\sigma}}{2\pi i} r_{x,y}(z) s_{x,y} (z) \,\frac{d \varepsilon}{d \kappa} \, + \,  \frac{e^{-\sigma}}{2\pi i} r_{x,y}(-z^*) s_{x,y} (-z^*) \, \frac{d \varepsilon}{d \kappa} (-z^*) \\\fl
	& =  & \frac{e^{-\sigma}}{\pi} \, {\rm Im} \left[  r_{x,y}(z) s_{x,y} (z) \, \frac{d \varepsilon}{d \kappa} (z) \right] \, .
\end{eqnarray*}
Then, we use Eq.~(\ref{eq:rsk_result}):
$$r_{x,y}(k) s_{x,y}(k)=
1\mp \frac{u}{w}\cos \kappa(k),
$$
where the upper (resp. lower) sign corresponds to $x-y\in 2\mathbb{Z}+1/2$ (resp. $x\in 2\mathbb{Z}-1/2$),
and recall $\frac{d\varepsilon}{d\kappa}=\frac{u w \sin \kappa}{w^2-u^2 \cos^2 \kappa}$. We obtain
\begin{eqnarray*}
 \frac{\partial \braket{\rho_{x,y}}}{\partial y} & = &\frac{1}{\pi\sqrt{(N/w)^2-(x/u)^2-(y/v)^2}}\times 
 {\rm Im}\,\left[\frac{\sin z(w\mp u\cos z)}{w^2-u^2 \cos^2 z}\right]\\
 & = &\frac{1}{\pi\sqrt{(N/w)^2-(x/u)^2-(y/v)^2}}\times 
 {\rm Im}\,\left[\frac{\sin z}{w\pm u \cos z}\right]\\
 &=&\frac{1}{\pi\sqrt{(N/w)^2-(x/u)^2-(y/v)^2}}\times \frac{-y(w^2 x\pm u^2 N)}{uvw((N\pm x)^2-y^2)}.
\end{eqnarray*}
This integrates to
\begin{equation}
 \braket{\rho_{x,y}}=\frac{1}{\pi}{\rm arccos} \left(\frac{w x/u\pm u N/w}{\sqrt{(N\pm x)^2-y^2}}\right).
\end{equation}
Now recalling that a fermion is a dimer on the even sublattice, but the absence of a dimer on the odd sublattice, the density of dimers on vertical links is given by $\rho_d(x,y)=\rho_{x,y}$ on the even sublattice, and $\rho_d(x,y)=1-\rho_{x,y}$ on the odd. Hence
\begin{equation}
\rho_d(x,y)=\frac{1}{\pi}{\rm arccos} \left(\frac{\pm w x/u+u N/w}{\sqrt{(N\pm x)^2-y^2}}\right).
\end{equation}
The results are summarized in Table.~\ref{tab:density_square}.
\begin{figure}[htbp]
\includegraphics{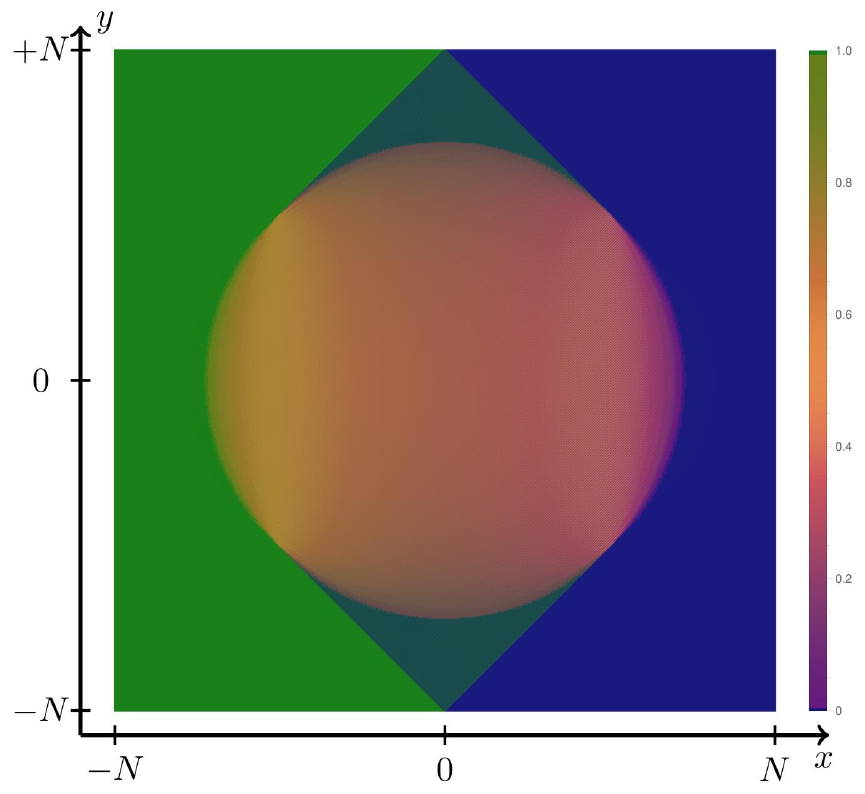}
\hfill
\includegraphics{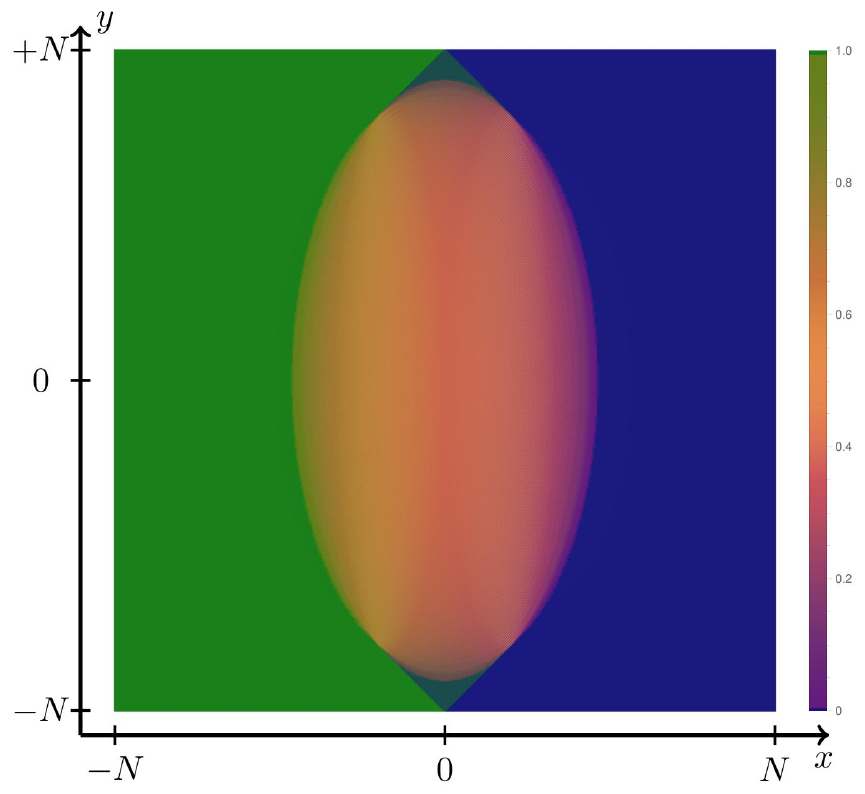}
\caption{Numerical fermionic density plots corresponding to an Aztec diamond of size $N=256$. \emph{Left:} Isotropic case $u/v=1$, where the arctic curve is a circle. \emph{Right:} $u/v=1/2$. The arctic curve is an ellipse with excentricity $e=\sqrt{3}/2$.}
 \label{fig:sqdensityplots}
\end{figure}
We finally note that the density of vertical dimers may also be expressed as
\begin{equation}
 \rho_d(x,y)=\frac{1}{2}-\frac{1}{\pi} \arctan\left(\frac{\frac{N u}{v w}\pm\frac{w x}{u v}}{\sqrt{N^2/w^2-x^2/u^2-y^2/v^2}}\right),
\end{equation}
which makes the arctic ellipse slightly more apparent. 
Setting $u=v=1$, $w=\sqrt{2}$ in this formula, we recover the result of Ref.~\cite{cohn1996local}. 
\begin{table}[ht]
	\begin{tabular}{c||c|c|}
		&  $\quad  x - y\in 2\mathbb{Z}-1/2\quad$   &   $\quad x - y \in 2\mathbb{Z}+1/2 \quad $ \\ &(odd sublattice)&(even sublattice)\\\hline \hline   &  & \\
		$\rho_{x,y}$  &  $\quad\frac{1}{\pi}{\rm arccos}\left(\frac{w x/u- u N/w}{\sqrt{(N-x)^2-y^2}}\right)\quad$   &  $\quad\frac{1}{\pi}{\rm arccos}\left(\frac{w x/u+ u N/w}{\sqrt{(N+x)^2-y^2}}\right)\quad$  \\
		\hline\hline
		&&\\
		$\rho_{d}(x,y)$ &$\quad\frac{1}{\pi}{\rm arccos}\left(\frac{-w x/u+ u N/w}{\sqrt{(N-x)^2-y^2}}\right)\quad$&$\quad\frac{1}{\pi}{\rm arccos}\left(\frac{w x/u+ u N/w}{\sqrt{(N+x)^2-y^2}}\right)\quad$\\
		\hline\hline
	\end{tabular}
	\caption{Average density of fermions $\rho(x,y)$ in the scaling limit, and corresponding density of dimers $\rho_{d}(x,y)$ on vertical edges.
	}
	\label{tab:density_square}
\end{table}

\newpage

\newpage

\section{Six-vertex model at \texorpdfstring{$\Delta = 0$}{} with domain-wall boundary conditions}

\label{part:6v}

\begin{figure}[htb]
$$
\begin{array}{lll}
	\begin{tikzpicture}[scale=0.55]
	\foreach \x in {0,1,...,10} \draw[red,dashed] (\x,0) -- (10,10-\x);
	\foreach \x in {1,2,...,10} \draw[red,dashed] (0,\x) -- (10-\x,10);
	\foreach \x in {0,1,...,10} \draw[red,dashed] (\x,10) -- (10,\x);
	\foreach \x in {1,2,...,10} \draw[red,dashed] (0,10-\x) -- (10-\x,0);
	\draw[line width=2pt] (0,0) -- ++(0.5,0.5) -- ++(0.5,-0.5) -- ++(0.5,0.5) -- ++(0.5,-0.5) -- ++(0.5,0.5) -- ++(0.5,-0.5) -- ++(0.5,0.5) -- ++(0.5,-0.5) -- ++(0.5,0.5) -- ++(0.5,-0.5);
	\draw (5,0) -- ++(0.5,0.5) -- ++(0.5,-0.5) -- ++(0.5,0.5) -- ++(0.5,-0.5) -- ++(0.5,0.5) -- ++(0.5,-0.5) -- ++(0.5,0.5) -- ++(0.5,-0.5) -- ++(0.5,0.5) -- ++(0.5,-0.5);
	\draw[line width=2pt] (0,10) -- ++(0.5,-0.5) -- ++(0.5,0.5) -- ++(0.5,-0.5) -- ++(0.5,0.5) -- ++(0.5,-0.5) -- ++(0.5,0.5) -- ++(0.5,-0.5) -- ++(0.5,0.5) -- ++(0.5,-0.5) -- ++(0.5,0.5);
	\draw (5,10) -- ++(0.5,-0.5) -- ++(0.5,0.5) -- ++(0.5,-0.5) -- ++(0.5,0.5) -- ++(0.5,-0.5) -- ++(0.5,0.5) -- ++(0.5,-0.5) -- ++(0.5,0.5) -- ++(0.5,-0.5) -- ++(0.5,0.5);
	\draw[thick,->]  (-0.1,5) -- (10,5) node[below]{$x$};
	\draw (9.2,4) node{$\in \mathbb{Z} + \frac{1}{2}$};
	\draw[thick,->]  (5,0) -- (5,10.4);
	\draw (5.8,10.6) node{$y \in \mathbb{Z} + \frac{1}{2}$};
	\foreach \y in {0.5,1,...,10} \draw (4.9,\y-0.25) -- ++ (0.2,0);
	\foreach \x in {0.5,1,...,9.5} \draw (\x-0.25,4.9) -- ++ (0,0.2);
	\draw (5.1,0.5) node[right]{$-N$};
	\draw (4.96,10) node[right]{$+N$};
	\draw (2.5,3) node{Imaginary};
	\draw (2.5,2.5) node{time};
	\draw[thick,->] (0.8,1.4) -- (0.8,3.6);
\end{tikzpicture}  \;
 &
\begin{tikzpicture}[scale=0.55,rotate=180]
	\foreach \x in {0,1,...,10} \draw[red,dashed] (\x,0) -- (10,10-\x);
	\foreach \x in {1,2,...,10} \draw[red,dashed] (0,\x) -- (10-\x,10);
	\foreach \x in {0,1,...,10} \draw[red,dashed] (\x,10) -- (10,\x);
	\foreach \x in {1,2,...,10} \draw[red,dashed] (0,10-\x) -- (10-\x,0);
	\begin{scope}
		\foreach \x in {0,1,...,5} \draw[line width=2pt] (5+\x,0) -- (10,5-\x);
		\foreach \x in {0,1,...,4} \draw[line width=2pt] (5+0.5*\x,1+0.5*\x) -- (6+\x,0);
		\foreach \x in {0,1,2,3} \draw[line width=2pt] (7.5+0.5*\x,3.5+0.5*\x) -- (10,1+\x);
	\end{scope}
	\begin{scope}[xshift=10cm,yshift=10cm,rotate=180]
		\foreach \x in {0,1,...,5} \draw (5+\x,0) -- (10,5-\x);
		\foreach \x in {0,1,...,4} \draw (5+0.5*\x,1+0.5*\x) -- (6+\x,0);
		\foreach \x in {0,1,2,3} \draw (7.5+0.5*\x,3.5+0.5*\x) -- (10,1+\x);
	\end{scope}
	\begin{scope}[xshift=10cm,rotate=90]
		\foreach \x in {0,1,...,5} \draw[line width=2pt] (5+\x,0) -- (10,5-\x);
		\foreach \x in {0,1,...,4} \draw[line width=2pt] (5+0.5*\x,1+0.5*\x) -- (6+\x,0);
		\foreach \x in {0,1,2,3} \draw[line width=2pt] (7.5+0.5*\x,3.5+0.5*\x) -- (10,1+\x);
	\end{scope}
	\begin{scope}[yshift=10cm,rotate=270]
		\foreach \x in {0,1,...,5} \draw (5+\x,0) -- (10,5-\x);
		\foreach \x in {0,1,...,4} \draw (5+0.5*\x,1+0.5*\x) -- (6+\x,0);
		\foreach \x in {0,1,2,3} \draw (7.5+0.5*\x,3.5+0.5*\x) -- (10,1+\x);
	\end{scope}
\end{tikzpicture} \quad
 &
\begin{tikzpicture}[scale=0.55,rotate=180]
	\foreach \x in {0,1,...,10} \draw (\x,0) -- (10,10-\x);
	\foreach \x in {1,2,...,10} \draw (0,\x) -- (10-\x,10);
	\foreach \x in {0,1,...,10} \draw (\x,10) -- (10,\x);
	\foreach \x in {1,2,...,10} \draw (0,10-\x) -- (10-\x,0);
	\begin{scope}
		\foreach \x in {0,1,...,5} \draw[line width=2pt] (5+\x,0) -- (10,5-\x);
		\foreach \x in {0,1,...,4} \draw[line width=2pt] (5+0.5*\x,1+0.5*\x) -- (6+\x,0);
		\foreach \x in {0,1,2,3} \draw[line width=2pt] (7.5+0.5*\x,3.5+0.5*\x) -- (10,1+\x);
	\end{scope}
	\begin{scope}[xshift=10cm,rotate=90]
		\foreach \x in {0,1,...,5} \draw[line width=2pt] (5+\x,0) -- (10,5-\x);
		\foreach \x in {0,1,...,4} \draw[line width=2pt] (5+0.5*\x,1+0.5*\x) -- (6+\x,0);
		\foreach \x in {0,1,2,3} \draw[line width=2pt] (7.5+0.5*\x,3.5+0.5*\x) -- (10,1+\x);
	\end{scope}
	 \draw[line width=2pt] (5,1) -- ++(-2,2) -- ++(0.5,0.5) -- ++(-1,1) -- ++(0.5,0.5) -- ++(-0.5,0.5) -- ++(1,1) -- ++ (-0.5,0.5) -- ++(2,2);
	 \draw[line width=2pt] (5.5,1.5) -- ++(-2,2) -- ++(0.5,0.5) -- ++(-0.5,0.5) -- ++(1,1) -- ++(-1,1) -- ++(2,2);
	 \draw[line width=2pt] (6,2) -- ++(-2,2) -- ++(1,1) -- ++(-0.5,0.5) -- ++(1,1) -- ++(-0.5,0.5) -- ++(1,1);
	 \draw[line width=2pt] (6.5,2.5) -- ++(-1,1) -- ++(0.5,0.5) -- ++(-1,1) -- ++(1,1) -- ++(-0.5,0.5) -- ++(1,1);
	 \draw[line width=2pt] (7,3) -- ++(-1,1) -- ++(0.5,0.5) -- ++(-0.5,0.5) -- ++(0.5,0.5) -- ++(-0.5,0.5) -- ++(1,1);
	 \draw[line width=2pt] (7.5,3.5) -- ++(-0.5,0.5) -- ++(0.5,0.5) -- ++(-1,1) -- ++(1,1);
	 \draw[line width=2pt] (8,4) -- ++(0.5,0.5) -- ++(-0.5,0.5) -- ++(0.5,0.5) -- ++(-0.5,0.5);
	 \draw[line width=2pt] (8.5,4.5) -- ++(0.5,0.5) -- ++(-0.5,0.5);
\end{tikzpicture}  \\
({\rm a}) & ({\rm b}) & ({\rm c})
\end{array}
$$
\caption{Six-vertex model on the square lattice. We regard the horizontal axis as space, and the vertical axis as imaginary time. (a) One starts form the domain-wall initial state (DWIS) at the bottom, then we apply the transfer matrix several times, and finally, we project back to the DWIS at the top. (b) Because of the DWIS, a large number of edges are constrained to be either empty or filled; the only vertices that are not fixed by the choice of boundary conditions are the ones inside the square region in the middle of the figure. This central square is what is known in the literature as the 'six-vertex model with domain-wall boundary conditions'. (c) A configuration of vertices compatible with the boundary conditions: note that one can interpret the filled edges as trajectories in space-time. This is the idea of the mapping to a fermion model.}
\label{fig:six-vertex}
\end{figure}
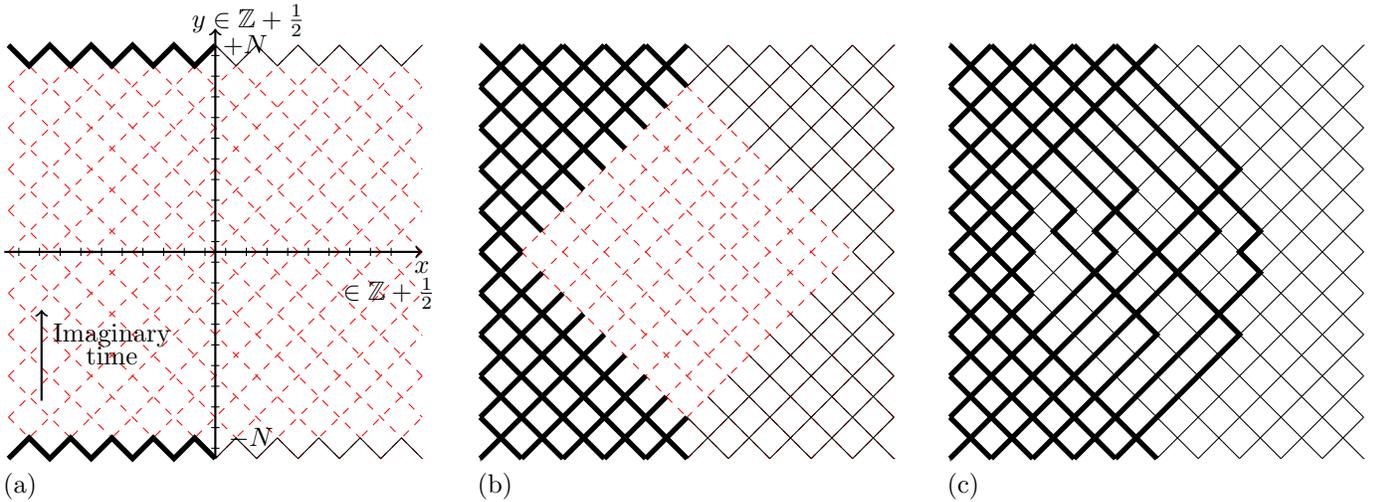

\noindent It is conventional to draw the vertices of the six-vertex model as
\begin{equation*}
	\label{eq:6v_config}
\begin{tikzpicture}
	\begin{scope}[scale=0.6,rotate=45]
		\draw[line width=2pt] (-1,0) -- (1,0);
		\draw[line width=2pt] (0,-1) -- (0,1);
		\draw (-1.3,-1.3) node {$a$};
	\end{scope}
	\begin{scope}[xshift=2cm,scale=0.6,rotate=45]
		\draw (-1,0) -- (1,0);
		\draw (0,-1) -- (0,1);
		\draw (-1.3,-1.3) node {$a$};
	\end{scope}
	\begin{scope}[xshift=5cm,scale=0.6,rotate=45]
		\draw (-1,0) -- (1,0);
		\draw (0,-1) -- (0,1);
		\draw[line width=2pt] (0,-1) -- (0,1);
		\draw (-1.3,-1.3) node {$b$};
	\end{scope}
	\begin{scope}[xshift=7cm,scale=0.6,rotate=45]
		\draw (-1,0) -- (1,0);
		\draw (0,-1) -- (0,1);
		\draw[line width=2pt] (-1,0) -- (1,0);
		\draw (-1.3,-1.3) node {$b$};
	\end{scope}
		\begin{scope}[xshift=10cm,scale=0.6,rotate=45]
		\draw (-1,0) -- (1,0);
		\draw (0,-1) -- (0,1);
		\draw[line width=2pt] (0,-1) -- (0,0) -- (1,0);
		\draw (-1.3,-1.3) node {$c$};
	\end{scope}
	\begin{scope}[xshift=12cm,scale=0.6,rotate=45]
		\draw (-1,0) -- (1,0);
		\draw (0,-1) -- (0,1);
		\draw[line width=2pt] (-1,0) -- (0,0) -- (0,1);
		\draw (-1.3,-1.3) node {$c$};
	\end{scope}	
\end{tikzpicture}
\end{equation*}
with weights $a,b,c \in \mathbb{R}$. The configurations of the six-vertex model are in one-to-one correspondence with (directed) self-avoiding trajectories on the square lattice as illustrated in Fig. \ref{fig:six-vertex}.(c). The trajectories can touch---a situation that corresponds to the first vertex with weight $a$---, but they cannot overlap along an edge. By convention, we decide that the trajectories never cross. This is convenient, because it means we do not have to be too careful about the fermionic statistics of the particles whose trajectories we are describing. [We would have to be more careful if we took periodic boundary conditions.] We view the system as a set of fermionic particles on the lattice $\mathbb{Z} + \frac{1}{2}$ that evolve in the $y$-direction, which is imaginary time. The six vertices drawn above are then operators acting on two neighboring sites $x$ and $x+1$, from bottom to top. They can be written respectively as
\begin{equation*}
	\begin{array}{rcl}
		{\rm weight }\; a \, : &&  \left\{ \begin{array}{c}
			c^\dagger_x c_x  \, c^\dagger_{x+1} c_{x+1} \\ \\
			(1-c^\dagger_x c_x ) \, ( 1-c^\dagger_{x+1} c_{x+1} )
		\end{array} \right. \\ \\ 
		{\rm weight }\; b \, : &&  \left\{ \begin{array}{c}
			c^\dagger_x c_{x+1}  \\ \\
			c^\dagger_{x+1} c_x
		\end{array} \right. \\ \\ 
		{\rm weight }\; c \, : &&  \left\{ \begin{array}{c}
			 ( 1-c^\dagger_x c_x ) c^\dagger_{x+1} c_{x+1} \\ \\
			c^\dagger_x c_x ( 1-c^\dagger_{x+1} c_{x+1 } ) \; .
		\end{array} \right. 
	\end{array}
\end{equation*}
Summing these six terms, we build the $R$-matrix of the six-vertex model,
\begin{eqnarray}
	\label{eq:R_matrix}
\nonumber	R_{x,x+1} & = & a \left[  c^\dagger_x c_x   c^\dagger_{x+1} c_{x+1} + (1-c^\dagger_x c_x ) \, ( 1-c^\dagger_{x+1} c_{x+1} )\right]  +   b \left[ c^\dagger_x c_{x+1}  + c^\dagger_{x+1} c_x \right]   \\
\nonumber && \qquad \qquad \quad  +  \, c \left[ c^\dagger_x c_x (1-c^\dagger_{x+1} c_{x+1} ) + ( 1-c^\dagger_x c_x ) c^\dagger_{x+1} c_{x+1}  \right] \\
	&=&   \, a \, \exp \left[ \log \left(\frac{b+c}{a}\right) \left( c^\dagger_x c_{x+1} \, + \, c^\dagger_{x+1} c_x \right) \right] \,.
\end{eqnarray}
This last equality holds provided that
\begin{equation}
	\Delta \, = \, \frac{a^{2}+b^{2}-c^{2}}{2ab} \, = \, 0 \,,
\end{equation}
and {\bf this is what we assume from now on}. Also, to keep formulas as simple as possible, {\bf we assume that $a>0$, $b>0$ and $c>0$}; this will save us the trouble of having to keep track of the signs.  Now, since the $R$-matrix is gaussian at $\Delta = 0$, the model is free: the transfer matrix is gaussian as well, and all observables can be obtained from the propagator using Wick's theorem. The techniques developed in previous parts apply, as we explain below.

However, before we turn to the technicalities about the transfer matrix and Hilbert transforms, let us briefly explain why, in our setting, we recover the six-vertex model with domain-wall boundary conditions---a model that has been studied extensively in the literature both analytically and numerically \cite{korepin_zinnjustin,zinn2000six,colomo2010arctic,colomo2005two,allison2005numerical}---. We focus once again on the DWIS $\ket{\Psi_0}$, which is completely filled with particles on the left, and completely empty on the right. The evolution of the system is generated by iterations of the transfer matrix. After a number of iterations, we project the one-dimensional system of particles back to the DWIS (Fig. \ref{fig:six-vertex}.(a)). Because of the initial and final states, most vertices on the left and right of the system are forced to be in the configurations with weight $a$, as illustrated in Fig. \ref{fig:six-vertex}.(b). Only a central region, which turns out to be a square, is not determined {\it a priori} by those initial and final states. The square in the middle hosts configurations  of the six-vertex model with boundary conditions (Fig. \ref{fig:six-vertex}.(b)) that are known traditionally as 'domain-wall boundary conditions'. Therefore, our setup, which combines the DWIS with the transfer matrix, is well suited to revisit this popular problem.

\subsection{The transfer matrix}

We need to define two transfer matrices, one for the even rows and another for the odd rows,
\begin{equation}
\left\{ 	\begin{array}{rcl}	\displaystyle T_{{\rm e}} & \equiv & \displaystyle \prod_{ x \in 2\mathbb{Z} + \frac{1}{2}  } R_{x,x+1} \\
	\displaystyle T_{{\rm o}} & \equiv & \displaystyle \prod_{x \in 2 \mathbb{Z} - \frac{1}{2}} R_{x,x+1} \, . \end{array} \right.
\end{equation}
Notice that the terms commute within each product; but, of course, $T_{{\rm e}} T_{{\rm o}} \neq T_{{\rm o}} T_{{\rm e}}$.  The weighted configurations of the model are obtained by applying iteratively these two operators, one after the other. One could focus on one of the double-row transfer matrices $T_{\rm e}T_{\rm o}$ or $T_{\rm o}T_{\rm e}$, which generate the imaginary time evolution. However the drawback of both those operators is that they are not hermitian. In order to apply the formalism developed in part \ref{sec:ktr}, it is more convenient to deal with a hermitian generator of translations in imaginary time. Thus, we define our (double-row) transfer matrix as the operator
\begin{equation}
	T^2 \, = \, T_{{\rm e}}^{\frac{1}{2}} \,T_{{\rm o}} \, T_{{\rm e}}^{\frac{1}{2}} 
\end{equation}
whose hermiticity follows from $T_{{\rm e}}^\dagger \, = \, T_{{\rm e}}$ and $T_{{\rm o}}^\dagger \, = \, T_{\rm o}$. Again, the upperscript/exponent '$2$' in $T^2$ is there to remind us that it is an operator that adds {\it two rows} to the system, instead of one. We will always deal with $T^2$, there will be no object '$T$'. Now let's diagonalize $T^2$. In $k$-space, we have
\begin{eqnarray*}\fl
	\left\{ \begin{array}{rcl} \displaystyle	T_{{\rm e}} &=& \displaystyle \exp \left[  \log \left(\frac{b+c}{a} \right) \int_{-\frac{\pi}{2}}^{\frac{\pi}{2}} \frac{dk}{2\pi} \left(  \begin{array}{cc}  c^\dagger(k) & c^\dagger (k+\pi)  \end{array}  \right)  \left(  \begin{array}{cc}  \cos k &  \sin k \\ \sin k & - \cos k \end{array}  \right)  \left(  \begin{array}{c}  c(k) \\  c(k+\pi)  \end{array}  \right)  \right] \\ \\
\displaystyle	T_{{\rm o}} &=& \displaystyle \exp \left[  \log \left(\frac{b+c}{a} \right) \int_{-\frac{\pi}{2}}^{\frac{\pi}{2}} \frac{dk}{2\pi} \left(  \begin{array}{cc}  c^\dagger(k) & c^\dagger (k+\pi)  \end{array}  \right)  \left(  \begin{array}{cc}  \cos k &  -\sin k \\ -\sin k & - \cos k \end{array}  \right)  \left(  \begin{array}{c}  c(k) \\  c(k+\pi)  \end{array}  \right)  \right]  \, . \end{array} \right.
\end{eqnarray*}
[Here and in what follows, we drop the unimportant normalization constant $a^{{\rm number \; of \; sites}}$ that comes from the factor $a$ in the r.h.s of (\ref{eq:R_matrix}).] Then the problem boils down to the diagonalization of a $2 \times 2$ matrix; after some easy algebra, we find 
\begin{equation*}
	T^2 \, = \, \exp \left[ -\int_{-\frac{\pi}{2}}^\frac{\pi}{2} \frac{dk}{2\pi}  \,2\varepsilon(k)\, \left( c_{+}^\dagger(k) c_{+}(k) \, - \,  c_{-}^\dagger(k) c_{-}(k)   \right) \right]
\end{equation*}
where
\begin{equation}\fl
\left\{	\begin{array}{rcl}
	\label{eq:diag_6v}
	2\varepsilon(k) &=& \displaystyle  - \log\left[ \frac{c + b \cos \kappa(k) }{c - b \cos \kappa(k) } \right] \\ \\
\displaystyle	\left( \begin{array}{c} c_{+}^\dagger(k) \\    c_{-}^\dagger(k)   \end{array} \right) &=&  \displaystyle  U(k) \left( \begin{array}{c}  c^\dagger(k) \\ c^\dagger(k+\pi)  \end{array}\right)   \, = \,   \displaystyle   \left( \begin{array}{cc}   \cos \left(\frac{k-\kappa(k)}{2}\right) &  \sin \left(\frac{k-\kappa(k)}{2}\right)   \\ \\  - \sin \left(\frac{k-\kappa(k)}{2}\right) &  \cos \left(\frac{k-\kappa(k)}{2}\right)   \end{array}   \right)  \left( \begin{array}{c}  c^\dagger(k) \\ c^\dagger(k+\pi)  \end{array} \right)  \, ,  
 \end{array} \right.
\end{equation}
and $\kappa(k)$ is the real-valued function $\kappa : [-\frac{\pi}{2},\frac{\pi}{2}] \rightarrow [-\frac{\pi}{2}, \frac{\pi}{2}]$ defined by
\begin{equation}
	\label{eq:pseudo_mom}
	\tan  \kappa(k)  \, =\,  \frac{a}{c} \, \tan k \, .
\end{equation}
The mapping $k\mapsto \kappa(k)$ is always bijective (since we assume $a > 0$, $c>0$), so it is possible to switch from the variable $k$ to the new variable $\kappa$ and to reparametrize all functions of $k$, like $\varepsilon(k) \rightarrow \varepsilon(\kappa)$ or $\tilde{\varepsilon}(k) \rightarrow \tilde{\varepsilon}(k)$ (introduced in the next paragraph). This change of variables $k \rightarrow \kappa$ is instrumental in the appearance of the Hilbert transform in the six-vertex model, as we explain in the next section. To keep notations as light as possible, it will be implicit in most formulas, unless specified otherwise. Let us make one last important remark about $\kappa$: the mapping $k \mapsto \kappa(k)$ can be analytically continued to $\mathbb{R}$, and it has the property $\kappa ( k + p \pi) \,= \, \kappa (k) + p \pi $ for any $p \in \mathbb{Z}$. In the formulas below, we use the function $\kappa : \mathbb{R} \rightarrow \mathbb{R}$ and this property without recalling it.

Finally, notice that the three assumptions of part \ref{sec:ktr} are satisfied: in particular, $\varepsilon(k)$ is an {\it anti-periodic} function of $k$ with period $\pi$ (namely, $\varepsilon(k+ \pi) \,=\, - \varepsilon(k) $), and the relation (\ref{eq:periodU}) holds, such that $c^\dagger_+(k+\pi) =  c^\dagger_-(k)$.

\subsection{\texorpdfstring{$k \mapsto \kappa$}{} and the Hilbert transform}

We are now in position to apply the formalism of part \ref{sec:ktr} to the transfer matrix $T^2$. First, we need the propagator in the DWIS, $ \bra{\Psi_0} c^\dagger_+(k) c_+(k') \ket{\Psi_0} $; it is computed very easily from Eqs. (\ref{eq:diag_6v})-(\ref{eq:pseudo_mom}), leading to the explicit result $ \cos (\frac{\kappa-\kappa'}{2} ) / [i \,\sin (k - k')]$, where $\kappa = \kappa(k)$ and $\kappa' = \kappa(k')$. For our purposes though, it is more convenient to write this equation in the form
\begin{equation}
	\label{eq:correl_cp_6v}
	\bra{\Psi_0} c^\dagger_+(k) c_+(k') \ket{\Psi_0} \, =  \,  \left( \frac{d\kappa}{d k} \right)^{\frac{1}{2}}   \left( \frac{d\kappa'}{d k'} \right)^{\frac{1}{2}}   \frac{ 1 }{2 i \sin (\frac{\kappa - \kappa'}{2})}  \, .
\end{equation}
As anticipated in part \ref{sec:ktr}, the transform $\tilde{\varepsilon}_{U}(k)$ that appears is nothing but the Hilbert transform, modulo conjugation by the change of variable $k \rightarrow \kappa$:
\begin{eqnarray}
\nonumber	\tilde{\varepsilon}_{U}(k) & = & {\rm p.v.} \int_{-\pi}^\pi \frac{dk'}{2\pi} \varepsilon(k')  2\partial_{k'} \log  \bra{\Psi_0} c^\dagger_+(k) c_+(k') \ket{\Psi_0}  \\
\nonumber				&=& {\rm p.v.} \int_{-\pi}^\pi \frac{d\kappa'}{2\pi} \varepsilon(k')  2\partial_{\kappa'}  \log \left[ \frac{1}{2 i \sin \left( \frac{\kappa - \kappa'}{2} \right)} \right] \, + \,  {\rm p.v.} \int_{-\pi}^\pi \frac{d k'}{2\pi} \varepsilon(k')  \partial_{k'}  \log \left[  \frac{d\kappa'}{d k'}  \right]    \\
\nonumber				&=& {\rm p.v.} \int_{-\pi}^\pi \frac{d\kappa'}{2\pi} \varepsilon(\kappa')  \cot \left( \frac{\kappa - \kappa'}{2} \right) \quad + \quad 0\, .		
\end{eqnarray}
In the second line, we have split the function $\log  \bra{\Psi_0} c^\dagger_+(k) c_+(k') \ket{\Psi_0} $ into a sum of three terms, one of which trivially vanishes; then in the first term we changed variables $k' \rightarrow \kappa'$, using the fact that the Jacobian $\frac{dk'}{d\kappa'}$ in the integration measure cancels the one coming from the chain rule for the derivative $\partial_{k'} = \frac{d\kappa'}{dk'} \partial_{\kappa'}$. In this term, we also use the notation $\varepsilon (\kappa') $ for $\varepsilon(k')$. In the third line, the second term is zero because of the {\it anti-periodicity} of $\varepsilon(k)$, $\varepsilon(k+\pi) = -\varepsilon (k)$. Indeed, the derivative $\frac{d\kappa}{dk}$ is {\it periodic} (with period $\pi$), therefore the integrand $\varepsilon(\kappa')   \partial_{k'}  \log \left[  \frac{d\kappa'}{d k'}  \right]$ is {\it anti-periodic}, and its integral from $-\pi$ to $0$ cancels the one from $0$ to $+\pi$.

\vspace{0.5cm}
We call $\tilde{\varepsilon}(\kappa) = \tilde{\varepsilon}_U(k(\kappa))$ the Hilbert transform of the function $\varepsilon(\kappa)$---viewed as a function of $\kappa$, not $k$---. For the function $\varepsilon (\kappa)$ of interest here, we have (see the appendix \ref{appx:hilb}):
\begin{equation}
	\label{eq:eps_epst_6v}
	2\varepsilon(\kappa) \, = \,  - \log\left[ \frac{c + b \cos \kappa}{c - b \cos \kappa } \right]  \quad  \Leftrightarrow \quad
	2\tilde{\varepsilon}_{U}(k) \,=\,  2 \tilde{\varepsilon}(\kappa)  \, = \,   i  \log\left[  \frac{a  \, + \, i \, b\,  \sin  \kappa }{ a  \, - \, i \, b\, \sin  \kappa   }   \right] \, .
\end{equation}
[Here we are using $a^2+b^2=c^2$, and also the assumption that $a>0$, $b>0$, $c>0$, which was announced earlier; there could be a minus sign in the r.h.s otherwise.] We note that the function $\tilde{\varepsilon}(\kappa)$ is automatically anti-periodic with period $\pi$, a property it inherits from $\varepsilon(\kappa)$.

\subsection{Fermionic correlators: exact finite-size formulas}
In this section we give the explicit formulas for the fermionic correlators. These formulas are exact in finite size. The most convenient way to express these results is, like in the introduction, to exhibit the relation between correlators $\left<  c^\dagger_{x_1,y_1} \dots c^\dagger_{x_n,y_n}   c_{x'_1,y'_1} \dots c_{x'_n,y'_n}  \right>$ in the two-dimensional geometry, and (known) correlators in the DWIS $\bra{\Psi_0} c^\dagger(k_1) \dots c^\dagger(k_n) c(k'_1) \dots c(k'_n) \ket{\Psi_0}$, based on the results of part \ref{sec:ktr}. Using the same notation as in the introduction---see Eq. (\ref{eq:notation_doteq})---this relation can be expressed through the identification
\begin{equation}
\label{eq:ident_6v}
	\left\{  \begin{array}{rcl} c^\dagger_{x,y}   & \doteq &  \displaystyle \int_{-\pi}^\pi \frac{dk}{2\pi} e^{-i k x + y \varepsilon (\kappa)  - i N \tilde{\varepsilon}(\kappa) }  \,  r_{x,y} (k)  \, c^\dagger_+(k) \\ \\
	c_{x,y}   & \doteq & \displaystyle  \int_{-\pi}^\pi \frac{dk}{2\pi} e^{i k x-y \varepsilon (\kappa)  + i N \tilde{\varepsilon}(\kappa) }  \,  s_{x,y} (k)  \, c_+(k) \, . \end{array} \right.
\end{equation}
The functions $r_{x,y}(k)$ and $s_{x,y}(k)$ can be computed by combining the results of the previous sections. The calculation itself is a little bit cumbersome, with various cases that need to be distinguished, so we defer it to the appendix \ref{sec:moredetails}. The final result, however, takes a simple form. We find
\begin{equation}\label{eq:rsk_vertex}
 r_{x,y}(k)=s_{x,y}(k)=\sqrt{1-i  \,(-1)^{x-y}\, \frac{b}{a}\sin \kappa} \, .
\end{equation}
The reason why various cases need to be distinguished is the alternation of $T_{\rm e}$ and $T_{\rm o}$ in our transfer matrix setup. For an observable $\mathcal{O}_{x,y}$ at point $(x,y) \in (\mathbb{Z} + \frac{1}{2})\times (\mathbb{Z} + \frac{1}{2})$, the formula for the expectation value depends on the parity of $y-\frac{1}{2}$. More explicitly, assuming that $N$ is {\it odd} as in the figure \ref{fig:six-vertex} [the case $N$ even would require small shifts in the powers of $T^2$], the formula reads
\begin{equation}\fl
(N \;{\rm odd}) \qquad  \quad	\left<  \mathcal{O}_{x,y}  \right> \, = \,  \left\{  \begin{array}{lll}
	\displaystyle \frac{  \bra{\Psi_0} T_{{\rm e}}^{1/2} T^{2(\frac{N-y-1/2}{2})} \, T_{{\rm e}}^{1/2} \mathcal{O}_x  T_{{\rm e}}^{-1/2}  \, T^{2(\frac{N+y+1/2}{2})} T_{{\rm e}}^{1/2}  \ket{\Psi_0}   }{\bra{\Psi_0} T_{{\rm e}}^{1/2} T^{2N} T_{{\rm e}}^{1/2}  \ket{\Psi_0} }  & \quad  & {\rm if} \quad y \in 2 \mathbb{Z} + \frac{1}{2}   \\ \\
	\displaystyle \frac{  \bra{\Psi_0} T_{{\rm e}}^{1/2} T^{2(\frac{N-y+1/2}{2})} \, T_{{\rm e}}^{-1/2} \mathcal{O}_x  T_{{\rm e}}^{1/2}  \, T^{2(\frac{N+y-1/2}{2})} T_{{\rm e}}^{1/2}  \ket{\Psi_0}   }{\bra{\Psi_0} T_{{\rm e}}^{1/2} T^{2N} T_{{\rm e}}^{1/2}  \ket{\Psi_0} }  & \quad  & {\rm if} \quad y \in 2 \mathbb{Z} - \frac{1}{2} \, .
	\end{array} \right.
\end{equation}
So it is clear that, in the derivation of the functions $r_{x,y}(k)$ and $s_{x,y}(k)$, one needs to treat the two cases $y \in 2 \mathbb{Z} \pm \frac{1}{2}$ separately, because they involve different conjugations by $T_{{\rm e}}^{\pm 1/2}$. The cases $x \in 2 \mathbb{Z} \pm \frac{1}{2}$ are distinguished for convenience; the formulas are more compact if one treats them separately.

\subsection{Fermionic correlators: scaling limit}

Once again, we are interested in the scaling limit of the corelators. These are obtained by the steepest descent method, exactly like in the introduction. We focus on the argument of the exponentials in Eq. (\ref{eq:ident_6v}). The stationary points are the solutions of
\begin{equation*}
	 \frac{x}{b} \, +\,  \left( -i \frac{y}{a} \sin \kappa  + \frac{N}{c} \cos \kappa \right)    \, = \, \frac{1}{b}\,\frac{d}{dk} \left[  k x  \,+\,   i y   \varepsilon(\kappa) +  N   \tilde{\varepsilon}(\kappa)     \right] \, = \, 0 \, .
\end{equation*}
Here we have used the chain rule $\frac{d}{d k} = \frac{d\kappa}{dk} \frac{d}{d\kappa}$ and the formulas
\begin{equation*}
	\frac{d \varepsilon(\kappa)}{d \kappa} \, = \, \frac{-  b c \, \sin \kappa }{c^2 - b^2  \cos^2 \kappa } \, , \qquad
	\frac{d \tilde{\varepsilon}(\kappa)}{d \kappa} \, = \, \frac{ a b \, \cos \kappa }{c^2 -  b^2   \cos^2 \kappa }  \, ,\qquad 	\frac{d k}{d \kappa} \, =\, \frac{a c}{ c^2 - b^2 \cos^2 \kappa } \, .
\end{equation*}
Generically, there are two non-degenerate stationary points
\begin{equation*}
	\kappa(x,y) \, =\,  z(x,y)   \qquad {\rm and}  \qquad \kappa(x,y) \, = \, -z^*(x,y)   \, ,
\end{equation*}
where
\begin{equation}
	\label{eq:z_6v}
	z(x,y) \, = \, - \frac{\pi}{2}  \,- \, {\rm arcsin} \frac{x/b}{\sqrt{ (\frac{N}{c} )^2 - (\frac{y}{a} )^2 }} \,-\, i \,{\rm arcth} \left( \frac{y/a}{N/c} \right) \, .
\end{equation}
By 'generically', we mean 'unless $\frac{x^2}{b^2} + \frac{y^2}{a^2} = \frac{N^2}{c^2}$'. This ellipse is, of course, the arctic curve for the six-vertex model with domain-wall boundary conditions. Now let us assume that $(x,y)$ is inside the ellipse; then around the two stationary points, the exponential is approximated by
\begin{equation*}\fl
	\left\{  \begin{array}{rcl}
		e^{ -i N \left[  k \frac{x}{N} +   i \frac{y}{N} \varepsilon(k) +  \tilde{\varepsilon}(k)    \right]  } &  \simeq  &  e^{ -i  \varphi(x,y )  - i \frac{1}{2} (k - k(z) )^2 \, e^{\sigma (x,y) } \left(\frac{dk}{d\kappa}\right)^{-1}    }  \qquad {\rm around} \quad k = k(z)  \\ \\
		e^{ -i N \left[  k \frac{x}{2N} +   i \frac{y}{N} \varepsilon(k) +  \tilde{\varepsilon}(k)    \right]  } &  \simeq  &  e^{ i  \varphi^*(x,y )  - i \frac{1}{2} (k - k(-z^*) )^2 \, e^{\sigma (x,y)}  \left(\frac{dk^*}{d\kappa^*}\right)^{-1}     }     \qquad {\rm around} \quad  k = k(-z^*)  \, ,
	\end{array} \right.
\end{equation*}
where
\begin{equation}\fl
	\label{eq:phi_sigma_6v}
	\left\{  \begin{array}{rcl}
		\varphi \,= \, \varphi(x,y) & \equiv &   k(z(x,y)) x  +   i  y \varepsilon(z(x,y)) +  N  \tilde{\varepsilon}(z(x,y))      \\  \\
		e^{\sigma} \, = \, e^{\sigma(x,y)} &\equiv& \left(\frac{dk}{d\kappa}\right)  \frac{d^2}{dk^2} \left[  i  y  \varepsilon(k) +  N \tilde{\varepsilon}(k)   \right] \, = \,  b  \, \frac{d}{d\kappa}  \left[  -i  \frac{y}{a} \, \sin \kappa  +  \frac{N}{c}  \, \cos \kappa  \right] \, =\,   b \, \sqrt{ \left(\frac{N}{c}\right)^2 - \left(\frac{x}{b}\right)^2 - \left(\frac{y}{a}\right)^2  } \, .
	\end{array}  \right. 
\end{equation}
Performing the gaussian integration, we find that the identification (\ref{eq:ident_6v}) becomes
\begin{equation}\fl
\label{eq:ident_6v_2}
	\left\{  \begin{array}{rcl} c^\dagger_{x,y}   & \underset{N \rightarrow \infty}{\doteq} &  \displaystyle  \frac{ e^{-i \frac{\pi}{4}} }{\sqrt{2\pi}}  e^{ -i  \varphi}   e^{-\frac{\sigma }{2}  }   \,  r_{x,y} (z)  \,    \left( \frac{dk}{d \kappa} \right)^{\frac{1}{2}} c^\dagger_+(z)  \;  + \;   \frac{ e^{i \frac{\pi}{4}} }{\sqrt{2\pi}}  e^{ i  \varphi^*}   e^{-\frac{\sigma }{2}  }   \,  r_{x,y} (-z^*)  \,    \left( \frac{dk^*}{d \kappa^*} \right)^{\frac{1}{2}} c^\dagger_+(-z^*)   \\ \\
	c_{x,y}   & \underset{N \rightarrow \infty}{\doteq}  &  \displaystyle  \frac{ e^{i \frac{\pi}{4}} }{\sqrt{2\pi}}  e^{ i  \varphi}   e^{-\frac{\sigma }{2}  }   \,  s_{x,y} (z)  \,    \left( \frac{dk}{d \kappa} \right)^{\frac{1}{2}} c_+(z)  \;  + \;   \frac{ e^{-i \frac{\pi}{4}} }{\sqrt{2\pi}}  e^{ -i  \varphi^*}   e^{-\frac{\sigma }{2}  }   \,  s_{x,y} (-z^*)  \,    \left( \frac{dk^*}{d \kappa^*} \right)^{\frac{1}{2}} c_+(-z^*)  \, . \end{array} \right.
\end{equation}
Notice that the appearance of $ \left( \frac{dk}{d \kappa} \right)^{\frac{1}{2}} c^\dagger_+(z) $ and $ \left( \frac{dk}{d \kappa} \right)^{\frac{1}{2}} c_+(z) $ is particularly convenient, since the roots of the Jacobians are exactly what is needed to cancel the ones in (\ref{eq:correl_cp_6v}). Their propagator is 
\begin{equation}
	\label{eq:prop_6v_wJac}
	\bra{\Psi_0} \left( \frac{dk}{d \kappa} \right)^{\frac{1}{2}} c^\dagger_+(z) \left( \frac{dk'}{d \kappa'} \right)^{\frac{1}{2}} c_+(z')  \ket{\Psi_0} \, =\, \frac{1}{2 i \sin \left( \frac{z-z'}{2}\right)} \, . 
\end{equation}

\subsection{Identification of the Dirac theory inside the arctic circle}

At this point, it is quite clear that the Dirac action in the six-vertex model is going to be the same as in the introduction, with the new $\varphi$ and $\sigma$ given in Eq. (\ref{eq:phi_sigma_6v}). The difference is rather in the relation between the lattice and continuous degrees of freedom, which is more complicated than in the case of the XX chain, see Eq. (\ref{eq:lattice_continuous_XX}). Indeed, we want a relation between the lattice fermions $c^\dagger_{x,y}$ and the two components of the Dirac spinor $\Psi^\dagger = \left( \begin{array}{cc}  \psi & \overline{\psi}  \end{array} \right)$ that is local, and that does not depend on position, apart from possible sub-lattice distinctions. Looking back at Eqs.~(\ref{eq:ident_6v})-(\ref{eq:rsk_vertex}), we see that this is obtained by expanding the factors $r_{x,y}(k) = s_{x,y}(k)$ in Fourier series:
\begin{equation*}
	r_{x,y}(k) \, = \,s_{x,y}(k) \, = \,  \sum_{n \in \mathbb{Z}} \alpha_{x,y,n} \, e^{i n k} \, .
\end{equation*}
The coefficients $\alpha_{x,y,n}$ decay exponentially with $|n|$ because $r_{x,y}(k)$ is analytic in $k$ (and, of course, periodic with period $2 \pi$). Plugging this expansion in Eq. (\ref{eq:ident_6v}) the coefficients given by Eq.~(\ref{eq:rsk_vertex}), and using the analysis of the previous section, we arrive at
\begin{equation}
	\label{eq:6v_c_psi}
	\left\{ \begin{array}{rcl}
		c_{x,y}^\dagger &=& \displaystyle  \sum_{n \in \mathbb{Z}}   \alpha_{x,y,n}  \left[  \frac{1}{\sqrt{2\pi}} \psi^\dagger(x-n,y) +  \frac{1}{\sqrt{2\pi}} \overline{\psi}^\dagger(x-n,y)  \right]    \\
		c_{x,y} &=& \displaystyle  \sum_{n \in \mathbb{Z}}   \alpha_{x,y,n}  \left[  \frac{1}{\sqrt{2\pi}} \psi(x+n,y) +  \frac{1}{\sqrt{2\pi}} \overline{\psi}(x+n,y)  \right]   \, .
	\end{array} \right.      \\ \\
\end{equation}
The propagators of $\psi^\dagger$, $\psi$ and $\overline{\psi}^\dagger$, $\overline{\psi}$ are
\begin{eqnarray*}\fl
	\left<  \psi^\dagger(x,y) \psi(x',y')  \right> & = & e^{-i (\varphi - \varphi')} e^{-\frac{\sigma+\sigma'}{2}}  \bra{\Psi_0} \left( \frac{dk}{d \kappa} \right)^{\frac{1}{2}} c^\dagger_+(z) \left( \frac{dk'}{d \kappa'} \right)^{\frac{1}{2}} c_+(z')  \ket{\Psi_0} \, = \, e^{-i (\varphi - \varphi')} e^{-\frac{\sigma+\sigma'}{2}}  \frac{1}{2i \sin \left( \frac{z-z'}{2} \right)}  \\\fl
	\left<  \overline{\psi}^\dagger(x,y) \overline{\psi}(x',y')  \right> & = & e^{-i (\varphi - \varphi')} e^{-\frac{\sigma+\sigma'}{2}}  \bra{\Psi_0} \left( \frac{dk}{d \kappa} \right)^{\frac{1}{2}} c^\dagger_+(z) \left( \frac{dk'}{d \kappa'} \right)^{\frac{1}{2}} c_+(z')  \ket{\Psi_0} \, = \, e^{-i (\varphi - \varphi')} e^{-\frac{\sigma+\sigma'}{2}}  \frac{1}{2i \sin \left( \frac{z-z'}{2} \right)} \, .
\end{eqnarray*}
These are exactly the same propagators as in the introduction, with functions $\varphi$ and $\sigma$ that are now given by (\ref{eq:phi_sigma_6v}). We thus find one again that the field theory inside the arctic circle is the generic Dirac theory with action (\ref{eq:Dirac}), with a diagonal tetrad $e^\mu_a \, = \, e^{-\sigma} \delta_{a}^\mu$, and with gauge fields $A^{({\rm a})}_\mu \, = \,- \partial_\mu {\rm Re}\, \varphi$ and $A^{({\rm v})}_\mu \, = \,- i\partial_\mu {\rm Im} \,\varphi$. The boundary conditions for the fields $\psi$, $\psi^\dagger$, $\overline{\psi}$, $\overline{\psi}^\dagger$ follow again from $1/\sin\left( \frac{z-z'}{2} \right) = \pm 1/\sin\left( \frac{-z^*-z'}{2} \right)$ at the two boundaries ${\rm Re} \,z = 0$ and ${\rm Re} \, z = \pi$ repectively; see the more detailed discussion in section \ref{sec:dirac_honey}.

\subsection{Density profile}

Let us conclude this part on the six-vertex model with the density profiles, which we can now derive easily, using the above formulas. We start from the definition of the density operator, and from the above formula with the propagator,
\begin{eqnarray*}\fl
		\left< \rho_{x,y} \right> &= &  \, \left< c_{x,y}^\dagger c_{x,y} \right> \\\fl
		& = &  \int_{-\pi}^\pi \frac{dk}{2\pi} \int_{-\pi}^\pi \frac{dk'}{2\pi}  e^{- i k x + y \varepsilon(\kappa) - i N \tilde{\varepsilon}(\kappa) } e^{ i k' x - y \varepsilon(\kappa') +i N \tilde{\varepsilon}(\kappa') }  r_{x,y} (k) s_{x,y} (k')  \left( \frac{d\kappa}{dk}  \right)^{\frac{1}{2}}  \left( \frac{d\kappa'}{dk'}  \right)^{\frac{1}{2}} \frac{1}{2 i  \sin\left( \frac{\kappa-\kappa'}{2} \right)}  \, .
\end{eqnarray*}
Again, we want to evaluate the double-integral with the steepest descent method. There is a problem with the denominator, due to the fact that the stationary points are going to be exactly the same for $k$ and $k'$ (or equivalently $\kappa$ and $\kappa'$). Therefore, we apply the same trick as in part \ref{sec:honeycomb}: we treat $y$ as a continuous variable and differentiate with respect to $y$ inside the integral. That way, we obtain
\begin{eqnarray*}\fl
	\partial_y \left< \rho_{x,y} \right> & = &  \int_{-\pi}^\pi \frac{dk}{2\pi} \int_{-\pi}^\pi \frac{dk'}{2\pi}  e^{- i k x + y \varepsilon(\kappa) - i N \tilde{\varepsilon}(\kappa) } e^{ i k' x - y \varepsilon(\kappa') +i N \tilde{\varepsilon}(\kappa') }  r_{x,y} (k) s_{x',y} (k')  \left( \frac{d\kappa}{dk}  \right)^{\frac{1}{2}}  \left( \frac{d\kappa'}{dk'}  \right)^{\frac{1}{2}} \frac{\varepsilon(\kappa)-\varepsilon(\kappa')}{2 i  \sin\left( \frac{\kappa-\kappa'}{2} \right)}  \\\fl
	& \underset{N \rightarrow \infty}{\simeq} & \frac{e^{-\sigma}}{2\pi i} r_{x,y}(z) s_{x,y} (z) \,\frac{d \varepsilon}{d \kappa} \, + \,  \frac{e^{-\sigma}}{2\pi i} r_{x,y}(-z^*) s_{x,y} (-z^*) \, \frac{d \varepsilon}{d \kappa} (-z^*) \\
	& =  & \frac{e^{-\sigma}}{\pi} \, {\rm Im} \left[  r_{x,y}(z) s_{x,y} (z) \, \frac{d \varepsilon}{d \kappa} (z) \right] \, .
\end{eqnarray*}
Now let's focus first on the case $x - y \in 2\mathbb{Z}$, for which $r_{x,y} = s_{x,y} = \sqrt{1 - i \frac{b}{a} \sin \kappa}$. Then, using the above formulas for $e^\sigma$ and for $\frac{d\varepsilon}{d\kappa}$, we get
\begin{eqnarray*}
	\partial_y \left< \rho_{x,y} \right> & = &  \frac{1/\pi}{b \sqrt{\left( \frac{N}{c} \right)^2 - \left( \frac{x}{b} \right)^2 - \left( \frac{y}{a} \right)^2 }} \, {\rm Im} \left[  \left( 1-i \frac{b}{a} \sin z \right)  \frac{- b c \sin z}{(a-i b \sin z) (a+ i b \sin z)}  \right] \\
	&=&  \frac{-1/\pi}{\frac{a}{c} \sqrt{\left( \frac{N}{c} \right)^2 - \left( \frac{x}{b} \right)^2 - \left( \frac{y}{a} \right)^2 }} \, {\rm Im} \left[  \frac{ \sin z}{a+ i b \sin z}  \right]  \\
	&=&  \frac{-1/\pi}{\frac{a}{bc} \sqrt{\left( \frac{N}{c} \right)^2 - \left( \frac{x}{b} \right)^2 - \left( \frac{y}{a} \right)^2 }}   \frac{ \frac{x (x+y)}{b^2} - \frac{N^2}{c^2} }{ N^2 - (x+y)^2 } \, .
\end{eqnarray*}
\begin{figure}[htbp]

\includegraphics{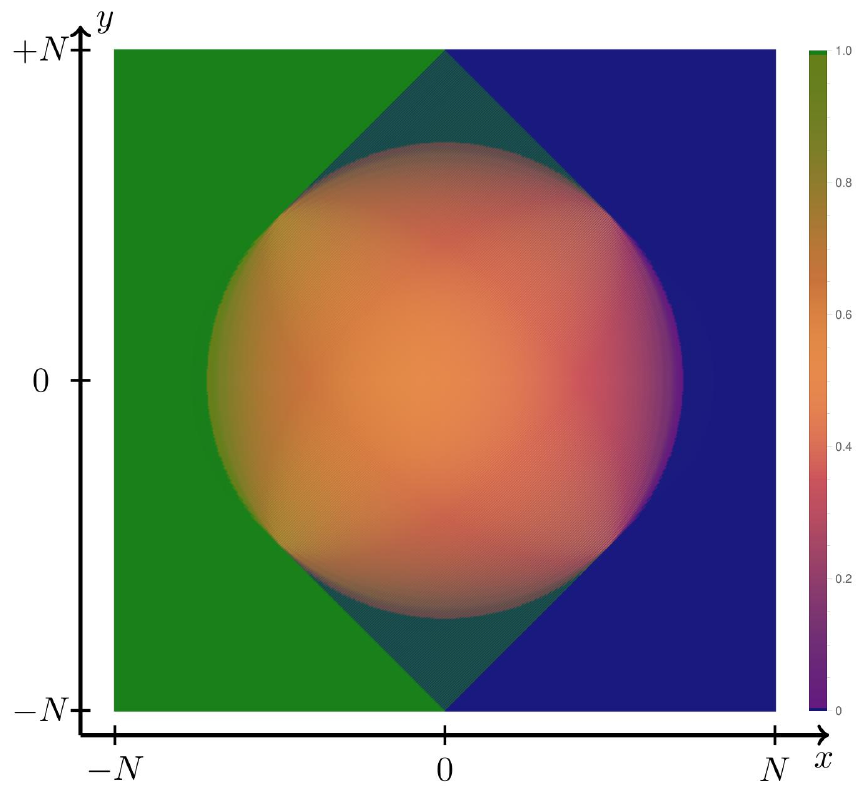}
\hfill
\includegraphics{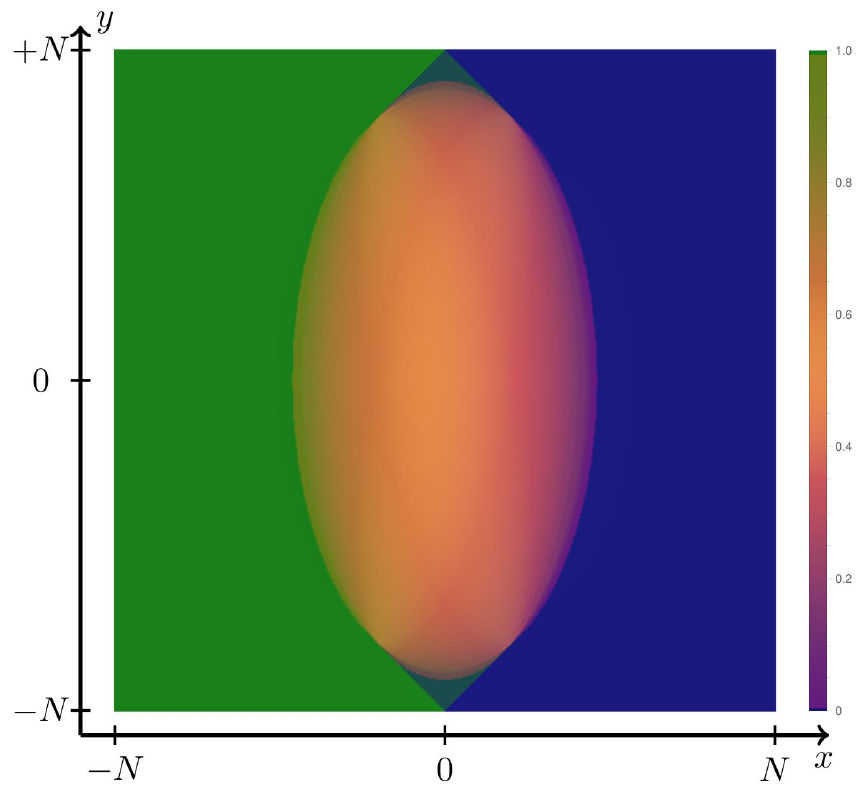}
 \caption{Left: Numerical density profile in a system of height $N=256$ at the free fermion point $\Delta \propto a^2+b^2- c^2 = 0$. \emph{Left}: $b/a=1$, the arctic curve is a circle. \emph{Right}: $b/a=1/2$, the arctic curve is an ellipse with excentricity $\sqrt{3}/2$. 
}
 \label{fig:densityplots6V}
\end{figure}

To go from the second to the last line, it is useful to notice that $\sin z \, = \, \frac{N}{|c|} \frac{  i \frac{c x y}{a b N}  -  \sqrt{  \left( \frac{N}{c} \right)^2 -  \left( \frac{x}{b} \right)^2 -  \left( \frac{y}{a} \right)^2  }   }{  \left(\frac{N}{c}\right)^2 - \left(\frac{y}{a}\right)^2   } $ and $\left|  a + i  b \sin z  \right|^2 \, = \, \frac{ N^2 - (x+y)^2 }{  \left(\frac{N}{c} \right)^2 - \left( \frac{y}{a} \right)^2 }$, and then use those two relations. This can be integrated; it gives
\begin{equation}
	\left< \rho_{x,y} \right> \, = \, \frac{1}{\pi} {\rm arccos} \left(  \frac{ a x/b - b y/a  }{\sqrt{ N^2 - (x+y)^2} }  \right) \, ,
\end{equation}
up to some additive function of $x$, which is argued to be zero by looking at the density along the arctic curve $\frac{x^2}{b^2}+\frac{y^2}{a^2}=\frac{N^2}{c^2}$, where it must be either one or vanish. The other case, namely $x-y \in 2\mathbb{Z}+1 $, is almost identical, up to a few sign changes. The results are summarized in table \ref{tab:density_6v}; they agree perfectly with our numerical checks displayed in Fig. \ref{fig:densityplots6V}.
\begin{table}[ht]
	\begin{tabular}{c||c|c}
		&  $x - y \in 2 \mathbb{Z}$   &   $x - y \in 2 \mathbb{Z}+1$  \\ 
		& (SW - NE  edges) &  (NW - SE  edges) \\ \hline \hline   & & \\
		$\left< \rho_{x,y} \right>$  &   $\displaystyle \quad  \frac{1}{\pi} {\rm arccos} \left(  \frac{ a x/b - b y/a  }{\sqrt{ N^2 - (x+y)^2} }  \right)  \quad $  &  $\displaystyle  \quad \frac{1}{\pi} {\rm arccos} \left(  \frac{ a x/b + b y/a  }{\sqrt{ N^2 - (x-y)^2} }  \right)  \quad$ \\ & & \\
	\end{tabular}
	\caption{Average occupation of the edges of the six-vertex model in the scaling limit. 
	}
	\label{tab:density_6v}
\end{table}

\newpage

\section{Height mapping and the gaussian free field}
\label{part:GFF}

In previous parts, we have made extensive use of fermionic degrees of freedom. The fact that all the above models can be mapped to free fermions is instrumental; it is the key that allows to compute correlators exactly on the lattice. Starting from the solutions of these models formulated in fermion language, it is not surprising that we end up with a rather generic free fermion field theory: the Dirac theory in curved two-dimensional euclidean space and in background gauge potentials. In this part we adopt a different point of view. As is well-known in the literature, it is possible to map dimer models and the six-vertex models onto {\it height models}, and to make the connection with bosonic free field theory, also called {\it gaussian free field} in the mathematical literature (see e.g.  \cite{sheffield2007gaussian,borodin2008anisotropic,petrov2015asymptotics,kuan2014gaussian,duits2015global}). In the physics literature, such methods are often called {\it Coulomb gas methods} (see Ref.~\cite{Nienhuis1987} for a review); those have played a very important role in the historical development of exact results and field theoretic methods in two-dimensional statistical mechanics. Therefore, we believe it is of some interest to reformulate some of the above results in the bosonic language, which may be more familiar to some readers. When formulated in bosonic language, we expect the continuous field theory describing the correlations inside the arctic circle to be of the form
\begin{equation}
	\label{eq:boson_curved}
	S \, = \, \frac{1}{8\pi} \int \sqrt{g} d^2 {\rm x} \,   g^{\mu \nu}  \, \partial_\mu  h \, \partial_\nu h \, ,
\end{equation}
Going from fermionic degrees of freedom to bosonic ones is a standard procedure in two-dimensional physics, called {\it bosonization} (the same procedure we used as a mathematical trick in part \ref{sec:ktr}). There are a few subtle points, however, that we find sufficiently non-standard to deserve further discussion. Among those, the fate of the background gauge potentials $A^{({\rm v})}$ and $A^{({\rm a})}$ and of the tetrad; those appear in the fermionic action, but are absent from the bosonic one. Thus, for completeness, this part focuses on the mapping to the height field, with some level of detail.

We shall illustrate the procedure with the dimer model on the honeycomb lattice. It is not difficult to adapt this discussion to deal with dimers on the square lattice or with the six-vertex model.

\subsection{From dimers to the height field: a quick review}

\begin{figure}[ht]
	$$
	\begin{array}{lll}
		\includegraphics[height=6cm]{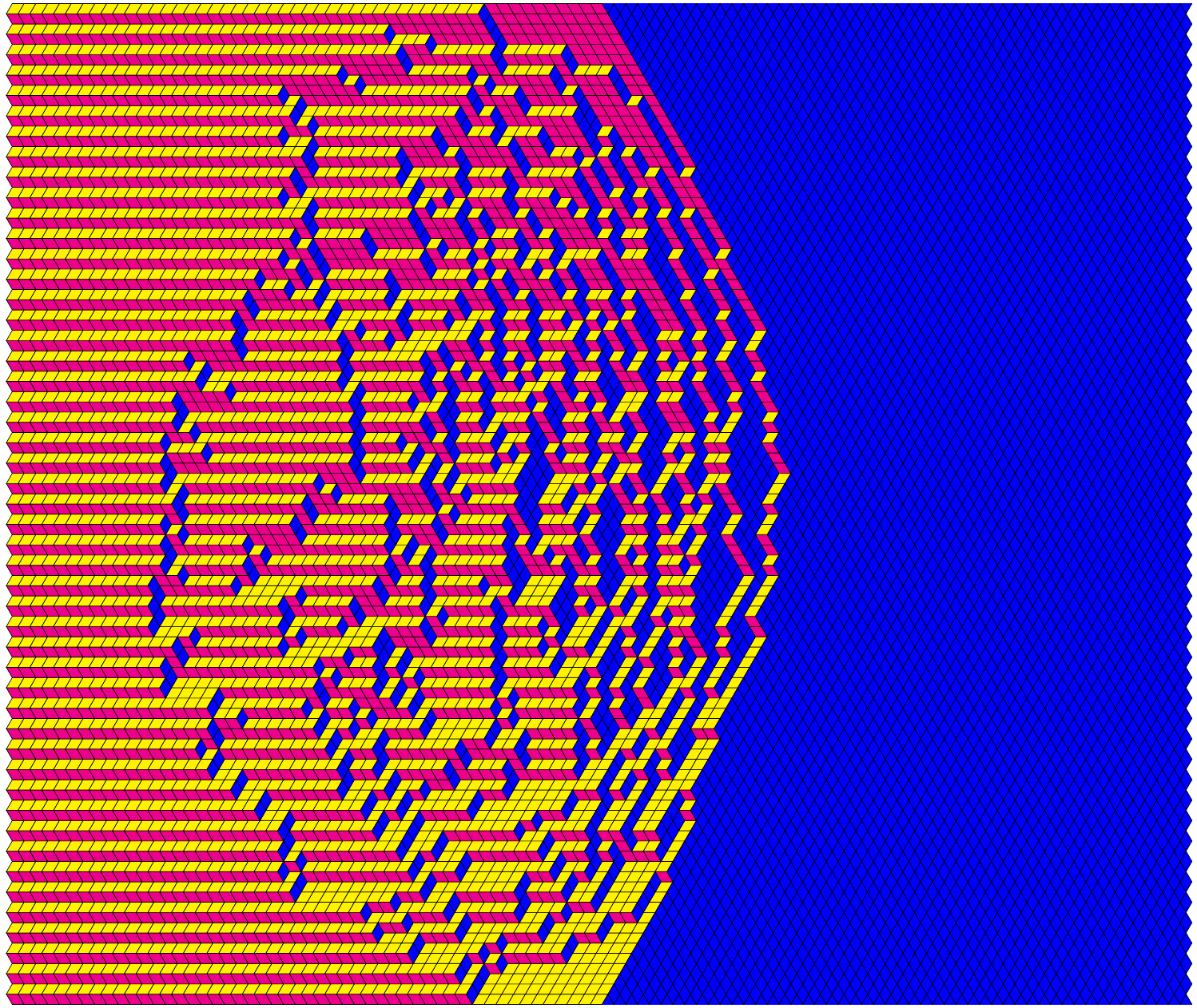} & \quad &  \includegraphics[height=6cm]{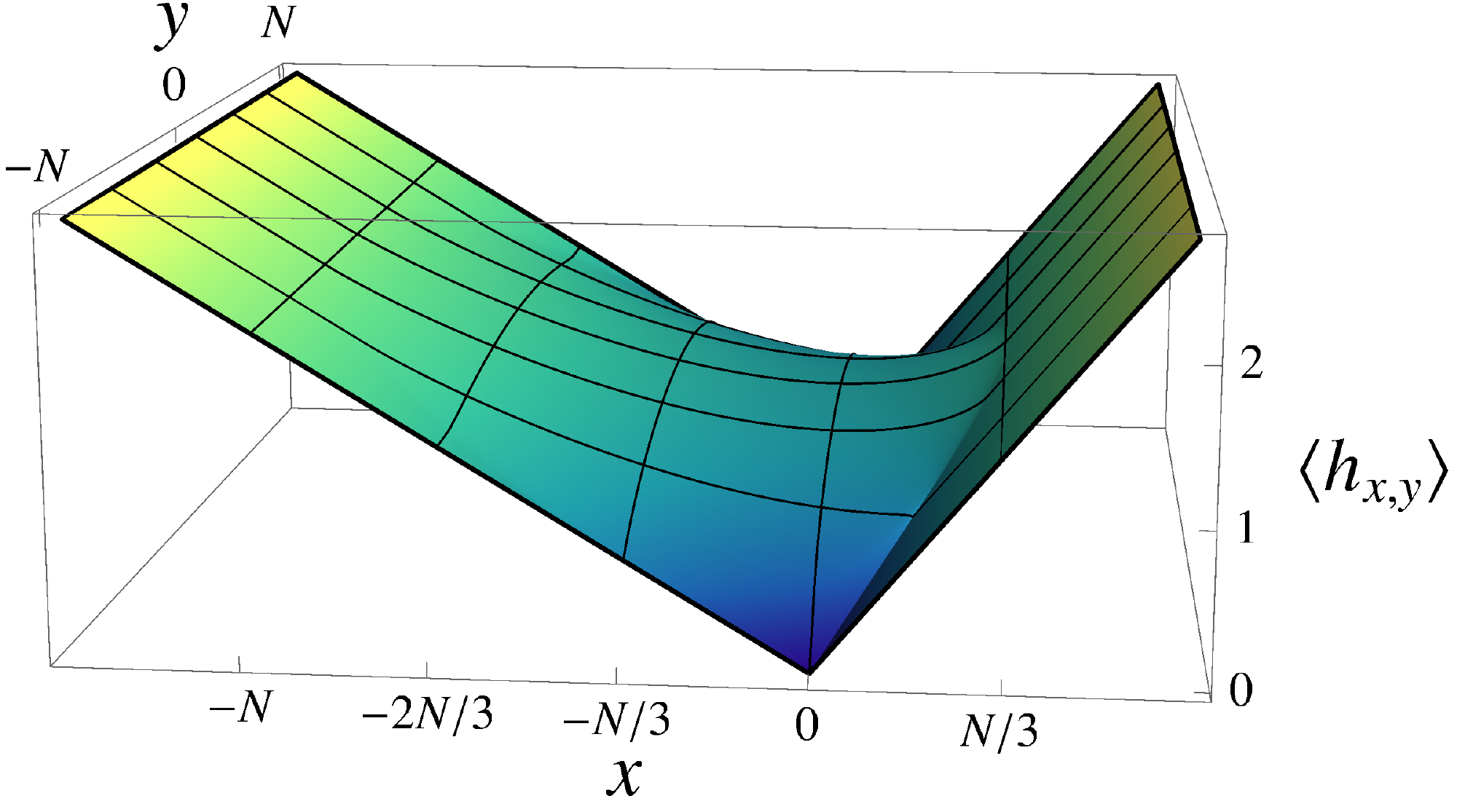}\\
		(a) && \quad(b)
	\end{array}
	$$
	\caption{(a) A typical configuration for a large width $2N=100$; the fluctuating region is an ellipse. (b) The average height, as given by Eq. (\ref{eq:average_height}), for $u=1/2$.}
	\label{fig:height_rombi2}
\end{figure}

Dimer configurations can be mapped on configurations of a discrete height field. The construction is well-known, see e.g. Refs.~\cite{blote1982roughening,nienhuis1984triangular}, so we shall go through it quickly.  
\begin{figure}[ht]
	$$
	\begin{array}{lll}
		\includegraphics[width=0.45\textwidth]{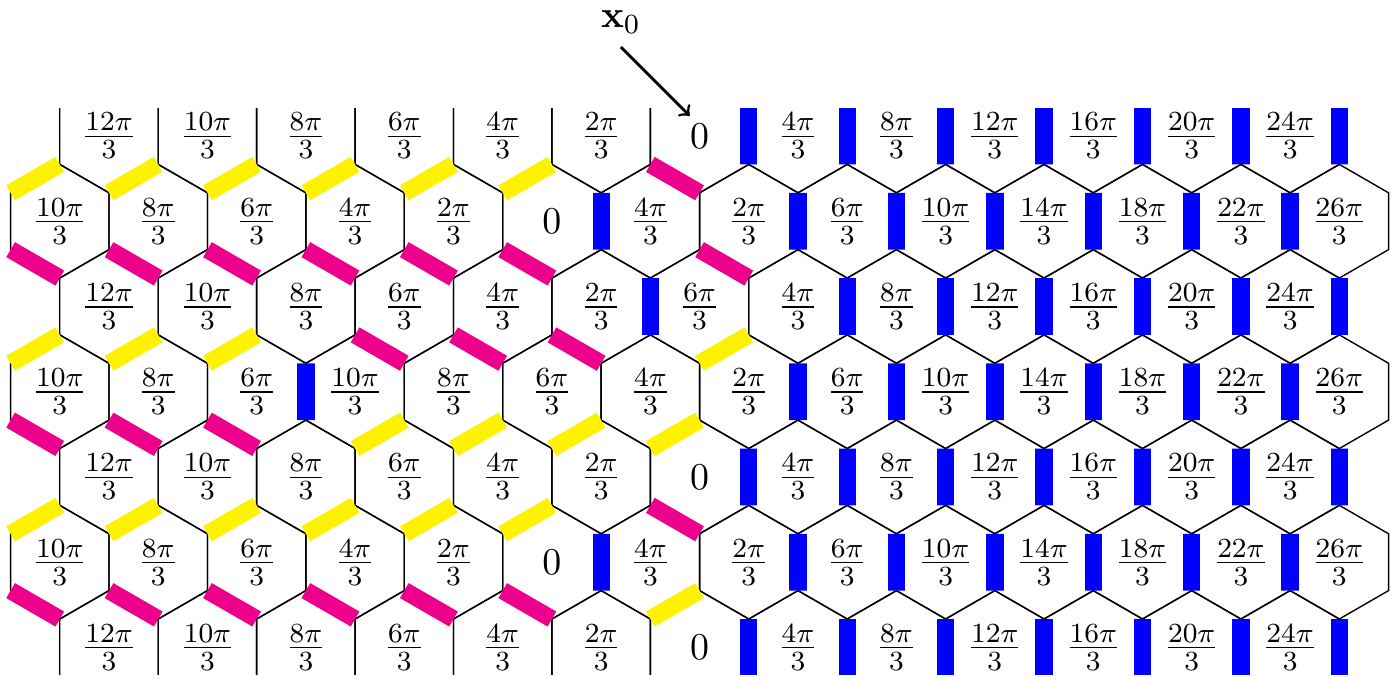} & \quad & \includegraphics[width=0.48\textwidth,height=3.3cm]{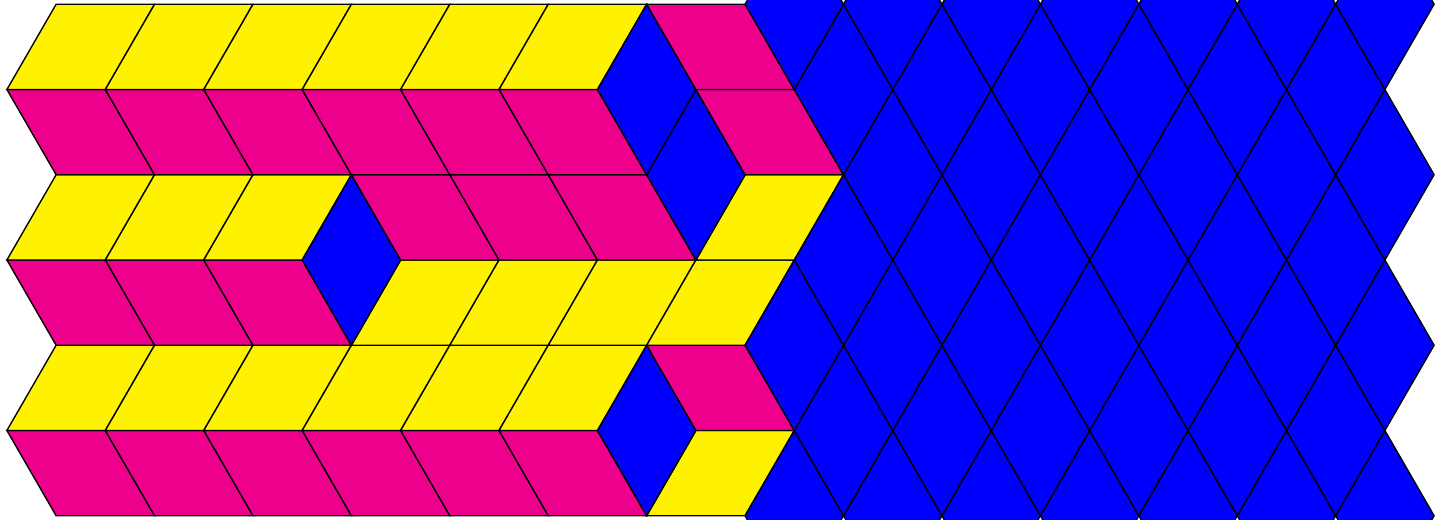} \\
		(a) && (b) \\ 
	\end{array}
	$$
	\caption{(a) A dimer configuration satisfying our boundary conditions, and the corresponding height configuration. ${\bf x}_0$ is the marked hexagon, where the height is set to $h_{{\bf x}_0} =0$. (b) Same configuration drawn with rhombi, which help visualize the height configuration.}
	\label{fig:height_rombi}
\end{figure}
Dimers live on the edges of the honeycomb lattice, while the height field lives on the faces. One needs to mark one of the faces, say at a chosen position ${\bf x}_0$, and decide that the height there is $h_{{\bf x}_0} =0$. Then the height is determined on all other faces by applying the following local rules. If two adjacent hexagons touch along a vertical edge, then, going through the edge in the East$\rightarrow$West direction, the height increases by $+\frac{4\pi}{3}$ if it is occupied by a dimer, and drops by $\frac{2\pi}{3}$ otherwise. The rule is the same for diagonal edges, going Northeast$\rightarrow$Southwest, and Southeast$\rightarrow$Northwest. See Fig. \ref{fig:height_rombi}.(a) for an illustration. The height field $h$ is related to the fermions as follows. Recall from Part \ref{sec:honeycomb} that a {\it fermion} on a vertical edge corresponds to the {\it absence of a dimer}. Thus the height operator is related to the fermion density through
\begin{equation*}
	h_{x+\frac{1}{2},y} - h_{x-\frac{1}{2},y}  \, = \,  \frac{4\pi}{3}(1- c^\dagger_{x,y} c_{x,y})  - \frac{2\pi}{3}   c^\dagger_{x,y} c_{x,y} \, .
\end{equation*}
The dimer configurations are frozen at infinity, so the height operator is a scalar as soon as $x$ is large enough: $h_{x,y} = \frac{4\pi}{3} x$ for any $x$ larger than some fixed $M$ (for instance, $M = N/2$ works, see Fig. \ref{fig:arctic_honey}). It follows that
\begin{eqnarray*}
	h_{x,y}   & = & h_{M,y} - \sum_{n=0}^{M-x-1} \left(  h_{x+n+1,y}  - h_{x+n,y}  \right)  \, = \,  \frac{4\pi}{3} x +  2\pi \sum_{n=0}^{M-x-1}  c^\dagger_{x+\frac{1}{2}+n,y} c_{x+\frac{1}{2}+n,y}  \, . 
\end{eqnarray*}
The average height $\left< h \right>$ can then be evaluated using the results of Part \ref{sec:honeycomb},
\begin{eqnarray}\fl
	\label{eq:average_height}
\nonumber	\left< h_{x,y} \right> 	& = &  \frac{4\pi}{3} x +  2\pi \sum_{n=0}^\infty  \left< \rho_{x+\frac{1}{2}+n,y} \right>  \\\fl
\nonumber	& \simeq &  \frac{4\pi}{3} x +  2\pi \int^\infty_x dx'  \left< \rho_{x',y} \right> \\\fl
& = & \left\{  \begin{array}{lll}
		\frac{4\pi }{3} x	 - 2 \,{\rm Re} \,\varphi (x,y)  &&	 {\rm if} \quad X^2 + y^2 \leq N^2 \, ,	\\
		-\frac{2\pi }{3} x  && {\rm if} \quad X  \leq - \sqrt{ N^2- y^2}  \, , \\
		-\frac{2\pi }{3} x + \pi (N - |y|)  && {\rm if} \quad X  \geq  \sqrt{ N^2- y^2}  \quad {\rm and}\quad  2x \leq N - |y| \quad {\rm and} \quad \frac{|y|}{N} \geq \frac{1-u^2}{1+u^2} \, ,	\\
		\frac{4\pi }{3} x  && {\rm otherwise} \, .
	\end{array}   \right.	
\end{eqnarray}
where the function ${\rm Re}\, \varphi(x,y)$ is the one given by Eq. (\ref{eq:Rephi_honeycomb}). This height profile is shown in Fig.~\ref{fig:height_rombi2}(b), together with a typical dimer configuration for a large system (Fig.~\ref{fig:height_rombi2}(a)).
\subsection{Large-scale fluctuations of the height field}

It is customary to decompose the height field $h$ into its classical part $\left< h \right>$, given by (\ref{eq:average_height}), and its zero-average fluctuating part $\tilde{h}$:
\begin{eqnarray*}
	h_{x,y} & = & \left<  h_{x,y} \right> \, + \, \tilde{h}_{x,y}  \, .
\end{eqnarray*}
Inside the fluctuating region, we know that we can make the replacements $c^\dagger_{x,y} = \frac{1}{\sqrt{2\pi}} (\psi^\dagger(x,y) + \overline{\psi}^\dagger(x,y))$ and $c_{x,y} = \frac{1}{\sqrt{2\pi}} (\psi(x,y) + \overline{\psi}(x,y))$; thus $\tilde{h}_{x,y}$ is a primitive integral of the form
\begin{eqnarray}
	\label{eq:tildeh}
\nonumber	\tilde{h}_{x,y} &\simeq & -\int^x dx' : [\psi^\dagger(x',y) + \overline{\psi}^\dagger(x',y)] [\psi(x',y) + \overline{\psi}(x',y)] : \\
	&\simeq & -\int^x dx' \left(  : \psi^\dagger(x',y)  \psi(x',y) : \, + \, : \overline{\psi}^\dagger(x',y)  \overline{\psi}(x',y)  : \right) \, .
\end{eqnarray}
Here, normal ordering is defined such that $\left< : \psi^\dagger_{x,y} \psi_{x',y'}: \right> \, = \, 0$. Going from the first to the second line, we have neglected the cross-terms, which are oscillating fast because of the chiral background gauge potential $A^{({\rm a})}$. Now, a crucial observation, that follows from the relation between $\tilde{h}$ and $:\psi^\dagger \psi:$ and $:\overline{\psi}^\dagger\overline{\psi}:$, is that the field $\tilde{h}_{x,y}$ is gaussian. Namely, all its connected correlators vanish:
\begin{equation}
	\left< \tilde{h}_{x_1,y_1}\tilde{h}_{x_2,y_2} \dots \tilde{h}_{x_n,y_n} \right>_{\rm conn.} \, = \, 0 \, , \quad \qquad {\rm if} \quad n > 2\, .
\end{equation}
This follows immediately from the fact that the connected correlators $\left< :\psi^\dagger_{x_1,y_1} \psi_{x_1,y_1}: \dots :\psi^\dagger_{x_n,y_n} \psi_{x_n,y_n}: \right>_{\rm conn.}$ themselves vanish if $n>2$. [One way to prove this is to observe that the only singular term in the operator product expansion of $:\psi^\dagger_{x_i,y_i} \psi_{x_i,y_i}:$ with $:\psi^\dagger_{x_j,y_j} \psi_{x_j,y_j}:$ is a scalar, so its connected correlation with any other terms (present if $n>2$) is zero.] The fluctuations of the height field around its mean value
$\left<  h_{x,y} \right>$ are therefore the ones of a gaussian field theory with the action (\ref{eq:boson_curved}); the only thing we need to do is to compute its propagator. In the process we expect to learn why the background gauge potential $A^{({\rm v})}$ and the tetrads are absent from the action. The derivative of the propagator is, locally,  
\begin{eqnarray*}
	\partial_x  \partial_{x'} \left< \tilde{h}_{x,y} \,  \tilde{h}_{x',y'}  \right> & \simeq &  \left< \left( : \psi^\dagger_{x,y} \psi_{x,y} : + : \overline{\psi}^\dagger_{x,y} \overline{\psi}_{x,y} :\right)  \left( : \psi^\dagger_{x',y'} \psi_{x',y'} : + : \overline{\psi}^\dagger_{x',y'} \overline{\psi}_{x',y'} :\right) \right>  \\
	&= &  - \left<:\psi^\dagger_{x,y}\psi_{x',y'}:\right>\left<:\psi^\dagger_{x',y'}\psi_{x,y}:\right>   \, + \, {\rm c.c.} \\
	& =&  -  \,\frac{e^{-\sigma - i \theta  - \sigma' - i \theta'}}{(z-z')^2}  \,   + \, {\rm c.c,}  
\end{eqnarray*}
where recall that $z=z(x,y)$ is given by Eq.~(\ref{eq:zhoney}). By locally we mean distances large compared to the lattice spacing ($|x-x'|$, $|y-y'|\gg 1$) but small compared to the full system ($|x-x'|$, $|y-y'|\ll N$). 
Using the fact that $e^{\sigma + i \theta} \partial_x z(x,y) \, = \, -1$, the previous equation can be easily integrated, leading to
\begin{eqnarray}\label{eq:proplog}
	\left< \tilde{h}_{x,y} \,  \tilde{h}_{x',y'}  \right> \, = \,  - \log |z-z'|^2 \qquad,\qquad 1\ll |z-z'|\ll N
\end{eqnarray}
up to some unimportant additive constant. This result is fully consistent with the action (\ref{eq:boson_curved}), with $h$ replaced by $\tilde{h}$. 
Indeed, written in the coordinate system provided by $z(x,y)$, the propagator for the action (\ref{eq:boson_curved}) is the Green's function for the following operator
\begin{equation}
 \partial_z \partial_{\bar{z}} \Braket{\tilde{h}(z,\bar{z})\tilde{h}(z',\bar{z}')}\;=\;4\pi\, \delta^{(2)}(z-z'),
\end{equation}
whose solution is exactly (\ref{eq:proplog}). Note also that 
$\tilde{h}_{x,y}$ is clearly gauge invariant, so the background gauge potential $A^{({\rm v})}$ cannot play any role. What about the tetrad ? Since $\tilde{h}_{x,y}$ is a scalar ({\it i.e.} spinless) field, it should not play any role either; we see that it has dropped from the propagator (\ref{eq:proplog}). 
In summary, the long distance correlations in the fluctuation region are fully described by a gaussian action in curved space, Eq.~(\ref{eq:boson_curved}). The tetrad as well as the vector gauge potential $A_\mu^{({\rm v})}$ essentially disappear from the bosonic formalism, as they should. However the axial gauge potential still plays a crucial role, as it fixes the classical part of the field $\braket{h}$, the average height (\ref{eq:average_height}).

\subsection{Probable connection with other works}
In other works (see e.g. Refs.~\cite{kenyon_okounkov,kenyon2006dimers,kenyon2009lectures}), the arctic curve is obtained by minimizing a functional of the form
\begin{equation}
	\mathcal{A} \, = \, \int_{{\rm strip}} dx dy \,  \sigma \left( \partial_x h , \partial_y h \right) \, ,
\end{equation}
where $\sigma$ is the so-called {\it surface tension}. It is the free energy per unit area of the homogeneous dimer model with a fixed average slope $(\partial_x h , \partial_y h)$. In that formalism, the average height configuration is the one that minimizes the functional $\mathcal{A}$. The fluctuations around that minimal configuration are given, at the leading order, by the hessian of $\sigma$. Thus, we expect that applying this formalism to our setting (the particular dimer model above on the strip) should lead to the formulas (\ref{eq:average_height}) and (\ref{eq:proplog}) above. Such a conjecture was also made in Ref.~\cite{complex_burgers}. Unfortunately, as far as we know this result is not directly available in the literature. We thus lead this particular exercise for future work.

\newpage
\section{Conclusion}

In summary: motivated by the quantum quench studied by Antal, R\'acz, R\'acoz and Sch\"utz in \cite{antal1999transport} and by the imaginary-time setup of Calabrese and Cardy \cite{calabrese2006time}, we studied the XX chain evolving from a domain-wall initial state (DWIS) in imaginary time. As it turns
out that this problem is naturally related to arctic circle problems, we revisited a number of those from the perspective of 1d quantum systems evolving with a transfer matrix.
Our main message is that the long-range correlations inside the fluctuating region are described by a massless field theory with a local action; since the system is inhomogeneous, the few parameters that enter this action must vary with position. In particular, the components of the metric themselves vary, so the field theory is, generically, a field theory in curved space-time.

All the models that we investigated in this paper could be mapped on free fermions. Obviously, this is a very strong restriction, and it is natural to ask whether our message is supposed to generalize to interacting problems. For instance, it would be very interesting to see what the action of the field theory becomes when one replaces the XX chain in the introduction by the XXZ chain (say in the critical regime, with $ -1 < \Delta \leq 1$) \cite{lancaster2010quantum,mossel2010relaxation,sabetta2013nonequilibrium}. There, the natural conjecture is that the continuous theory is a gaussian free field (like the one in part \ref{part:GFF}), but this time with a stiffness that depends on position as well. The reason is that, as is well-known, the density fluctuations of one-dimensional quantum systems with conserved particle number can always be described by a scalar field theory which is gaussian up to irrelevant perturbations. These systems are known as 'Luttinger liquids' (see \cite{senechal2004introduction} for a review), and they are all described by a free boson action, with (at least) two free parameters: the velocity $v$ of the low-energy excitations, and the Luttinger parameter $K$, related to the stiffness of the gaussian free field. The variation of the velocity with position in space-time is, in the language of this paper, equivalent to the variation of the metric: $ds^2 \sim  v(x,t)^2 dt^2 - dx^2$. What is new compared to free fermion systems is the variation of $K$ with position in space-time. Importantly, since the Luttinger liquid is only an effective theory in which the continuous free fields correspond to some renormalized version of the lattice operators, we should expect that the relation between the lattice degrees of freedom and the continuous fields should vary with positions as well, in contrast with (\ref{eq:lattice_continuous_XX}) and other similar relations in this paper. We hope to come back to these questions in future papers. Of course, there already are some known results on arctic circle problems in interacting systems, such as the six-vertex model at $\Delta \neq 0$.
For instance, the partition function is known thanks to the work of Korepin and Zinn-Justin \cite{korepin_zinnjustin} and the equation that determines the arctic curve itself was conjectured by Colomo and Pronko in \cite{colomo2010arctic}. To our knowledge, however, not much is known about the long-range correlations inside the fluctuating region, and we hope that the present paper, together with the perspective of possible applications to quantum quenches, motivates further work in that direction.

\begin{figure}[ht]\centering
	\includegraphics[width=0.6\textwidth]{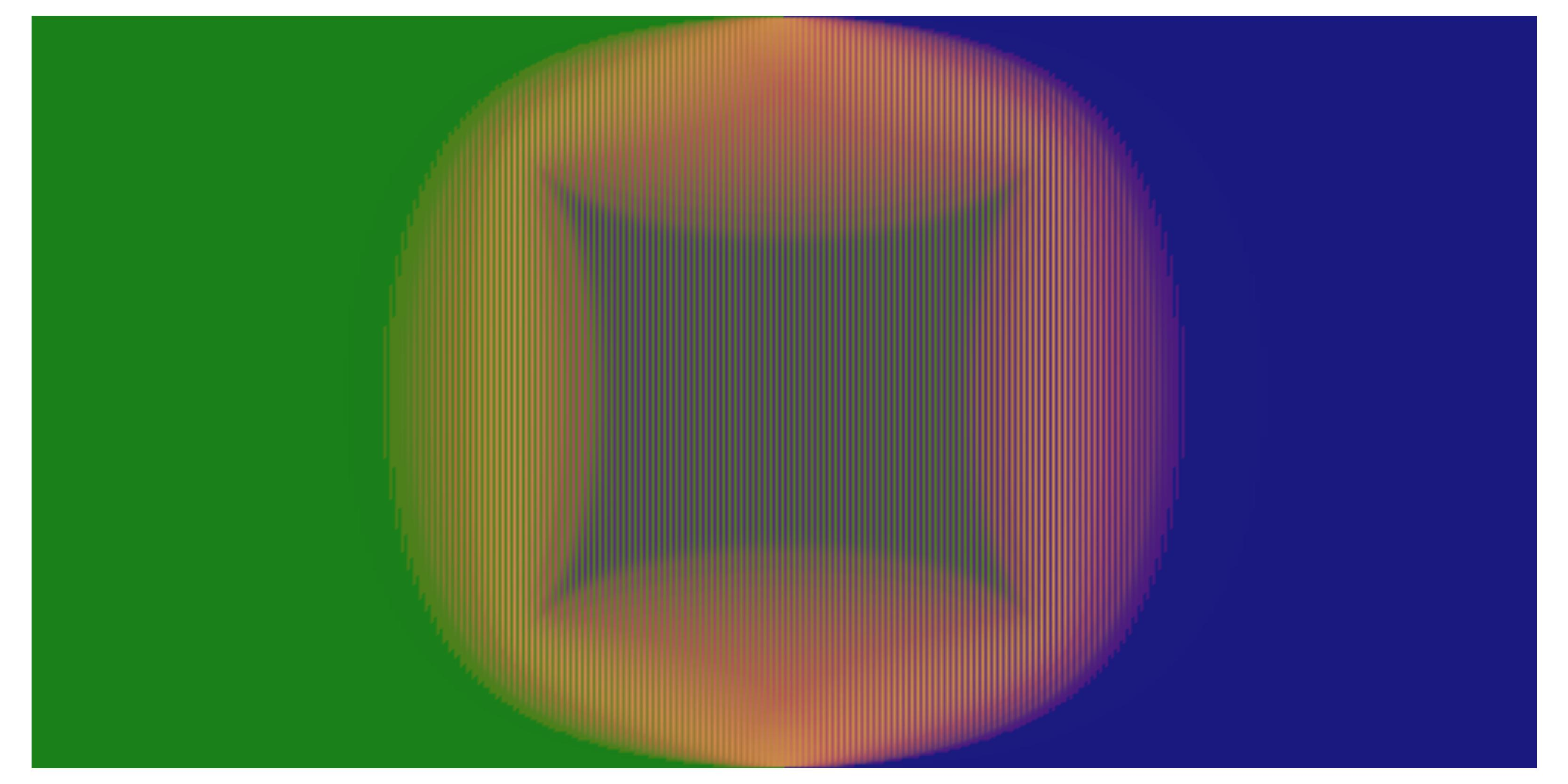}
	\caption{Typical density profile in imaginary time for a free fermion chain with increased lattice periodicity ($N=512$ sites, and $R = 128$; the conventions are the same as in the introduction). Here we added a staggered chemical potential term $\sum_x (-1)^x c^\dagger_x c_x$ to the XX Hamiltonian. This model is gapped at half-filling. The density profile now contains more phases: the central region is half-filled, and is gapped. The colored region around it is critical, and the green and blue regions on the sides are again completely frozen.}
	\label{fig:staggered}
\end{figure}

Another possible extension of the present paper, not quite as ambitious as trading the XX chain for the XXZ one, is to look at free fermion models with more bands and with energy gaps. This is  achieved by increasing the size of the unit cell in the model, namely by breaking the symmetry group of translations $\mathbb{Z}$ to a subgroup $n \mathbb{Z}$. For instance, one could replace the XX chain (which has a single band) by the so-called Su-Schrieffer-Heeger model (with two bands separated by a gap) \cite{su1979solitons}, or by a model with a staggered chemical potential term. The ground state of either model, at half-filling, is gapped, so it has short-range correlations. On the other hand, at other filling fractions, it is gapless. So we expect that, evolving in imaginary time from the DWIS, we should see a more complex density profile in space-time, with a central region that is massive, some intermediate region that is critical, and frozen regions outside. Some of our preliminary numerical results are displayed in Fig. \ref{fig:staggered}. This model is a particularly simple realization of a situation with three phases, a phenomenon already studied by \cite{nienhuis1984triangular, kenyon_okounkov}, and also revisited recently in \cite{chhita2014domino, di2014arctic}. This phenomenon appears to be quite generic when one starts from a quantum 1d system that is gapped; this is the case for free fermion cases, as the previous references illustrate, but also when the gap is due to interactions \cite{allison2005numerical} rather than to broken translation symmetry. 
Another interesting project would be to better connect our work to semiclassical methods which can be used \cite{Abanov} to determine arctic curves and density profiles.
We plan to come back to these questions elsewhere.

\ack

We thank Pasquale Calabrese for discussions in 2012 that motivated this paper,
and Leonid Glazman for discussions about Ref.~\cite{bettelheim2012quantum} which
inspired the setup of part \ref{sec:ktr}. We are grateful to J. Bouttier, F. Colomo, S. Klevtsov, V. Korepin, P. Krapivsky, N. Reshetikhin, G. Sch\"utz, H. Spohn and A. Sportiello for their encouragements to publish these results, and for providing guidance through the literature on the arctic circle. We also thank M. Haque and G. Misguich for stimulating discussions about real time dynamics. We acknowledge hospitality and financial support from the GGI in Florence in 2012 (JD, JV) and in 2015 (NA, JMS, JV), and from the Visitors Program of the MPIPKS Dresden (JD, 2014). Part of this work was supported by the Mission Interdisciplinaire du CNRS (JD) and by the Conseil R\'egional de Lorraine (JD).

\appendix

\newpage
\section[ \hspace{2cm}Analytic continuation and quantum quenches]{Analytic continuation and quantum quenches}
\label{appx:real}

We discuss here the analytic continuation to real time of our main results for the XX chain. As emphasized in the introduction, the connection between imaginary and real-time dynamics is one of the motivations behind the present paper. We explain how this can be done, and check that one gets results consistent with previous studies \cite{antal1999transport,antal2008,real_time}. This means our results can be used to determine the real time dynamics as a biproduct. We mainly focus on partition functions and correlators. It is our hope that the above connection can be used to make non trivial predictions about more complicated quantities such as particle fluctuations or entanglement entropies; this will be discussed elsewhere, however.

The present appendix also serves as an opportunity to discuss a well known \cite{Abanov,antal2008,BettelheimWiegmann,real_time} heuristic method to obtain results about the real time dynamics, called semi-classics. Here we will go a bit further that that, and try to reverse the logic. Is it possible to interpret imaginary time correlations using semiclassical arguments ? We will see that this can be done in the XX chain.

The section is organized as follows. In \ref{sec:real_asymptotic} we check the analytic continuation of partition functions, asymptotic forms of simple correlations such as the density profile, as well as the metric of the underlying Dirac action. We then explain the logic behind (real-time) semiclassical methods in \ref{sec:real_semiclassics}, and extend those to determine the analogous quantities in imaginary time. Finally, we check in \ref{sec:real_correlations} that the analytic continuation may also be consistently performed at the level of exact finite-size correlators.

\subsection[ \hspace{2cm}Partition functions, asymptotics of the correlation functions and the metric]{Partition functions, asymptotics of the correlation functions and the metric}
\label{sec:real_asymptotic} 
Let us start with the simplest quantity one can compute, the partition function $Z(R)=\braket{\Psi_0|e^{-2RH}|\Psi_0}=e^{R^2/2}$ (the proof of this formula may be found in appendix.~\ref{sec:partitionsfunctions}). Setting $R=it/2$ in the previous formula, one recovers the correct return probability $R(t)=\left|\braket{\Psi(0)|\Psi(t)}\right|=e^{-t^2/8}$ after the quench \cite{real_time}. Another simple physical quantity is the density profile in the slab. Recall that it is given by the formula
\begin{equation}\label{eq:xxdensity}
\braket{\rho_{x,y}}=\frac{1}{\pi}{\rm arccos} \left(\frac{x}{\sqrt{R^2-y^2}}\right)\qquad,\qquad x^2+y^2\leq R^2,
\end{equation}
inside the fluctuating region in the scaling limit. To obtain the real time density profile, it is necessary to first replace $y$ by $it$, and then take the limit $R\to 0^+$. We obtain
\begin{equation}\label{eq:exactdensityt}
 \rho_x(t)=\frac{1}{\pi}{\rm arccos} \left(\frac{x}{t}\right)\qquad,\qquad x\leq t,
\end{equation}
which is the result of Ref.~\cite{antal1999transport}. Here by scaling limit we mean large $t$ and $x$, but with $x/t$ finite. More generally, one can easily check that all equal (real) time correlators can be recovered from their imaginary time counterpart. Said differently,
\begin{equation}
 \Braket{c_x^\dag(t) c_{x'}(t)}_{\rm real}\;=\;\lim_{R\to 0^+}\,\left.\Braket{c_{x,y}^\dag c_{x',y}}\right|_{y=it}\,.
\end{equation}
Let us finally comment on the metric. We have seen that it is given in imaginary time by
\begin{eqnarray}
 ds^2&=&e^{2\sigma} dz d\bar{z}\\
 &=&\left(dx+\frac{xy}{R^2-y^2}dy\right)^2+\frac{R^2\left(R^2-x^2-y^2\right) }{\left(R^2-y^2\right)^2}dy^2.
\end{eqnarray}
Now performing the analytic continuation as explained above, we obtain
\begin{equation}
 ds^2\,=\,\left(dx-v(x,t)dt\right)^2,
\end{equation}
where $v(x,t)=x/t$ is the speed of propagation.

\subsection[ \hspace{2cm}Semi-classical interpretation]{Semi-classical interpretation}
\label{sec:real_semiclassics}
\begin{figure}[htbp]
\begin{tikzpicture}[scale=1.04]
 \begin{scope}[xshift=0cm,yscale=0.6,xscale=0.7,opacity=1]
\fill[fill=gray!50] (-5,-pi)--plot[domain=-pi:pi, samples=500] ({0*sin(\x r)}, \x)
--(-5,pi)--cycle;
\draw[semithick](-5,pi)--(5,pi);
\draw[semithick](-5,-pi)--(5,-pi);
\draw[->,semithick](0,-pi-0.5)--(0,pi+0.5);
\draw[->,semithick](-5-0.5,0)--(5+0.2,0);
\draw(5+0.2,0) node[right] {\large{$x$}};
\draw(0.3, 3.9*pi/3) node[left] {\large{$k$}};
\draw(-5,pi) node[left]{\large{$\pi$}};
\draw(-5,-pi) node[left]{\large{$-\pi$}};
\draw (2.5,-pi-0.9) node {\large{$\braket{\rho}=0$}};
\draw (-2.5,-pi-0.9) node {\large{$\braket{\rho}=1$}};
\draw (0,-5.5) node {\large{$t=0$}};
\end{scope}
 \begin{scope}[xshift=9cm,yscale=0.6,xscale=0.7,opacity=1]
\fill[fill=gray!50] (-5,-pi)--plot[domain=-pi:pi, samples=500] ({3*sin(\x r)}, \x)--(-5,pi)--cycle;
\draw[ultra thick,blue] plot[domain=-pi:pi, samples=500] ({3*sin(\x r)}, \x);
\draw[dashed, red,very thick](3,-pi-1.2)--(3,pi+0.5);
\draw[dashed, red,very thick](-3,-pi-1.2)--(-3,pi+0.5);
\draw[semithick](-5,pi)--(5,pi);
\draw[semithick](-5,-pi)--(5,-pi);
\draw[->,semithick](0,-pi-0.5)--(0,pi+0.5);
\draw[->,semithick](-5-0.5,0)--(5+0.2,0);
\begin{scope}[yshift=0.3cm]
\draw (0, -pi-1.2) node {\large{Inhomogeneous region}};
\draw (4.15, -pi-1.2) node {\large{$\braket{\rho}=0$}};
\draw (-4.15, -pi-1.2) node {\large{$\braket{\rho}=1$}};
\end{scope}
\draw(5+0.2,0) node[right] {\large{$x$}};
\draw(0.3, 3.9*pi/3) node[left] {\large{$k$}};
\draw(-5,pi) node[left]{\large{$\pi$}};
\draw(-5,-pi) node[left]{\large{$-\pi$}};
\draw (0,-5.5) node {\large{$t>0$}};
\draw[color=blue] (2,-1.5) node {\Large{$t v(k)$}};
\draw[color=blue,->,thick] (1.9,-1) -- (1.5,0.35); 
\end{scope}
\end{tikzpicture}
\caption{Semiclassical picture for the real time dynamics. Left: Domain wall initial state. The ``probability density'' (Wigner function $W(x,k)$) to find a fermion at position $x$ and momentum $k$ is uniform equal to one for $x<0$ and zero for $x>0$. Right: same picture at time $t>0$, assuming that each ``fermion'' at momentum $k$ moves freely with speed $v(k)=\sin k$.}
\label{fig:semiclassics}
\end{figure}
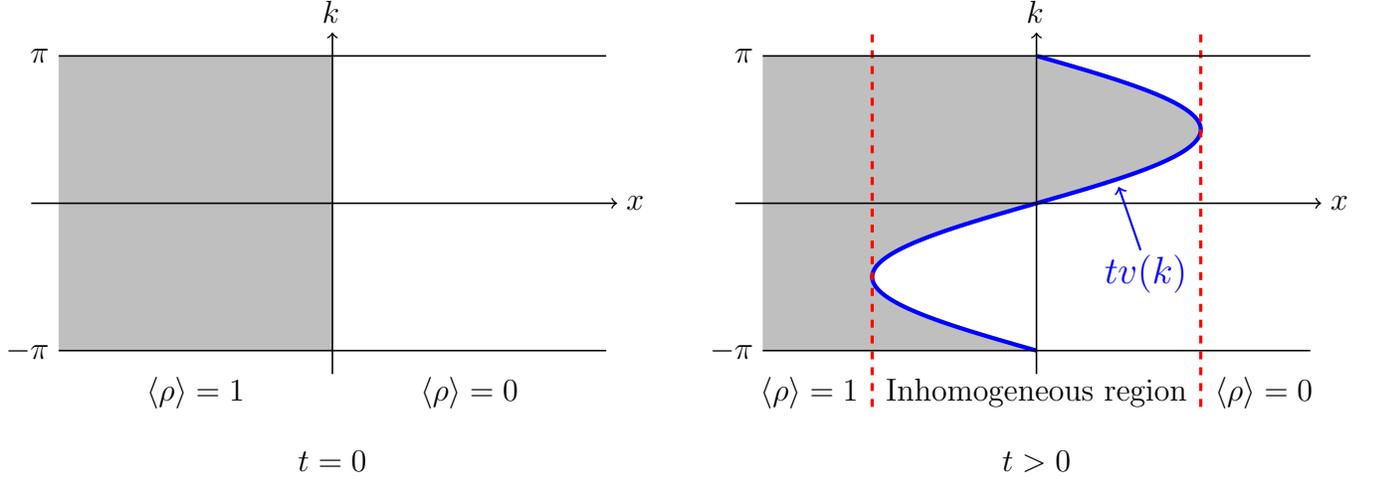

The XX chain has dispersion relation $\varepsilon(k)=-\cos k$, and group velocity $v(k)=\frac{d\varepsilon(k)}{dk}=\sin k$. However in this subsection, we will try to keep the notations $\varepsilon(k)$ and $v(k)$ for as long as possible, as it makes the physical explanations much more transparent. 
One particularly nice feature of the real time results is that they can be interpreted using a semiclassical picture, as pointed out in \cite{antal2008}, and studied in detail in \cite{real_time}. The key idea behind such methods is to assign to each ``particle'' \emph{both} a well-defined position $x$ and a well-defined momentum $k\in [-\pi,\pi]$. Doing that violates a basic principle of quantum mechanics, the Heisenberg's uncertainty relations. Therefore, such an assumption can only be approximate; it is often called semiclassical (or hydrodynamic) approximation in the physics literature. To see how this can be done, let us write the occupation number of a single $k$ mode in real space. We have
\begin{eqnarray}
 \braket{c^\dag (k) c(k)}=\sum_{j \in \mathbb{Z}} \sum_{j'\in \mathbb{Z}} e^{ik(j-j')}\braket{c_j^\dag c_{j'}}.
\end{eqnarray}
[In the following we use different conventions than in the main text, and label the sites by integers. The DWIS becomes $\ket{\Psi_0}=\prod_{x\leq 0}c_x^\dag \ket{0}$.]
The idea is to rewrite this as sums of terms centered at positions $x=\frac{j+j'}{2}$ (or $x=\frac{j+j'+1}{2}$, depending on the parity of $j+j'$). Writing $j=x+y$ and $j'=x-y$, we obtain
\begin{eqnarray}
 \braket{c^\dag (k) c(k)}&=&\sum_{x\in \mathbb{Z}}\left(\sum_{y \in \mathbb{Z}} e^{ik 2y}\braket{c^\dag_{x+y}c_{x-y}}+e^{ik (2y+1)}\braket{c^\dag_{x+y+1}c_{x-y}}\right)\\\label{eq:wigner}
 &\equiv& \sum_{x\in \mathbb{Z}} W(x,k).
\end{eqnarray}
The function $W(x,k)$ defined by Eq.~(\ref{eq:wigner}) is a discrete analog of the \emph{Wigner function}; in the following we will also use that name for the discrete case. $W(x,k)$ is, roughly speaking, the probability of finding a fermion at position $x$ and momentum $k$ in the phase space. All correlation functions can be, in principle, expressed in terms of $W$. For example, the mean density is $\braket{\rho_x}=\int_{-\pi}^{\pi} \frac{dk}{2\pi}W(x,k)$. For our quench protocol we want to know how $W(x,k,t)$ evolves in time. 
In the domain wall initial state we have
\begin{equation}
 W(x,k,t=0)=\Theta(-x),
\end{equation}
where we use the slightly unusual convention $\Theta(x\geq 0)=1$ and $\Theta(x<0)=0$, for the Heaviside theta function. 
This yields $\braket{\rho_x}=\Theta(-x)$, as expected. 
We then assume that each particle at momentum $k$ propagates ballistically (classically) with group velocity $v(k)=\frac{d\varepsilon(k)}{dk}$. This means the Wigner function at time $t$ is simply given by
\begin{equation}
 W(x,k,t)=W(x-v(k)t,k,0)=\Theta(-x+v(k)t).
\end{equation}
The density profile is then $\braket{\rho_x(t)}=\int_{-\pi}^{\pi} \frac{dk}{2\pi}\Theta(-x+v(k)t)$, which reproduces Eq.~(\ref{eq:exactdensityt}) exactly for dispersion relation $\varepsilon(k)=-\cos k$. Said differently, semiclassics is \emph{exact} in the scaling limit $x,t\to \infty$ with $x/t$ fixed. For this particular quench, such an observation can be proved by expressing the correlators in terms of Bessel functions \cite{antal2008}. We refer also to Ref.~\cite{real_time} for a general discussion of these methods in free fermions systems. All this essentially follows from the stationary phase equation, which is here given by
\begin{equation}\label{eq:statphasereal}
 x-t v(k)=0,
\end{equation}
and can be shown to dominate the correlation functions. The ballistic spreading of the particles essentially follows from (\ref{eq:statphasereal}) (see Fig.~\ref{fig:semiclassics} for an illustration).

Now can such (heuristic) pictures be used to determine the density profile in imaginary time, at least for cosine dispersion relation? As explained in the introduction the stationary phase equation now reads
\begin{equation}
 x+i y v(k)+R\tilde{v}(k)=0,
\end{equation}
where $\tilde{v}(k)=\frac{d\tilde{\varepsilon(k)}}{dk}$ is the derivative of the Hilbert transform of the dispersion relation. We now use the following trick. We suppose for a moment that $y$ may be replaced by $it$ with $t$ \emph{real}. We then assume that the particle move at speed $v(k)$, with an extra contribution proportional to $\tilde{v}(k)$ in the imaginary-time direction.
 Therefore, the new ``Wigner function'' $W_R(x,k,t)$ corresponding to the imaginary time problem satisfies
\begin{equation}
 W_R(x,k,t)=W_R(x-v(k)t-R\tilde{v}(k),k,0),
\end{equation}
and the density is given by
\begin{eqnarray}
\braket{ \rho_{x,t} }&=& \int_{-\pi}^{\pi} \frac{dk}{2\pi} W_R(x,k,t)\\
&=&  \int_{-\pi}^{\pi} \frac{dk}{2\pi} \Theta \left(-x+t\sin k +R \cos k \right)\\
&=&\frac{1}{\pi}{\rm arccos} \left(\frac{x}{\sqrt{R^2+t^2}}\right).
\end{eqnarray}
Now plugging $t=-iy$ in the above formula, we recover the exact result (\ref{eq:xxdensity}).
\subsection[ \hspace{2cm}Finite-size correlation functions]{Finite-size correlation functions}
\label{sec:real_correlations}
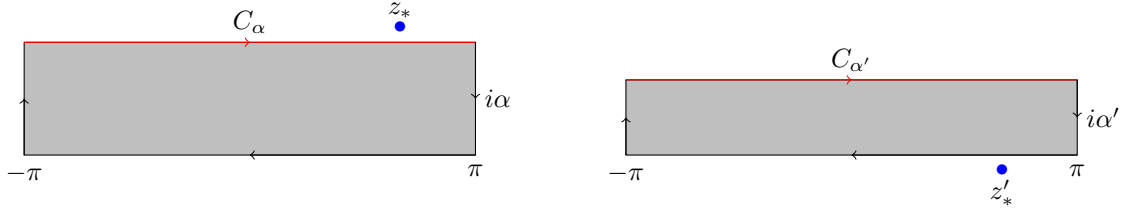
\begin{figure}[htbp]
 \begin{tikzpicture}
 \draw[fill=gray!50] (-3,0) rectangle (3,1.5); 
 \draw[ ->, red](-3,1.5)--(0,1.5);
 \draw[ red](0,1.5)--(3,1.5);
  \draw[ ->](3,1.5)--(3,1.5/2);
 \draw[ ->](3,0)--(0,0); 
 \draw[ ->](-3,0)--(-3,0.75);
 \draw(0,1.5) node[above]{$C_{\alpha}$};
 \draw(-3,0) node[below]{$-\pi$};
 \draw(3,0) node[below]{$\pi$};
 \draw(3,0.75) node[right]{$i\alpha$};
 \draw[blue](2,1.7) node{$\bullet$};
 \draw(2,1.7) node[above]{$z_*$};
 \begin{scope}[xshift=8cm]
 \draw[fill=gray!50] (-3,0) rectangle (3,1); 
 \draw[ ->, red](-3,1)--(0,1);
 \draw[ red](0,1)--(3,1);
  \draw[ ->](3,1)--(3,1/2);
 \draw[ ->](3,0)--(0,0); 
 \draw[ ->](-3,0)--(-3,0.5);
 \draw(0,1) node[above]{$C_{\alpha'}$};
 \draw(-3,0) node[below]{$-\pi$};
 \draw(3,0) node[below]{$\pi$};
 \draw(3,0.5) node[right]{$i\alpha'$};
 \draw[blue](2,-0.2) node{$\bullet$};
 \draw(2,-0.2) node[below]{$z'_{*}$};
 \end{scope}

 \end{tikzpicture}
\caption{\textit{Left.} The contour $C_{\alpha}$ is denoted in red. The contour can be lowered back to the real axis
provided the gray area is free of singularity. The pole $z=k'+i\alpha+i0$ for $k'\in[-\pi,\pi]$ is also denoted.
\textit{Right.} Contour of integration $C_{\alpha'}$ and pole in the variable $z'=k+i(\alpha'-\alpha)-i0$.}
\label{fig_contour}
\end{figure}

We now present more general arguments about the analytic continuation. Let $\mathcal{O}_k (x_k)$ be
local operators at positions $x_k\in\mathbb R$,
with expectation value in time given by
\begin{equation}\fl
\label{QC}
\langle\mathcal{O}_1(x_1,t_1)\dots\mathcal{O}_n(x_n,t_n)
\rangle\equiv\lim_{R\rightarrow 0}
\frac{\langle\psi_0|\mathcal{T}\bigl[(e^{iH(t_1+iR)}\mathcal{O}(x_1)e^{-iH(t_1-iR)})\dots
(e^{iH(t_n+iR)}
\mathcal{O}(x_n)e^{-iH(t_n-iR)})\bigr]|\psi_0\rangle}
{\langle\psi_0|e^{-2RH}|\psi_0\rangle},  
\end{equation}
where $R>0$ is a dumping factor introduced to ensure convergence and
$\mathcal{T}$ the time-ordering prescription. Explicit manipulations
of (\ref{QC}) are performed in imaginary time, {\it i.e.} by setting $t_k=-i y_k$, and eventually
taking the limit $R\rightarrow 0$ \textit{after} analytic continuation back in real time.
We should stress that along the procedure analyticity is assumed \textit{a priori}.
Notice also that the effective geometry of the time-evolution in (\ref{QC}) -- before taking the limit
$R\rightarrow 0$ -- is the slab in shown Fig. \ref{fig:param2}. Computations in the slab geometry and analytic continuation back to real time
were performed in~\cite{calabrese2006time, calabrese2007quantum} to  extract the large-time behavior of correlation functions
in a protocol satisfying the conditions
\begin{itemize}
 \item the bulk of the slab is described by a gapless quantum field theory (CFT),
 \item the boundary state $|\psi_0\rangle$ renormalizes to a conformal invariant boundary condition.
\end{itemize}
As emphasized in the introduction, the dynamics from the domain wall initial state in the XX model provides a particularly neat  
example of violation of the second condition above (see also \cite{cardy2015quantum} for a discussion) that is also exactly solvable in the slab geometry. 
We start by
establishing under which conditions the two-point correlation functions can be brought into a form 
suitable for analytic continuation to real
time. We have
\begin{equation}
\label{corryy1}
 \langle c^{\dagger}(x,y)c(x',y')\rangle=
 \int_{-\pi}^{\pi}\frac{dk}{2\pi}\int_{-\pi}^{\pi}\frac{dk'}{2\pi}\frac{e^{-ixk-y\cos k+iR\sin k+ix'k'+
 y'\cos k'-iR\sin k'}}{2i\sin\left(\frac{k-k'}{2}-i0\right)}.
\end{equation}
Introducing the notations $\alpha={\rm arcth}(y/R)$ (resp. $\alpha'$ ) and $\tilde{R}=\sqrt{R^2-y^2}$
(resp. $\tilde{R'}$), we can rewrite (\ref{corryy1}) as
\begin{equation}
\label{Cf_tilde}
 \langle c^{\dagger}(x,y)c(x',y')\rangle
 =\int_{-\pi}^{\pi}\frac{dk}{2\pi}\int_{-\pi}^{\pi}\frac{dk'}{2\pi}\frac{e^{-ixk+i\tilde{R}\sin(k+i\alpha)+ix'k'
 -i\tilde{R}'\sin(k'+i\alpha')}}{2i\sin\left(\frac{k-k'}{2}-i0\right)}.
\end{equation}
To show the main idea of the calculation we assume $\alpha, \alpha'\geq 0$,
the other cases can be worked out in a similar fashion. First we look at the correlation functions as contour integrals in
the complex plane of $z$ and $z'$ defined as follows
\begin{equation}
\label{Cf_an}
 \langle c^{\dagger}(x,y)c(x',y')\rangle
 =e^{-\alpha x+\alpha'x'}\int_{C_{\alpha}}\frac{dz}{2\pi}\int_{C_{\alpha'}}\frac{dz'}{2\pi}
 \frac{e^{-ixz+i\tilde{R}\sin z+ix'z'
 -i\tilde{R}'\sin z'}}{2i\sin\left(\frac{z-z'-i\alpha+i\alpha'}{2}-i0\right)}.
\end{equation}
The contours $C_{\alpha}$, $C_{\alpha'}$ are shifted along the imaginary axis by a constant amount $i\alpha$ and
$i\alpha'$ with respect to the segment $[-\pi,\pi]$, see Fig.~\ref{fig_contour}.

The contributions of the vertical edges in Fig.~\ref{fig_contour} cancel due to periodicity of the integrand
under $z\rightarrow z+2\pi$ and $z'\rightarrow z'+2\pi$. Therefore we can lower
the contours back to the real axis provided no singularities are encountered in the process. Let us first look 
at the integral over $z$: The only pole with $|{\rm Re}\, z|\leq \pi$ is located at $z_*=z'+i\alpha-i\alpha'+i0$, but remembering
that $z'$ belongs to the contour $C_{\alpha'}$ we have $z_*=k'+i\alpha+i0$, with $k'\in[-\pi,\pi]$. Therefore for
$\alpha\geq 0$ we can deform the contour $C_{\alpha}$ into the real interval $[-\pi,\pi]$. Consider now the
integral over $z'$. The only singularity we can encounter is located at $z'_*=z-i(\alpha-\alpha')-i0$, but now the variable
$z$ lies on the real axis and $z'_*=k-i(\alpha-\alpha')-i0$, for $k\in[-\pi,\pi]$. The point $z'_*$ will be outside the gray region of
Fig.~\ref{fig_contour} if and only if $\alpha-\alpha'\geq 0$ since we have
assumed $\alpha$ and $\alpha'$ non negative. Summarizing, in the case $y$ and $y'$ non-negative we can
analytically continue the integral (\ref{Cf_an}) to
\begin{equation}
 \label{Cf_fin}
 \langle c^{\dagger}(x,y)c(x',y')\rangle
 =e^{-\alpha x+\alpha'x'}\int_{-\pi}^{\pi}\frac{dk}{2\pi}\int_{-\pi}^{\pi}\frac{dk'}{2\pi}
 \frac{e^{-ixk+i\tilde{R}\sin k+ix'k'
 -i\tilde{R}'\sin k'}}{2i\sin\left(\frac{k-k'-i\alpha+i\alpha'}{2}-i0\right)}.
\end{equation}
if and only if $y'\leq y$. A case where (\ref{Cf_fin}) can be obtained in closed form is $y=y'$, one gets
\begin{equation}
 \label{samey}
 \langle c^{\dagger}(x,y)c(x',y)\rangle=\frac{\tilde{R} e^{-\alpha(r-s)}}{2(r-s)}
 [J_{r-1}(\tilde{R})J_{s}(\tilde{R})-J_{r}(\tilde{R})J_{s-1}(\tilde{R})]
\end{equation}
where $x=r-1/2$, $x'=s-1/2$ for $r,s\in\mathbb Z$ and $J_{\nu}(t)=\int_{-\pi}^{-\pi}\frac{dk}{2\pi}
e^{-i\nu x+i\sin k t }$ is the Bessel function, in agreement with \cite{prahofer2002scale}. One can prove (\ref{samey}) by taking the derivative with respect to
$\tilde{R}$ in (\ref{Cf_fin}).  Now we are allowed to analytically continue  directly (\ref{samey}) for $y=it$ and
set $\tilde{R}$ to zero; in this limit $\alpha=i\pi/2$ and we are back with exactly the same correlation functions
of the real time quench problem first found in~\cite{antal1999transport}. 
\newpage
\section[ \hspace{2cm}Airy kernel, Tracy-Widom distribution and Airy process]{Airy kernel, Tracy-Widom distribution and Airy process} 
\label{sec:airy}
The goal of this appendix is to provide a concise, non-rigorous, derivation of the boundary scaling behavior of the propagator, and its link with the celebrated Tracy-Widom distribution~\cite{tracy1994}. This behavior at the edge
is obtained in a scaling regime that is different from the one in which one sees the bulk
correlations---recall (\ref{eq:scaling_regime})---therefore it is a topic that is not really the focus of this paper.
However, once the fermion propagator is calculated, it is really a minor effort to get to the Tracy-Widom distribution. We therefore discuss these aspects now, for completeness. We also briefly review
the appearance of the {\it Airy process} introduced by Pr\"ahofer and Spohn in~\cite{prahofer2000universal}.

\subsection[ \hspace{2cm}Fermion propagator close to the boundary: generic appearance of the Airy kernel]{Fermion propagator close to the boundary: generic appearance of the Airy kernel}

Once again, we want to analyze the behavior of some integral using the stationary phase approximation. To illustrate the general features,
consider the following warm-up exercise: let us determine the behavior of 
\begin{equation}
I(x,y)=\int_{-\pi}^{\pi}\frac{dk}{2\pi} e^{i(kx+R\tilde{\varepsilon}(k)+iy\varepsilon(k))}
\end{equation}
for a point that is near the arctic curve. [To fix the ideas, one can focus on the XX chain of the introduction, for which $\varepsilon(k) = -\cos k$ and $\tilde{\varepsilon}(k) = -\sin k$;  then the arctic curve is the circle $x^2 + y^2 = R^2$. We will however try to keep the notations more general, in order to emphasize that the discussion of this appendix applies more generally, and in particular it can be adapted straightforwardly to all the models treated in the main text, by chosing the appropriate dispersion $\varepsilon(k)$, or $\varepsilon(\kappa)$.] Let us parametrize the arctic curve locally as $(x,y) = (x_{\rm a.c}(y), y)$. Then when we say 'near the arctic curve', we mean, more precisely, that we are interested in the following limit:
\begin{equation}
	\label{eq:edge_scaling}
	\frac{y}{R} \quad {\rm fixed} ,  \qquad  \quad  \frac{x-x_{{\rm a.c.}}(y)}{R^{1/3}} \quad {\rm fixed} , \qquad \qquad {\rm and} \quad R \rightarrow \infty .
\end{equation}
The reason why the power $R^{1/3}$ shows up will become clear below. First, notice that such points are 'near the boundary' in the sense that the distance from the point $(\frac{x}{R},\frac{y}{R})$ to the arctic curve goes to zero, so in the bulk scaling regime (\ref{eq:scaling_regime}), these are really points on the boundary. Now, the crucial difference with bulk scaling is that, in the limit (\ref{eq:edge_scaling}), the two solutions to the stationary phase equation $x + R \frac{d}{dk} \tilde{\varepsilon}+iy \frac{d}{dk}\varepsilon$, which we called $k=z$ and $k=-z^*$ throughout the paper, are not distinct; instead, the two stationary points coalesce at a degenerate point $z_d(y) = z = -z^*$. Around this critical point, the argument in the exponential vanishes not only at first order, but also at second order. Therefore, one needs to expand it to third order \cite{erdelyi_book}, leading to
\begin{eqnarray}
\label{Airy_analytic}
\nonumber I(x,y)  & \underset{{\rm near the b.}}{\longrightarrow} &
e^{i z_d(y) (x-x_{{\rm a.c.}})+ i\varphi(z_d(y) )}
\int_{-\infty}^{\infty}\frac{d\tilde{k}}{2\pi}
e^{i\left[\tilde{k}\left(x-x_{\rm a.c.}(y) \right)+
\frac{\tilde{k}^3 C(y) }{3}\right]} \\
&& = \;  e^{i z_d(y) (x-x_{{\rm a.c.}})+ i\varphi(z_d(y) )}~
C(y)^{-1/3}
{\rm Ai}(X) \, ,
\end{eqnarray}
where $\tilde{k} \equiv k-z_d(y)$ and $C(y) \equiv \frac{1}{2} \frac{d^3}{dk^3} \left( R\tilde{\varepsilon}+iy \varepsilon \right)_{|_{k = z_d(y)}}$. Notice that $C(y)$ must be of order $O(R)$. [One can check that $C(y)$ is real and non-zero, for all the models we looked at in the main text. Moreover, we can assume without loss of generality that $C(y) > 0$ (if $C(y) <0$, all formulas are identical up to a change $C(y) \rightarrow -C(y)$ and $X \rightarrow -X$). For instance, for the XX chain, $C(y)=\sqrt{R^2-y^2}$.] In the first line, we extended the integral to the real line, since the integrand is oscillating fast at large $\tilde{k}$ and therefore gives only
subleading contributions. In going from the first to the second line, we recognized the Airy function,
${\rm Ai}(X)=\int_{\mathbb R}\frac{dK}{2\pi} e^{iKX+iK^3/3}$, and used the rescaled variables $X= C(y)^{-1/3}(x-x_{\rm a.c.}(y))$ and $K =  C(y)^{1/3} \tilde{k}$.

\vspace{4mm}

After this instructive exercise, let us turn to the task of determining the behavior of the
propagator for two points near the boundary. Recall that the propagator takes the form
\begin{equation}
\label{cf_exact}
\langle c^{\dagger}_{x,y} c_{x',y'}\rangle=\int_{-\pi}^{\pi}
 \int_{-\pi}^{\pi}\frac{dk~dq}{(2\pi)^2}~
 \frac{e^{-i(kx+R\tilde{\varepsilon}(k)+iy\varepsilon(k))+i(q x'+R\tilde{\varepsilon}(q)
 +iy'\varepsilon(q))}}{2i\sin\left(\frac{k-q}{2}-i0\right)}.
\end{equation}
Since we want the two points $(x,y)$ and $(x',y')$ to be near
the arctic curve, the stationary points in $k$ and $q$ are the degenerate critical points above $z_d(y)$ and $z_d(y')$. 
When $y\not=y'$, $z_d(y) \neq z_d(y')$, so the denominator in (\ref{cf_exact}) does not vanish, 
and the asymptotics of the two-point functions immediately follows from the result of the previous exercise. It is given by the product of Airy functions
\begin{equation}
\label{Airy_1}
\langle c^{\dagger}_{x,y}c_{x',y'}\rangle_{|_{y\not = y'}}
\underset{{\rm near\, the\, b.}}{\longrightarrow}
\frac{[I(x,y)]^* I(x',y')}{2i\sin\left(\frac{z_c(y)-z_c(y')}{2}\right)}.
\end{equation}
The case $y\rightarrow y'$ is more interesting. Although it could be derived directly
from (\ref{Airy_1}), taking the limit $y' \rightarrow y$, we prefer to
follow a more direct path. When $y=y'$ we can expand the denominator in
(\ref{cf_exact}) to the first order both in $k$ and $q$, 
close to $z_c(y)$. This gives
\begin{equation}\fl
\langle c^{\dagger}_{x,y}c_{x',y}\rangle\underset{{\rm near \,the\, b.}}
  {\longrightarrow}e^{-i z_d(y) (x-x')}
  \int_{-\infty}^{\infty}\int_{-\infty}^{\infty}\frac{d\tilde{k}d\tilde{q}}{(2\pi)^2}
  ~\frac{e^{-i\left[\tilde{k}\left(x-x_{{\rm a.c.}}(y)\right)+
\frac{\tilde{k}^3C(y)}{3}\right]+i\left[\tilde{q}\left(x'-x_{{\rm a.c.}}(y)\right)+
\frac{\tilde{q}^3 C(y)}{3}\right]}}{i(\tilde{k}-\tilde{q}-i0)} \, ,
\end{equation}
where $\tilde{k}=k-z_d(y)$ and $\tilde{q}=q-z_d(y)$. Finally, with the same
rescaled variables $X$ and $K$ (and $X'$ and $Q$) as above, we obtain
\begin{equation}
\label{airy_stat}
\langle c^{\dagger}_{x,y}c_{x',y}\rangle\underset{{\rm near\, the\, b.}}
  {\longrightarrow
  }e^{-i z_d(y) (x-x')} C(y)^{-1/3} \,   K(X, X') \, ,
  \end{equation}
where $K(X,X')$ is the Airy kernel~\cite{tracy1994}
\begin{equation}
\label{Airy_k}
 K(X,X')=
 \int_{0}^{\infty} d\lambda {\rm Ai}(X+\lambda){\rm Ai}(X'+\lambda).
\end{equation}
The result (\ref{airy_stat}) can be easily proved by observing
that $[i(\tilde{k}-\tilde{q}-i0)]^{-1}=\int_{0}^{\infty}d\lambda
e^{-i\lambda(\tilde{k}-\tilde{q}-i0)}$ and
recalling the definition of the Airy function given above. In passing, notice that one
equivalent and somewhat more explicit way of writing the Airy kernel is
$K(X,X')=[{\rm Ai}(X){\rm Ai}'(X')-{\rm Ai}(X'){\rm Ai}'(X)]/(X-X')$. For later purposes though, 
the integral form (\ref{Airy_k}) is slightly more convenient.

In summary, the appearance of the Airy kernel close to the arctic curve is very generic: it comes from a degenerate critical point when one performs the stationary phase approximation \cite{erdelyi_book}. Once the propagator is expressed in the form of the Airy kernel (\ref{airy_stat}), the Tracy-Widom distribution follows straightforwardly, as we review briefly now.

\subsection[ \hspace{2cm}Tracy-Widom distribution]{Tracy-Widom distribution}

To understand the emergence of the Tracy-Widom distribution and the Airy process
we adapt the original arguments 
of~\cite{prahofer2002scale} (see also the more recent Ref. \cite{PhysRevLett.110.060602}).
Let us consider the interval of the $x$-axis $E_s$ defined by the points $x$ such that  $s\leq X< \infty$,
with $X$ the scaling edge variable and $y$ fixed ($y^2\not=1$).
We then ask the question: what is the generating function
$\chi(\lambda,s)$ for
the particle number $N(E_s)$ in the region $E_s$ when $x$ is near the boundary (in the sense discussed in the previous paragraph) ? Formally,
\begin{equation}
 \chi(\lambda,s)\,=\,\braket{\Psi_0|e^{i\lambda N(E_s)}|\Psi_0}\,=\,\sum_{n=0}^{\infty}e^{i\lambda n}~\mathbb P[N(E_s)=n],
\end{equation}
where $\mathbb P[N(E_s)=n]$ is the probability that exactly $n$ particles are in the interval $E_s$.
By standard free fermion manipulations, one can recast the function $\chi(\lambda,s)$ in the form of a
determinant, $\det \left[ \left(  \delta_{x,x'} + (e^{i \lambda} -1) \left< c^\dagger_x c_{x'} \right>  \right)  \right]$,
where $x$ and $x'$ are integers both restricted to the semi-infinite interval where $X, X' \in E_s$.
Then, replacing the lattice propagator $\left< c^\dagger_x c_{x'} \right> $ by its asymptotic form (\ref{airy_stat}),
and considering $x$, $x'$ (or equivalently $X$, $X'$) as continuous variables, $\chi (\lambda,s)$ becomes
the Fredholm determinant (see also \cite{its1990differential, PhysRevLett.70.1704})
\begin{equation}
 \chi(\lambda,s)=\det[1+(e^{i\lambda}-1)K(X,X')],
\end{equation}
where it is understood that $X, X'$ both belong to the interval $E_s$.
To see this, notice that the phase in (\ref{airy_stat}) may be removed by a gauge transformation $c^\dagger_x \rightarrow e^{i z_d x} c^\dagger_x $, which leaves the determinant invariant. Also, notice that the Jacobian
$\frac{\partial x}{\partial X}= C(y)^{1/3}$ exactly compensates the prefactor in
(\ref{airy_stat}). It follows that all probabilities $\mathbb P[N(E_s)=n]$ can be expressed as
\begin{equation}
 \mathbb P[N(E_s)=n]=\left.\frac{(-1)^n}{n!}\frac{d^n}{du^n}\det(1-uK)\right|_{u=1}.
\end{equation}
Now let us define $F_2(X)dX$ as the probability that the rightmost particle is in the interval $[X,X+dX]$. Then the cumulative distribution of $F_2(X)$
is $\mathbb P[N(E_s)=0]=\int_{-\infty}^s d X~F_2(X)$, which is equal to $\det(1-K)$, where $1-K$ is an operator on the interval $E_s$ with $s=X$, as above.
This is the gaussian unitary Tracy-Widom distribution.

\subsection[ \hspace{2cm}Airy process]{Airy process}

The last step is to introduce the stochastic process \cite{prahofer2000universal} associated with the motion of the right-most particle. Since we are dealing with free fermion problems, we know already that all multipoint correlation functions may be obtained from the fermion correlators
$\langle c^{\dagger}_{x_1,y_1}\dots c^{\dagger}_{x_n,y_n}c_{x_1',y_1'}\dots c_{x_n',y_n'} \rangle$, which themselves are obtained from Wick's theorem. The only relevant question is: what is the propagator? In mathematical language, this means that the process one is trying to define must be a determinantal point process. The question is: what is its kernel ?

As usual in this paper, the propagator (or kernel) is given by (\ref{cf_exact}), and we need to analyze its scaling behavior. Here, the scaling behavior we want to focus on is, following \cite{prahofer2000universal},
\begin{equation}\fl
	\label{eq:airy_process_scaling}
	\frac{y_0}{R} \; {\rm fixed} ,  \quad \frac{x-x_{{\rm a.c.}}(y_0)}{R^{1/3}} \quad {\rm fixed} , \quad  \frac{x'-x_{{\rm a.c.}}(y_0)}{R^{1/3}} \quad {\rm fixed}, \quad  \frac{y-y_{0}}{R^{2/3}} \quad {\rm fixed} , \quad  \frac{y'-y_{0}}{R^{2/3}} \quad {\rm fixed}  , \quad {\rm and} \; R \rightarrow \infty .
\end{equation}
That is, we zoom in on a region around the point $(x_{{\rm a.c.}}(y_0), y_0)$ on the arctic curve. The reason why the powers $R^{1/3}$ and $R^{2/3}$ show up is again related to the way we perform the stationary phase approximation. This time, in (\ref{cf_exact}), we are going to integrate on $k$ around the stationary point $z_d (y)$, which is itself close to $z_d(y_0)$, so we write
\begin{eqnarray*}
	k -z_d(y_0) & = & (k-z_d(y)) + (z_d(y)-z_d(y_0))   \,  \simeq \,  \frac{1}{C(y_0)^{1/3}} \left(   K    - i \tau \right) \, ,
\end{eqnarray*}
with $K = (k-z_d(y)) C(y_0)^{1/3}$ and $\tau  \, = \, i (y-y_0) \, C(y_0)^{1/3} \, \left(\left.\frac{dz_d(y)}{dy}\right|_{y=y_0}\right)$. [Notice that $\tau$ is of order $O(1)$ in the regime (\ref{eq:airy_process_scaling}); one can also check that $\tau$ is always real.] The same relation holds for the rescaled variables $Q$ and $\tau'$ in terms of $q$ and $y'$. Then, 
expanding the argument of the exponential in (\ref{cf_exact}) as previously, one finds that,
\begin{eqnarray}
\label{extended_airy}
 \nonumber \langle c^{\dagger}_{x,y}c_{x',y'}\rangle   &  \underset{{\rm near\, the\, b.}}{\longrightarrow} &
  e^{-i z_d(y) (x-x_{{\rm a.c.}}(y))- i\varphi(z_d(y) )} e^{i z_d(y') (x'-x_{{\rm a.c.}}(y'))+ i\varphi(z_d(y') )}   \, C(y_0)^{-1/3}  \\
  	&& \quad \times \quad  \int_{-\infty}^{\infty}\frac{d K}{2\pi}\int_{-\infty}^{\infty}\frac{dK'}{2\pi}
 \frac{e^{-i K X-i\frac{K^3}{3}+iX'K'+i\frac{K'^3}{3}}}{i(K-i\tau-K'+i\tau'-i0)},
\end{eqnarray}
where $X= C(y_0)^{-1/3} (x-x_{\rm a.c.} (y))$, and the same relation holds for $X'$, $\tau'$ in terms of $x'$, $y'$. The phases may be removed by a gauge transformation as above, and one is left with $\langle c^{\dagger}_{x,y}c_{x',y'}\rangle\rightarrow C(y_0)^{-1/3} K_{Ai}(X,\tau; X', \tau')$, where $K_{Ai}$ is the {\it extended Airy kernel},
\begin{equation}
\label{ex_Airy}
  K_{Ai}(X,\tau; X' \tau') \; = \;  \left\{  \begin{array}{lcl}  \displaystyle \int_{0}^{\infty}d\lambda \ e^{-\lambda(\tau-\tau')}{\rm Ai}(X+\lambda){\rm Ai}(X'+\lambda)   &&  {\rm if} \quad  \tau\geq\tau' \\
  \displaystyle  -\int_{\infty}^{0}d\lambda~e^{-\lambda(\tau-\tau')} {\rm Ai}(X+\lambda){\rm Ai}(X'+\lambda)   && {\rm if}  \quad \tau < \tau'  \, .
  \end{array} \right.
\end{equation}
The determinantal point process defined by this extended Airy kernel is called the {\it Airy process} \cite{prahofer2000universal,prahofer2002scale, johansson2005}.

\newpage
\section[ \hspace{2cm}Useful Hilbert transforms]{Useful Hilbert transforms}

\label{appx:hilb}

In this appendix we compute the Hilbert transforms needed in the main text. We always work on the space of periodic, real, analytic functions on $\mathbb{R}$, with period $2\pi$, and with zero average: $\int_{-\pi}^\pi dk f(k) \, = \, 0$. On this space, the Hilbert transform $\tilde{f}$ of the function $f$ may be defined in (at least) three equivalent ways:
\begin{enumerate}
	\item as the principal value of the integral
		$$
		\tilde{f}(k) \, = \, {\rm p.v.} \int_{-\pi}^\pi \frac{dk'}{2\pi} f(k') {\rm cot} \left( \frac{k-k'}{2} \right) 
		$$
	\item using the Fourier series of $f(k)$ to define the one of $\tilde{f}(k)$:
		$$
		f(k) \, = \, \sum_{p \geq 1} \left[ a_p  \cos (p k) +  b_p \sin (p k) \right] \quad \Rightarrow  \quad \tilde{f}(k) \, = \, \sum_{p \geq 1} \left[ a_p  \sin (p k) -  b_p \cos (p k) \right]
		$$
	\item as the unique periodic, real-valued function on $\mathbb{R}$, which is such that there exists an analytic continuation of $k \mapsto f(k) + i \tilde{f}(k)$ to a complex neighborhood of $\mathbb{R}$ that includes the upper half-plane ${\rm Im}\, k >0$. 
\end{enumerate}
We are free to pick any of those three equivalent definitions when we perform concrete calculations. While the Hilbert transform appeared in part \ref{sec:ktr} in the first form (the principal value), the two other definitions are also very useful. For example the Hilbert transform in the toy model immediately follows from the second definition.

All other models may be treated by making use of the third definition.
The basic ingredient is the complex function (for some real parameter $u$ with $0<u<1$):
$$
k \mapsto \log (1+ u e^{i k}) \, .
$$
This function is analytic in the upper half-plane ${\rm Im}\, k >0$. Therefore, according to the third definition above, if we look at the pair of real-valued functions $f$ and $\tilde{f}$ defined for $k \in \mathbb{R}$ as
$$
f(k) \, = \, {\rm Re } \left[ \log (1+ u e^{i k}) \right]  \qquad \qquad   \tilde{f}(k) \, = \, {\rm Im } \left[ \log (1+ u e^{i k}) \right] \, ,
$$
then $\tilde{f}$ is the Hilbert transform of $f$. This is all we need for the dimer model on the honeycomb lattice, since the dispersion relation in this model turns out to be exactly $\varepsilon_{{\rm honey.}}(k) \, =\,  -f(k)$. In the six-vertex and square dimer models, the situation is similar (we use the notations of parts \ref{sec:square}, \ref{part:6v}, and write $\kappa$ instead of $k$ for the argument of the functions): 
\begin{eqnarray*}
2\varepsilon_{{\rm 6v.}} (\kappa) & = &  - \log \left( \frac{c + b \cos \kappa  }{c   - b \cos \kappa } \right) \, = \,  -2\,  {\rm Re} \left[ \log ( 1 + u e^{i \kappa} )  \right]\, + \,2 \,{\rm Re} \left[ \log ( 1 - u e^{i \kappa} )  \right] \, ,
\end{eqnarray*}
with $\frac{b}{c} = \frac{2u}{1+u^2}$.
Then the Hilbert transform is
\begin{eqnarray*}
2\tilde{\varepsilon}_{{\rm 6v.}} (\kappa) & = &  -2\,  {\rm Im} \left[ \log ( 1 +u e^{i \kappa} )  \right]\, + \,2 \,{\rm Im} \left[ \log ( 1 - u e^{i \kappa} )  \right] \\
&=&   i \,   \log \left(\frac{  1+ i \frac{ 2u}{1-u^2} \sin \kappa }{  1 - i  \frac{2u}{1-u^2} \sin \kappa } \right)\\
& =&   i \,   \log \left(\frac{  a+ i b \sin \kappa }{  a - i b \sin \kappa } \right)  \, ,
\end{eqnarray*}
as claimed in the main text. In the last line, we have used the assumptions of part \ref{part:6v}, namely $a^2+b^2=c^2$, and $a>0$, $b>0$, $c>0$. The Hilbert transform for the dimer model is exactly the same, provided one makes the substitutions $a=v$, $b=u$, and $c=w$.

\newpage

\section[ \hspace{2cm}Partition functions from bosonization]{Partition functions from bosonization}\label{sec:partitionsfunctions}

In this appendix, we recover the partition functions of all the models studied in this paper, using the bosonization procedure presented in section \ref{sec:ktr}. We start by introducing the general formalism of bosonization and applying it to our toy model, {\it i.e.} the XX chain. In the rest of the appendix, we shall see how this procedure can be modified to compute the partition function of the honeycomb and square lattice dimer model as well as the 6-vertex model. 

\subsection[ \hspace{2cm}Bosonization and the XX chain]{Bosonization and the XX chain}

The idea is to think for a moment of the real time analog of the partition function, the return probability $\tilde{Z}(t)=\braket{\Psi_0|e^{-i H t}|\Psi_0}$. In bosonized form, the following substitution holds
\begin{equation}
	e^{-i t H} \ket{\Psi_0 }  \; \rightarrow \;  \mathcal{N}: e^{-i t \int \frac{dk}{2\pi} \varepsilon (k) \partial \varphi (k) } : \ket{\Psi_0 } \,,
\end{equation}
up to some unknown prefactor $\mathcal{N}$. It is obtained by looking at the norm of that state, which is one due to the unitarity of the real time evolution:
\begin{eqnarray*}
\bra{\Psi_0} e^{i t H}	e^{-i t H} \ket{\Psi_0}  &=&1\\
 &=&\mathcal{N}^2  \bra{\Psi_0} : e^{i t \int \frac{dq'}{2\pi} \varepsilon (q') \partial \varphi (q'- i \epsilon) } :  : e^{-i t \int \frac{dq}{2\pi} \varepsilon (q) \partial \varphi (q) } : \ket{\Psi_0 }\\
 &=&\mathcal{N}^2\; \exp \left(  t^2    \int \frac{dq}{2\pi} \int \frac{dq'}{2\pi} \varepsilon (q) \varepsilon (q') \partial_q \partial_{q'} \bra{\Psi_0} \varphi (q' -i \epsilon)\varphi (q) \ket{\Psi_0}   \right)\,, 
\end{eqnarray*}
By definition the expectation value of a single time-ordered exponential is one, so the return probability is given by $Z(t)=\mathcal{N}$. Therefore, we have
\begin{equation*}
 \tilde{Z}(t)=\exp \left( -\frac{t^2}{2}    \int \frac{dq}{2\pi} \int \frac{dq'}{2\pi} \varepsilon (q) \varepsilon (q') \partial_q \partial_{q'} \bra{\Psi_0} \varphi (q' -i \epsilon)\varphi (q) \ket{\Psi_0}   \right). 
\end{equation*}
The partition function we are looking for, $Z(R)=\braket{\Psi_0 |e^{-2RH}|\Psi_0}$, may be obtained by Wick-rotating back $t\to 2iR$:
\begin{equation}\label{eq:prescription}
 Z(R)=\tilde{Z}(2iR)=\exp \left( 2R^2    \int \frac{dq}{2\pi} \int \frac{dq'}{2\pi} \varepsilon (q) \varepsilon (q') \partial_q \partial_{q'} \bra{\Psi_0} \varphi (q' -i \epsilon)\varphi (q) \ket{\Psi_0}   \right) .
\end{equation}
As we shall see, all the partition functions may be obtained by applying the formula (\ref{eq:prescription}). Before doing so, let us however mention that this formula can be put on a more rigorous footing using the so-called infinite wedge formalism (see \cite{KacBook,OkounkovReshetikhin}). We explain the method on the toy model Hamiltonian
\begin{equation*}
 H=\int_{-\pi}^{\pi} \frac{dk}{2\pi}\varepsilon(k) c^\dag(k)c(k)
\end{equation*}
for some dispersion relation $\varepsilon(k)$ that has zero mean value. What we really have in mind is the particular case $\varepsilon(k)=-\cos k$, but the following argument is general. 
$\varepsilon(k)$ may be written in Fourier series:
\begin{equation}
 \varepsilon(k)=\sum_{p\geq 1}\, [\varepsilon]_{p} e^{ipk}+\sum_{p\geq 1}\, [\varepsilon]_{-p} e^{-ipk}=\Hr{\varepsilon}(k)+\Hl{\varepsilon}(k),
\end{equation}
where we have separated the positive (right) Fourier modes from the negative (left) ones. Similarly, we also write $H=\Hr{H} +\Hl{H}$, and make the observation that $\Hl{H}\ket{\Psi_0}=0$ and $\bra{\Psi_0} \Hr{H}=0$. As is also explained in Ref.~\cite{OkounkovReshetikhin}, the result is a priori sentitive to the choice regularization. Here we choose to consider a system of large even size $L$ with open boundary conditions, with sites running from $-L/2+1/2$ to $L/2-1/2$, and only then take the limit $L\to \infty$. This is clearly the most natural choice for the XX chain; for the dimer and vertex models, it will actually be imposed. 

Let us specify a cosine dispersion relation for simplicity. The idea now is to apply the Baker-Campbell-Hausdorff (BCH) formula. The commutator between $\Hl{H}$ and $\Hr{H}$ is given by
\begin{equation}\label{eq:somecommutator}
[\Hr{H},\Hl{H}]=\frac{1}{4}\sum_{x=-L/2+1/2}^{L/2-3/2}\left(c_{x+1}^\dag c_{x+1}-c_{x}^\dag c_x\right).
\end{equation}
The crucial point is that $\ket{\Psi_0}$ is an eigenstate of the commutator with eigenvalue $-1/4$, independent on $L$. One can also check that all higher order nested commutators up to order $L/2$ have expectation value zero in the domain wall initial state. Therefore, if $L$ is sent to infinity first, the commutator (\ref{eq:somecommutator}) may be considered as a scalar when taking expectation values in the DWIS. Using the BCH formula we obtain
\begin{eqnarray}
 Z(R)&=&\braket{\Psi_0 |e^{-2R \Hr{H}} e^{-2R^2[\Hr{H},\Hl{H}]} e^{-2R \Hl{H}}|\Psi_0}\\
 &=&e^{-2 R^2 [\Hr{H},\Hl{H}]}\\
 &=&e^{R^2/2}.
\end{eqnarray}
The above can be reformulated in bosonic language. To do that, we introduce a set of operators $a_n$, defined as
\begin{equation}
a_n=  \sum_{x\in \mathbb{Z}+1/2} c^\dag_{x-n} c_x
\end{equation}
These operators satisfy the Heisenberg commutation relations
\begin{equation}\label{eq:comm}
[a_n,a_m]=n \delta_{n,-m},
\end{equation}
in the sense described above, that is, when taking expectation values in the DWIS. The $a_n$ are nothing but the modes of the bosonic field $\varphi$. Using (\ref{eq:comm}), computing the commutator between $\Hr{H}$ and $\Hl{H}$ trivialises, and we find
\begin{equation}\label{eq:genpartitionfunction}
 Z(R)=\exp\left(2R^2 \sum_{p\geq 1} p [\varepsilon]_p [\varepsilon]_{-p}\right).
\end{equation}
The equality between (\ref{eq:genpartitionfunction}) and (\ref{eq:prescription}) can easily be shown by expanding $\varepsilon(k)$ in Fourier series. With this at hand, we are able to compute all the partition functions of interest in the present paper. The only complication will be that the mean energy $\braket{\Psi_0|H|\Psi_0}$ will not necessary be zero.

\subsection[ \hspace{2cm}Dimers on the honeycomb lattice]{Dimers on the honeycomb lattice}

We discuss now dimers on the honeycomb lattice, and regularize by looking at a finite system of size $L$ finite and even, and then take the limit $L\to \infty$. The diagonalization of the transfer matrix is similar to the text, the only difference being  that the momenta are now discrete. For simplicity we label the sites from $1$ to $L$ here, so that the DWIS is $\ket{\Psi_0}=\prod_{j=1}^{L/2} c_j^\dag \ket{0}$. We still have $T=e^{-H}$, with
\begin{equation}\fl
 H=-\frac{1}{2}\sum_{m=1}^L \log \left[1+u^2+2u \cos \left(\frac{m}{L+1}\right) \right] c_{m}^\dag c_m \qquad,\qquad c_m^\dag=\sqrt{\frac{2}{L+1}}\sum_{j=1}^{L} \sin\left( \frac{m\pi j}{L+1} \right)c_{j}^\dag
\end{equation}
Let us now show that the expectation value of $H$ in the DWIS is not zero. A simple calculation shows $\braket{c_m^\dag c_m}=1/2$ for all $m$. Using this and expanding the logarithm in Fourier series, we obtain
\begin{eqnarray}\nonumber
 \braket{H}&=& \sum_{p=1}^{\infty} \frac{(-u)^p}{2p} \sum_{m=1}^{L}\cos \left(\frac{p m\pi}{L+1}\right)\\\nonumber
 &=& \sum_{p=1}^{\infty} \frac{(-u)^p}{2p} \left[-\frac{1+(-1)^p}{2}+(L+1)\delta_{p,2L+2}\right]\\\label{eq:energy0}
 &=&\frac{1}{4}\log (1-u^2)-\frac{1}{4}\log \left(1-u^{2L+2}\right)
\end{eqnarray}
Since $u<1$, the second term can be safely ignored, but not the first one, which is finite. We now use the following trick. We write the partition function as
\begin{equation*}
 Z=\braket{T^{N}}=\braket{e^{-2N H}}=e^{-2N \braket{H}}\braket{e^{-2N \left(H-\braket{H}\right)}}
\end{equation*}
and bosonize $H-\braket{H}$ -- which has average exactly zero -- according to (\ref{eq:prescription}). Combining (\ref{eq:energy0}) with (\ref{eq:genpartitionfunction}), we obtain
\begin{eqnarray}\nonumber
 Z&=& e^{-\frac{N}{2}\log (1-u^2)}\times e^{-\frac{2N^2}{4}\log (1-u^2)}\\
 &=&\left(\frac{1}{1-u^2}\right)^{\frac{N}{2}(N+1)}
\end{eqnarray}
This result is exact in the limit $N \to \infty$ and in agreement with the results of \cite{prahofer2002scale} for the partition function of the {\it multilayer PNG droplet} model.

\subsection[ \hspace{2cm}Dimers on the square lattice and 6-vertex model]{Dimers on the square lattice and 6-vertex model}

We apply exactly the same method to the dimer model. For convenience, we set the vertical weights to $v=1$; this keeps the partition function finite. The Hamiltonian for a finite size system is
\begin{equation}
 H=-\sum_{m=1}^{L} {\rm arcsinh}\left(u \cos \frac{m\pi}{L+1}\right) c_{+,m}^\dag c_{+,m}.
\end{equation}
Note that we have used the fact that the two bands are the analytic continuations of each other. Expanding again in Fourier modes, we find 
\begin{equation}
 \braket{H}=-\frac{1}{4}\log \left(1+u^2\right).
\end{equation}
Using the same two band bosonization method explained in Section.~\ref{sec:ktr} and \ref{sec:square}, we find that the result (\ref{eq:genpartitionfunction}) can still be used, albeit with the Fourier coefficients in terms of the $\kappa$ variable.
 The Fourier coefficients of the dispersion relation $\varepsilon(\kappa)=-\frac{1}{2}\log\left(\frac{w+u\cos \kappa}{w-u \cos \kappa}\right)$ are given by $[\varepsilon]_{2p+1}=(\frac{w-v}{u})^{2p+1}/(2p+1)$ and $[\varepsilon]_{2p}=0$ otherwise. We finally obtain $\sum_{p\geq 1}p[\varepsilon]_p [\varepsilon]_{-p}={\rm arctanh}\left(\left[\frac{w-v}{u}\right]^2\right)=\frac{1}{2}\log(w/v)= \frac{1}{4}\log(1+u^2)$. Hence the partition function reads
 \begin{eqnarray}\nonumber
  Z&=&e^{\frac{2N}{4}\log (1+u^2)}e^{\frac{2N^2}{4}\log(1+u^2)}\\
  &=&(1+u^2)^{\frac{N}{2}(N+1)}
 \end{eqnarray}
which is the celebrated Aztec diamond partition function \cite{elkies1992alternating}. We switch now to the 6-vertex model with domain-wall boundary conditions and start by setting $a=1$, which makes the partition function finite. 
The calculation is very similar to that of the dimer model, the only real difference being that $\braket{H}=0$. The contribution coming from the bosonized part can be deduced from that of the dimers, we simply need to replace $u$ by $b$ and $w$ by $c$. We recover the well known result \cite{korepin1982calculation,izergin1987partition,bogoliubov2002boundary}
\begin{equation}
 Z_{\rm dwbc}=c^{N^2}.
\end{equation}

\newpage

\section[\hspace{2cm}Additional details about the finite-size correlators for two-band models]{Additional details about the finite-size correlators for two-band models}
\label{sec:moredetails}
The purpose of this appendix is to provide some additional technical details regarding the derivation of the coefficients $r_{x,y}(k)$ and $s_{x,y}(k)$ (\ref{eq:ident_square}) and (\ref{eq:ident_6v}) needed to obtain exact finite-size correlations for models with two bands. We treat separately the dimer model on the square lattice 
 and the six vertex model with domain wall boundary conditions. 
 
 \vspace{0.5cm}
\subsection[\hspace{2cm} Dimers on the square lattice]{Dimers on the square lattice}
\label{sec:moredetails_dimers}
Here we derive the result
\begin{eqnarray}\label{eq:odd_action_2bis}
&&T_{\rm o}\ c_{2j+1/2}\ T_{\rm o}^{-1}=\frac{1}{v}\, c_{2j+1/2},\nonumber \\
&&T_{\rm o}\ c_{2j-1/2}\ T_{\rm o}^{-1}=\frac{1}{v^2}\left(-u c_{2j-3/2}+ v c_{2j-1/2}-u c_{2j+1/2}\right),
\end{eqnarray}
for the action of the odd transfer matrix $T_{\rm o}$ on the annihilation operators. The easiest is to go to momentum space. Recall the action on the creation operators is given by
\begin{equation}
  T_{\rm o} \ccvec T_{\rm o}^{-1}=M_{\rm o}(k)\qquad,\qquad M_{\rm o}(k)= \twomat{v+u\cos k}{iu \cos k}{iu \cos k}{v-u\cos k}\ccvec.
\end{equation}
Now diagonalizing the matrix $M_{\rm o}(k)$, one can define new creation operators $\tilde{c}^\dag_+(k), \tilde{c}_+(k)$, so that the transfer matrix reads $T_{\rm o}=\exp\left(-\int_{-\pi}^{\pi}\frac{dk}{2\pi}\tilde{\varepsilon}(k)\tilde{c}^\dag_+(k) \tilde{c}_+(k)\right)$. Hence $T_{\rm o} \tilde{c}_+^\dag (k)T_{\rm o}^{-1}=e^{-\tilde{\varepsilon}(k)}\tilde{c}_+^\dag(k)$ and also $T_{\rm o} \tilde{c}_+ (k)T_{\rm o}^{-1}=e^{\tilde{\varepsilon}(k)}\tilde{c}_+(k)$. Said differently, $T_{\rm o}$ acts on the creation operators as a two by two matrix $M_{\rm o}(k)$, and acts on the annihilation operators as the inverse of $M_{\rm o}(k)$:
\begin{equation}\label{eq:someinversestuff}\fl
 T_{\rm o} \left(\begin{array}{cc}
                  c(k)&c(k+\pi)
                 \end{array}
\right) T_{\rm o}^{-1}=\left(\begin{array}{cc}
                  c(k)&c(k+\pi)
                 \end{array}
\right) M_{\rm o}^{-1}(k)\qquad,\qquad M_{\rm o}^{-1}(k)= \frac{1}{v^2}\twomat{v-u\cos k}{-iu \cos k}{-iu \cos k}{v+u\cos k}
\end{equation}
Going back to real space, the equation (\ref{eq:someinversestuff}) becomes (\ref{eq:odd_action_2bis}), which was used in the main text.

\vspace{1cm}

\subsection[\hspace{2cm}Six-vertex model]{Six-vertex model}\label{sec:moredetails_6v}
We now focus on the six vertex model, for which the determination of the coefficient $r_{x,y}(k)$ and $s_{x,y}(k)$ is slightly more involved. 
The propagator of the modes $c^\dagger_+(k)$, $c_+(k')$ is
\begin{equation}
	\label{eq:prop_pm_6v_a}
	\left< c^\dagger_{+,y}(k) c_{+,y'}(k')  \right> \, = \, e^{y \varepsilon (\kappa)  - i N \tilde{\varepsilon}(\kappa) }  e^{-y' \varepsilon (\kappa')  + i N \tilde{\varepsilon}(\kappa') }    \bra{\Psi_0} c^\dagger_+(k) c_+(k') \ket{\Psi_0}    \, ,
\end{equation}
again with $\kappa = \kappa(k)$ and $\kappa' = \kappa(k')$. Below, we use once again the change of basis  between the modes $c^\dagger (k)$, $c^\dagger (k+\pi)$, and $c^\dagger_+ (k)$, $c^\dagger_- (k)$, given by $U(k)$. It will then be convenient to use (\ref{eq:prop_pm_6v_a}) in the following matrix form
\begin{eqnarray}\fl
	\label{eq:prop_pm_6v_b}
\nonumber		&& \left( \begin{array}{cc} \left<  c^\dagger_{+,y} (k)  c_{+,y'} (k')    \right>  &   \left<  c^\dagger_{+,y} (k)  c_{-,y'} (k')    \right> \\  \\ \left<  c^\dagger_{-,y} (k)  c_{+,y'} (k')    \right>  &   \left<  c^\dagger_{-,y} (k)  c_{-,y'} (k')  \right> \end{array}  \right) \, = \, \left( \begin{array}{cc} \left<  c^\dagger_{+,y} (k)  c_{+,y'} (k')    \right>  &   \left<  c^\dagger_{+,y} (k)  c_{+,y'} (k'+\pi)    \right> \\  \left<  c^\dagger_{+,y} (k+\pi)  c_{+,y'} (k')    \right>  &   \left<  c^\dagger_{+,y} (k+\pi)  c_{+,y'+\pi} (k')  \right> \end{array}  \right) \\\fl
		  && = \,  \bra{\Psi_0}  \left( \begin{array}{c}  e^{E(\kappa)} c^\dagger_+(k)  \\  e^{E(\kappa+\pi)} c^\dagger_+(k+\pi) \end{array}  \right)   \left( \begin{array}{cc} e^{-E(\kappa')} c_+(k')  & e^{-E(\kappa'+\pi)} c_+(k'+\pi) \end{array}  \right)  \ket{\Psi_0}
\end{eqnarray}
where $E(\kappa) \, \equiv \, y  \varepsilon (\kappa) - i N \tilde{\varepsilon}(\kappa)$; notice that we have used $c^\dagger_-(k) = c^\dagger_+(k+\pi)$.

Thus, focusing on the fermion creation/annihilation operators, it is clear that one needs to be able to take into account the conjugation by $T_{\rm e}^{\pm 1/2}$. A little calculation shows that
\begin{equation}\fl
	\begin{array}{ccc}
		T_{\rm e}^{1/2}  c^\dagger(k) T_{\rm e}^{-1/2}  \, = \, v^\dagger(k) \left( \begin{array}{c}  c^\dagger(k) \\ c^\dagger(k+\pi)   \end{array} \right)      &    &     T_{\rm e}^{-1/2}  c^\dagger(k) T_{\rm e}^{1/2}  \, = \, v^\dagger(k+\pi) \left( \begin{array}{c}  c^\dagger(k) \\ c^\dagger(k+\pi)   \end{array} \right)    \\
		T_{\rm e}^{1/2}  c(k) T_{\rm e}^{-1/2}  \, = \, \left( \begin{array}{cc}  c(k) & c(k+\pi)   \end{array} \right)  v(k+\pi)     &    &     T_{\rm e}^{-1/2}  c(k) T_{\rm e}^{1/2}  \, = \,  \left( \begin{array}{cc}  c(k) & c(k+\pi)   \end{array} \right)  v(k)  \\ \\
		{\rm with } \quad v(k) \, = \, \left(  \begin{array}{c}   \frac{a+b+c}{2 \sqrt{a(b+c)}} +\frac{c+b-a}{2 \sqrt{a(b+c)}}  \cos k    \\   \frac{c+b-a}{2 \sqrt{a(b+c)}} \sin k  \end{array} \right) \, .
	\end{array}
\end{equation}
Let us focus on an example: the case $y \in 2\mathbb{Z} + \frac{1}{2}$, $y' \in 2\mathbb{Z} + \frac{1}{2}$ with $y \geq y'$. The propagator in $k$-space is
\begin{eqnarray}\fl
\nonumber	\left<  c^\dagger_{y} (k) c^\dagger_{y'} (k')  \right> & = & \frac{  \bra{\Psi_0} T_{\rm e}^{1/2} T^{2(\frac{N-y-1/2}{2})} \, T_{\rm e}^{1/2} c^\dagger(k)  T_{\rm e}^{-1/2}  \, T^{2(\frac{y-y'}{2})} \, T_{\rm e}^{1/2} c(k')  T_{\rm e}^{-1/2}  \, T^{2(\frac{N+y'+1/2}{2})}   T_{\rm e}^{1/2}  \ket{\Psi_0}   }{\bra{\Psi_0} T_{\rm e}^{1/2} T^{2N} T_{\rm e}^{1/2}  \ket{\Psi_0} } \\
\nonumber	&=&  v^\dagger (k) U^\dagger(k)   \left(  \begin{array}{cc}    \left<  c_{+,y}^\dagger(k) c_{+,y'}(k')  \right>  &  \left< c_{+,y}^\dagger(k) c_{-,y'}(k')  \right>  \\  \\ \left<  c_{-,y}^\dagger(k) c_{+,y'}(k')  \right>  &  \left<  c_{-,y}^\dagger(k) c_{-,y'}(k')  \right>    \end{array}  \right) U(k') v(k'+\pi) \\
\nonumber		&=&    \bra{\Psi_0} v^\dagger (k) U^\dagger(k)  \left( \begin{array}{c}  e^{E(\kappa)} c^\dagger_+(k)  \\  e^{E(\kappa+\pi)} c^\dagger_+(k+\pi) \end{array}  \right)   \left( \begin{array}{cc} e^{-E(\kappa')} c_+(k')  & e^{-E(\kappa'+\pi)} c_+(k'+\pi) \end{array}  \right)U(k') v(k'+\pi)  \ket{\Psi_0}  \, .
\end{eqnarray}
In the second line, we used the change of basis given by $U(k)$ in Eq. (\ref{eq:diag_6v}), and in the third line we used (\ref{eq:prop_pm_6v_b}), with $E(\kappa) \equiv (y+\frac{1}{2}) \varepsilon(\kappa) - i  N \tilde{\varepsilon}(\kappa) $ (the shift $+1/2$ comes from the power of the transfer matrix, which involves $y+\frac{1}{2}$ instead of $y$). Note also  that $U(k+\pi)=U(k)$ for the six vertex model. We then have the following 'identity'. Whenever the fields $c^\dagger_y(k)$ and $c_y(k)$ appear in a bracket $\left< . \right>$, we may replace it by the combination
\begin{equation}
	\left\{  \begin{array}{rcl}   	c^\dagger_y (k) & \rightarrow &  v^\dagger (k) U^\dagger(k)  \left( \begin{array}{c}  e^{E(\kappa)} c^\dagger_+(k)  \\  e^{E(\kappa+\pi)} c^\dagger_+(k+\pi) \end{array}  \right)    \\ \\
		c_y (k') & \rightarrow &   \left( \begin{array}{cc} e^{-E(\kappa')} c_+(k')  & e^{-E(\kappa'+\pi)} c_+(k'+\pi) \end{array}  \right)U(k') v(k'+\pi) 
	\end{array} \right.
\end{equation}
and evaluate the r.h.s in the DWIS $\ket{\Psi_0}$. 
When we come back to real-space, the 'identity' can be simplified,
\begin{eqnarray}\fl
	\nonumber (y \in 2 \mathbb{Z}+ \frac{1}{2}) \qquad  c^\dagger_{x,y} & \rightarrow &  \int_{-\pi}^\pi \frac{dk}{2\pi}  e^{-i k x} v^\dagger (k) U^\dagger(k)  \left( \begin{array}{c}  e^{E(\kappa)} c^\dagger_+(k)  \\  e^{E(\kappa+\pi)} c^\dagger_+(k+\pi) \end{array}  \right)   \\
	\nonumber	  &&= \,  \int_{-\pi}^\pi \frac{dk}{2\pi} e^{- i k x} \, v^\dagger(k) U^\dagger (k)  \left( \begin{array}{c}  e^{E(k)}   c^\dagger_+(k)  \\ 0  \end{array} \right)  +  \int_{-\pi}^\pi \frac{dk}{2\pi} e^{- i k x} \, v^\dagger(k) U^\dagger (k)  \left( \begin{array}{c}  0 \\  e^{E(k+\pi)} c^\dagger_+(k+\pi)  \end{array} \right)    \\	
	\nonumber	  &&= \,  \int_{-\pi}^\pi \frac{dk}{2\pi} e^{- i k x} \, v^\dagger(k) U^\dagger (k)  \left( \begin{array}{c}  e^{E(k)}   c^\dagger_+(k)  \\ 0  \end{array} \right)  +  \int_{-\pi}^\pi \frac{dk}{2\pi} e^{- i (k-\pi) x} \, v^\dagger(k-\pi) U^\dagger (k)  \left( \begin{array}{c}  0 \\  e^{E(k)} c^\dagger_+(k)  \end{array} \right)    \\
	  &&= \,  \int_{-\pi}^\pi \frac{dk}{2\pi} e^{- i k x} \, \left[ v^\dagger(k) U^\dagger (k)  \left( \begin{array}{c} 1  \\ 0 \end{array} \right) + v^\dagger(k-\pi) U^\dagger (k)  \left( \begin{array}{c} 0  \\ e^{i \pi x}  \end{array} \right)  \right]     e^{E(\kappa)}   c^\dagger_+(k) \, .
\end{eqnarray}
It is convenient to distinguish the cases $x \in 2 \mathbb{Z}+ \frac{1}{2}$ and $x \in 2 \mathbb{Z}- \frac{1}{2}$, for which $e^{i \pi x} = \pm i$. All the possible cases are summarized in table \ref{tab:table_6v_cp_appendix}.  

\begin{table}
	\begin{tabular}{c||c|c}
	$r_{x,y}(k)$	&  $y \in 2\mathbb{Z} + \frac{1}{2}   $		&   $y \in 2\mathbb{Z} - \frac{1}{2} $  \\  \hline \hline  & & \\
	$x \in 2\mathbb{Z} + \frac{1}{2}$	& $ e^{\frac{\varepsilon (\kappa)}{2}}  \left[ v^\dagger(k) U^\dagger (k)  \left( \begin{array}{c} 1  \\ 0 \end{array} \right) + v^\dagger(k-\pi) U^\dagger (k)  \left( \begin{array}{c} 0  \\ e^{i \pi x}  \end{array} \right) \right] $ &  $ e^{-\frac{\varepsilon (\kappa)}{2}} \left[  v^\dagger(k+\pi) U^\dagger (k)  \left( \begin{array}{c} 1  \\ 0 \end{array} \right) + v^\dagger(k) U^\dagger (k)  \left( \begin{array}{c} 0  \\ e^{i \pi x}  \end{array} \right) \right]$  \\
		 &  $\displaystyle = \, e^{\frac{\varepsilon (\kappa)}{2}}  \left[ \frac{a+b+c}{2 \sqrt{a(b+c)}}  e^{ -i  \frac{k-\kappa}{2} } +  \frac{c+b-a}{2 \sqrt{a(b+c)}} e^{-i \frac{k+\kappa}{2} } \right] $    &  $\displaystyle = \, e^{-\frac{\varepsilon (\kappa)}{2}}  \left[ \frac{a+b+c}{2 \sqrt{a(b+c)}}  e^{ -i  \frac{k-\kappa}{2} } -  \frac{c+b-a}{2 \sqrt{a(b+c)}} e^{-i \frac{k+\kappa}{2} } \right]$  \\  &  &  \\  \hline  & &  \\
		$x \in 2\mathbb{Z} - \frac{1}{2}$ &  $e^{\frac{\varepsilon (\kappa)}{2}} \left[ v^\dagger(k) U^\dagger (k)  \left( \begin{array}{c} 1  \\ 0 \end{array} \right) + v^\dagger(k-\pi) U^\dagger (k)  \left( \begin{array}{c} 0  \\ e^{i \pi x}  \end{array} \right) \right] $ &  $ e^{-\frac{\varepsilon (\kappa)}{2}}  \left[ v^\dagger(k+\pi) U^\dagger (k)  \left( \begin{array}{c} 1  \\ 0 \end{array} \right) + v^\dagger(k) U^\dagger (k)  \left( \begin{array}{c} 0  \\ e^{i \pi x}  \end{array} \right) \right] $ \\
		 &  $\displaystyle = \,e^{\frac{\varepsilon (\kappa)}{2}}  \left[ \frac{a+b+c}{2 \sqrt{a(b+c)}}  e^{ i  \frac{k-\kappa}{2} } +  \frac{c+b-a}{2 \sqrt{a(b+c)}} e^{i \frac{k+\kappa}{2} } \right] $    &  $\displaystyle = \, e^{-\frac{\varepsilon (\kappa)}{2}}  \left[ \frac{a+b+c}{2 \sqrt{a(b+c)}}  e^{ i  \frac{k-\kappa}{2} } -  \frac{c+b-a}{2 \sqrt{a(b+c)}} e^{i \frac{k+\kappa}{2} } \right]$  \\ 
	\end{tabular} \\ \vspace{0.8cm}

		\begin{tabular}{c||c|c}
	$s_{x,y}(k)$	&  $y \in 2\mathbb{Z} + \frac{1}{2}   $		&   $y \in 2\mathbb{Z} - \frac{1}{2} $  \\  \hline \hline  & & \\
	$x \in 2\mathbb{Z} + \frac{1}{2}$	& $ e^{-\frac{\varepsilon (\kappa)}{2}}  \left[  \left( \begin{array}{cc}  1 & 0 \end{array} \right)  U(k)  v(k+\pi) + \left( \begin{array}{cc}  0 & e^{-i \pi x} \end{array} \right)  U(k)  v(k)     \right]  $ &  $ e^{\frac{\varepsilon (\kappa)}{2}} \left[  \left( \begin{array}{cc}  1 & 0 \end{array} \right)  U(k)  v(k) + \left( \begin{array}{cc}  0 & e^{-i \pi x} \end{array} \right)  U(k)  v(k-\pi)     \right]   $  \\
		 &  $\displaystyle = \, e^{-\frac{\varepsilon (\kappa)}{2}}  \left[ \frac{a+b+c}{2 \sqrt{a(b+c)}}  e^{ i  \frac{k-\kappa}{2} } -  \frac{c+b-a}{2 \sqrt{a(b+c)}} e^{i \frac{k+\kappa}{2} } \right] $    &  $\displaystyle = \, e^{\frac{\varepsilon (\kappa)}{2}}  \left[ \frac{a+b+c}{2 \sqrt{a(b+c)}}  e^{ i  \frac{k-\kappa}{2} } +  \frac{c+b-a}{2 \sqrt{a(b+c)}} e^{i \frac{k+\kappa}{2} } \right]$  \\  &  &  \\  \hline  & &  \\
		$x \in 2\mathbb{Z} - \frac{1}{2}$ &  $e^{-\frac{\varepsilon (\kappa)}{2}} \left[  \left( \begin{array}{cc}  1 & 0 \end{array} \right)  U(k)  v(k+\pi) + \left( \begin{array}{cc}  0 & e^{-i \pi x} \end{array} \right)  U(k)  v(k)     \right]  $ &  $ e^{\frac{\varepsilon (\kappa)}{2}}  \left[  \left( \begin{array}{cc}  1 & 0 \end{array} \right)  U(k)  v(k) + \left( \begin{array}{cc}  0 & e^{-i \pi x} \end{array} \right)  U(k)  v(k-\pi)     \right]   $ \\
		 &  $\displaystyle = \,e^{-\frac{\varepsilon (\kappa)}{2}}  \left[ \frac{a+b+c}{2 \sqrt{a(b+c)}}  e^{ -i  \frac{k-\kappa}{2} } -  \frac{c+b-a}{2 \sqrt{a(b+c)}} e^{-i \frac{k+\kappa}{2} } \right] $    &  $\displaystyle = \, e^{\frac{\varepsilon (\kappa)}{2}}  \left[ \frac{a+b+c}{2 \sqrt{a(b+c)}}  e^{- i  \frac{k-\kappa}{2} } +  \frac{c+b-a}{2 \sqrt{a(b+c)}} e^{-i \frac{k+\kappa}{2} } \right]$  \\ 
	\end{tabular}

	\caption{Functions $r_{x,y}(k)$ and $s_{x,y}(k)$ appearing in Eq. (\ref{eq:ident_6v}).}
	\label{tab:table_6v_cp_appendix}
\end{table}

It turns out the formulas can be simplified even further. Let us demonstrate how so for $r_{x,y}$ in the case $x,y\in 2\mathbb{Z}+1/2$ (all other cases will be slight variants of this). First, we rewrite $r_{x,y}(k)$ as
\begin{equation*}
 r_{x,y}(k)=\frac{1}{\sqrt{a}} e^{\frac{\varepsilon(\kappa)-ik}{2}}\left[\cos \frac{\kappa}{2}+i \frac{a}{\sqrt{b+c}}\sin \frac{\kappa}{2}\right],
\end{equation*}
and then compute its square. Using $e^{\varepsilon(\kappa)-ik}=\frac{a\cos \kappa-i c \sin \kappa}{c+b\cos \kappa}$, we obtain
\begin{eqnarray*}
 \left[r_{x,y}(k)\right]^2&=&\frac{a\cos \kappa-i c \sin \kappa}{c+b\cos \kappa} \frac{\left[(b+c)\cos \frac{\kappa}{2}+i a \sin \frac{\kappa}{2}\right]^2}{a(b+c)}\\
 &=&\frac{a\cos \kappa-i c \sin \kappa}{c+b\cos \kappa} \frac{b+c\cos \kappa+i a \sin \kappa}{a}\\
 &=&1-i\frac{b}{a}\sin \kappa
\end{eqnarray*}
It is then easy to see that the correct square root is
\begin{equation}
 r_{x,y}(k)=\sqrt{1-i\frac{b}{a}\sin \kappa}.
\end{equation}
Repeating the same procedure on all possible cases, we find in the end the simple result
\begin{equation}
 r_{x,y}(k)=s_{x,y}(k)=\sqrt{1-i \sigma \frac{b}{a}\sin \kappa}\qquad,\qquad \sigma=(-1)^{x-y}.
\end{equation}

\newpage
 \section*{References}
\bibliography{arctic.bib}{}
 \bibliographystyle{h-physrev}
\end{document}